%% file: Thesis_Reset_January_2022.tex
\documentclass[12pt, oneside]{report}
\usepackage[utf8]{inputenc}
\usepackage{graphicx}
\usepackage{anyfontsize}
\usepackage[letterpaper,width=160mm,top=25mm,bottom=25mm,left=30mm]{geometry}
\graphicspath{{Imagenes/}}
\usepackage{caption}
\usepackage{tabularx}
\usepackage{subcaption}
\usepackage[labelfont=bf, font = footnotesize]{caption}
\usepackage{fancyhdr}
\usepackage{mathrsfs}
\usepackage{amsmath}
\usepackage{url}
\usepackage{booktabs}
\usepackage{hyperref}
\hypersetup{
	colorlinks,
	linktocpage,
	citecolor=blue,
	filecolor=black,
	linkcolor=blue,
	urlcolor=black
}
\usepackage{setspace}
\usepackage{color}
\usepackage{floatrow}
\usepackage{amssymb} %For real symbol
\usepackage{placeins} %For FloatBarrier
\usepackage[toc]{appendix} %for appendix

\begin{document}
\thispagestyle{empty}
\begin{figure}
	\centering
	\includegraphics[width=0.2\linewidth]{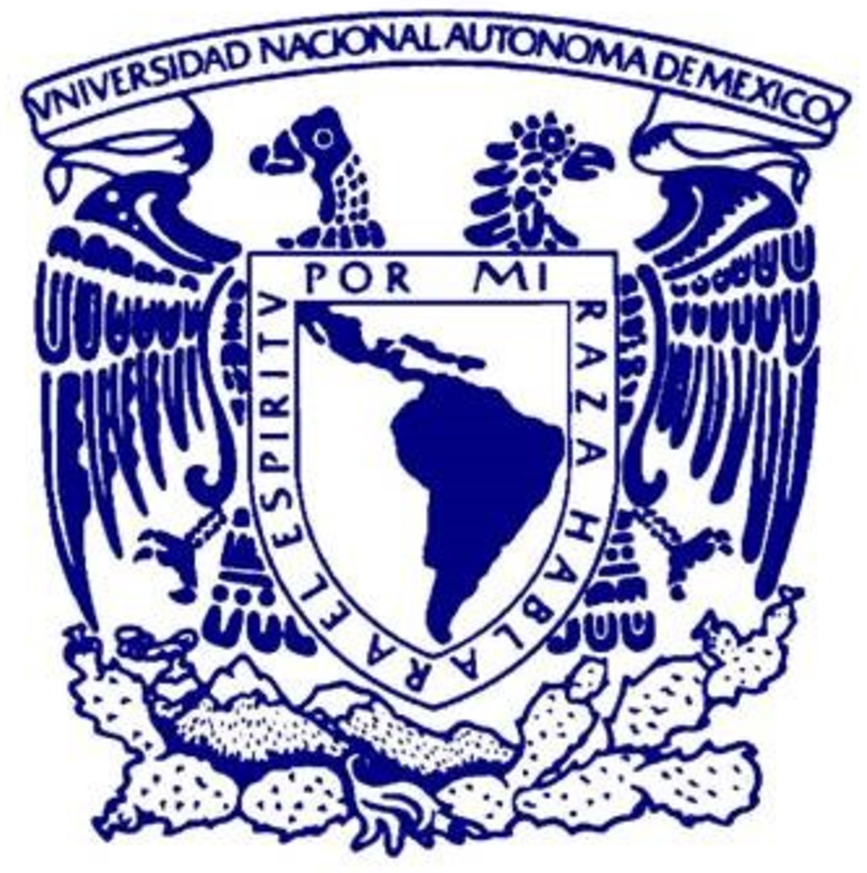}
	\label{fig:escudo}
\end{figure}

\begin{center}
	\textbf{{\large UNIVERSIDAD NACIONAL AUTÓNOMA DE MÉXICO}}\\
\medskip	{\large Maestría en Ciencias (Física) }\\%Obligatorio
\medskip	{\large Instituto de Física}\\%Opcional
%\medskip	{\large FÍSICA ESTADÍSTICA Y SISTEMAS COMPLEJOS}\\%Opcional
\vspace{1.5cm}
{\Large Random walks on networks with stochastic reset to multiple nodes}\\
\vspace{1cm} 
	\bigskip Tesis\\
QUE PARA OPTAR AL GRADO DE:\\
Maestra en Ciencias (Física)
	\\
	\bigskip PRESENTA:\\
Fís. Fernanda Hern\'andez Gonz\'alez
	\\
\vspace{1cm}
	\medskip DIRECTOR DE TESIS:\\
Dr. Alejandro P\'erez Riascos\\
	Instituto de Física\\
\vspace{1cm}
	\medskip COMITÉ TUTOR:\\
	Dr. Denis Boyer\\
	Instituto de Física\\
	Dr. Hernán Larralde Ridaura\\
	Instituto de Ciencias Físicas
\\
\vspace{1cm}
CIUDAD DE MÉXICO, MÉXICO, FEBRERO 2022
\end{center}
\newpage
\input{Chapters/abstract}
%\chapter*{Acknoledgments}

\input{Chapters/acknowledgments}

\tableofcontents
%\listoffigures
%\listoftables

\newpage

\input{Chapters/long_intro}

\input{Chapters/Chap_Intro.tex}

\input{Chapters/Chap_Reset1.tex}
\input{Chapters/Chap_Reset2.tex}
\input{Chapters/Chap_ResetM.tex}
\input{Chapters/Chap_Conc.tex}

%\input{Chapters/appendix.tex}

%\input{}

%\appendix
%\chapter{Appendices}
\input{Chapters/appendix.tex}

%\bibliographystyle{myunsrt}
%\bibliography{References}

%\appendix
%\chapter{Apendix}
%\input{Chapters/appendix.tex}

\end{document}

%% file: Chapters/abstract.tex
\chapter*{Abstract}
\addcontentsline{toc}{chapter}{Abstract}

In this research, we study Markovian random walks with resetting performed in networks. We deduce different analytical results for the stationary distribution and mean first passage time (MFPT) for different random walk models, these quantities help us explore the efficiency and the effect of resetting to multiple nodes in networks. We present a method to obtain these values in terms of the eigenvalues and eigenvectors of the transition matrix without resetting.

The formalism is applied to one, two, and an arbitrary number of resetting nodes. To test the method, for one and two resetting nodes, we study the MFPT for finite and infinite rings using exact known spectral properties. We also explore L\'evy flights on rings and the tendency of values towards the infinity limit. For an arbitrary number of resetting nodes, we analyze the dynamics on Cayley trees, a random walk to visit a distribution of points in a continuous space, and the \textit{Google} search strategy in interacting cycles.

%First, the fundamentals of stochastic processes and networks is introduced considering the concepts that are relevant to this approach. The dynamics of random walks in networks is explained in terms of the master equation, deducing the stationary distribution and MFPT expressed with the eigenvalues, eigenvectors left and right of the transition matrix.

%In consecutive chapters, the stochastic resetting is reviewed for one and two nodes, calculating the properties and validating the method using circulant graphs. Finally, the notation and method is generalized to resetting to $\mathcal{M}$ number of nodes and applied to Cayley trees and randomly distributed points. On interacting cycles we implement the Google search strategy and compare the global MFPT.

%% file: Chapters/acknowledgments.tex
\chapter*{Acknowledgments}
\addcontentsline{toc}{chapter}{Acknowledgments}

I would like to thank CONACYT for the scholarship which meant a huge economical support during difficult times and  acknowledge Ciencia de Frontera
2019 (CONACYT) project, “Sistemas complejos estocásticos:
Agentes móviles, difusión de partículas, y dinámica de
espines” (Grant No. 10872).

I'll be forever thankful to my principal supervisor, Dr. Alejandro Pérez Riascos for his constant support and encouragement, same goes for the tutoring committee, Dr. Denis Boyer and Dr. Hernán Larralde Ridaura, and the evaluating jury: Dr. Thomas Gorin, Dr. Octavio Reymundo Miramontes Vidal, Dra. Yuriria Cortés Poza and Dr. Leonardo Dagdug Lima  for their comments aimed at improving this thesis and their time.

To my family and friends, for their support and advice.

%% file: Chapters/long_intro.tex
\chapter*{Introduction}
\addcontentsline{toc}{chapter}{Introduction}

Complex systems can be found everywhere in nature, and its accurate description includes many scientific areas. Powerful and abstract tools are necessary to study the emergent properties of these systems. When the system is described as elements and interactions, we can use networks to represent it.  In this framework, the adjacency matrix encodes all the information about the nodes and links connecting pairs of nodes. In addition, different network properties are used to classify and determine the main features of the system. Furthermore, we can define dynamical processes on networks, such as random walks and Lévy flights for the analysis of transport in discrete structures.

Something as intricate as the human mobility in a city can be abstracted and modeled by a network using these principles and analyze them. We can, for example, optimize the exploration of the entire network, reduce the time it takes to move from one location to another using a strategy, or find a particular target node in the shortest time possible, among many others.

A particular dynamic procedure to optimize network exploration is stochastic resetting. This implies returning to a certain location with probability $\gamma$. In this document, we apply this strategy to standard random walks and Lévy flights, restoring to one and two nodes initially. Afterwards, this procedure is generalized to consider $\mathcal{M}$ resetting nodes, finding a simple equation for the mean first passage time in terms of the eigenvalues and eigenvectors of the transition matrix without resetting, starting from the master equation of the process.

First, in Chapter \ref{Chapter_StoNet}, we review the fundamentals of stochastic processes and networks, an essential part of this work. We begin with the basis of random processes and quantities that measure globally their behaviors, such as mean, variance, and stationary distribution. We also review the statistical background for random walks and Lévy flights. Next, we introduce the adjacency matrix and several properties that characterize networks. We present a detailed study of circulant matrices, with rings being a particular case. The formalism of random walks on networks in terms of the master equation is also discussed.

In Chapter \ref{RefChapterReset1}, we present the method to study random walks on networks with resetting to one node. We review rings, as, in this case, the exact eigenvalues and eigenvectors of the transition matrix are known. We consider resetting to one node, both for the classical random walk and Lévy flights. To assess the efficiency of network exploration, we calculate the stationary distribution and mean first passage time. Also, we explore the limit where the number of nodes in the ring tends to infinity. Following an analogous procedure, using the same structures and taking similar limits, in Chapter \ref{ch:two_nodes} we analyze resetting to two nodes, calculating the stationary distribution and the mean first passage time between pairs of nodes. The principal objective is to introduce the reader to a method that can be extended to include multiple resetting nodes.

Finally, in Chapter \ref{ch:M_reset} we generalize the method so that resetting to $\mathcal{M}$ nodes is possible and apply it to Cayley trees, random distributions of points in space, and interacting cycles. In the latter case, we implement the \textit{Google} search strategy, where the resetting is made to all the nodes in the network.

%% file: Chapters/Chap_Intro.tex
\chapter{Stochastic processes and networks}
\label{Chapter_StoNet}
\section{Introduction}
We are surrounded by complex systems, from the intricate  structure of the internet to the brain composed by a web of interconnected neurons, from the multi-factorial stock exchange to the organization of a bee-hive. This is a motivation to study and model their behaviors and understand these phenomena, with the ultimate goal of obtaining accurate simulations and predictions. In this case, graph theory and network science have an important role because in this formalism it is possible to abstract the structure of a complex system and represent it in the language of mathematics. Furthermore, it is possible with numerical simulations to study dynamical processes on networks and analyze their evolution in time. In this introductory chapter, we review the definition of stochastic processes and some examples such as random walks and Lévy flights; also, we will explore the basis of graph theory, and finally discuss diffusive transport processes taking place on networks and their mathematical formalism.

\section{Stochastic processes}

Randomness permeates all natural processes at all scales, therefore is important to study and understand its effect in different models. To describe such behavior, we need to introduce the concept of \textit{random variable}, denoted by $X$. This is a mathematical object defined by a set of possible values and a probability distribution $P_{X}(x)$ \cite{kampen2007stochastic}. If we apply a mapping $f$ to the random variable, we obtain other random variables.

Given a probability space $(\Omega, \mathcal{F},P_{X})$, where $\Omega$ is the set of all possible outcomes of the random variable, $\mathcal{F}$ is the event space and $P_{X}$ the probability function, we define a \textit{stochastic process} as any collection of random variables defined on the probability space \cite{florescu2014probability_stochastic}. We can denote it as $\{X(t) : t\in \mathcal{I}\}$ where $\mathcal{I}$ is the index set, which orders the succession of the random variables \cite{florescu2014probability_stochastic}.

We can apply a mapping $f$ to the random variable $X$ and let $t$ be the time,  such that $Y_{X}(t)=f(X,t)$. A \textit{realization} is when the random variable takes a single value $X=x$ and we have a function only of time $Y_{x}(t)=f(x,t)$ and then the process is formed by an ensemble of realizations \cite{kampen2007stochastic}.

Since stochastic processes have a random component, different realizations of the process take different values, thus we need quantities that summarize information and give an insight into the behavior in time. For instance, the \textit{average} is defined as \cite{kampen2007stochastic}
\begin{equation}
    \langle Y(t) \rangle=\int Y_{x}(t)P_{X}(x)dx
\end{equation}
which is integrated over the random variable, alternatively we can have also an average over time. Fluctuations also can occur, a measure of such variations is the variance calculated as \cite{kampen2007stochastic}
\begin{equation}
    \sigma^2(t)=\langle Y^2(t) \rangle-\langle Y(t)\rangle^2.
\end{equation}
A stochastic process which is said to be Markovian asserts that the distribution at time $t+1$ depends only on the state $t$ \cite{kijima1997markov_asseertsMarkov}. Particularly, for discrete time steps, we take a set of successive times $t_{1}<t_{2}<\dots<t_{n}$ then the Markovian property is expressed as \cite{kampen2007stochastic}
\begin{equation}
    P_{1|n-1}(y_{n},t_{n}|y_{1},t_{1};\dots;y_{n-1},t_{n-1})=P_{1|1}(y_{n},t_{n}|y_{n-1},t_{n-1})
\end{equation}
which formally states that the conditional  probability density at time $t_{n}$ of the random variable $Y=y_{n}$ depends only on the values of the immediate previous step $y_{n-1}$ and $t_{n-1}$. Taking a sequence of Markov steps, a Markov chain is formed. This conditional probability is directly linked to transition probabilities, which will be addressed later on. This type of processes are said to be \textit{memoryless} since the future state will depend only on the latest information available on the state of the system, not on the path followed to get there \cite{motwani1995randomized_rw}. The stationary distribution of a Markov chain $\mathbf{P}^\infty$ with transition probability matrix $\mathbf{W}$ is a row vector that satisfies $\mathbf{P}^\infty=\mathbf{P}^\infty\mathbf{W}$ \cite{kampen2007stochastic}, so if the process follows the stationary distribution at a given time $t$, in consecutive times the same behavior remains, so it is interpreted as a steady-state condition \cite{motwani1995randomized_rw}.

Several physical phenomena can be described by a time-dependent random process. Some examples are diffusion \cite{pavliotis2014stochastic,kallianpur2014stochastic_diffusion}, Brownian motion, surface growth \cite{PhysRevE.55.6501_surface,Cuerno2004STOCHASTICDE_surface}, stochastic noise \cite{10.1117/12.2087559_stocastic_noise}, bounded drift \cite{mahnke2009physics}, nuclear scattering by simple liquids \cite{shuler2009stochastic_liquids}, linear response \cite{pavliotis2014stochastic}, nucleation in supersaturated vapors \cite{mahnke2009physics}, turbulent dispersion \cite{Durbin1983StochasticDE_turbulence} among many others. In the next sections we will analyze in detail two particular examples.

\subsection{Random walks}

A very important concept that will be widely discussed throughout this thesis is the random walk. The main idea is intuitive and can lead to interesting results. A one-dimensional random walk in discrete time consists of a walker that flips a fair coin and moves one step to the right or to the left, depending on the result of the coin toss \cite{lawler2010random}. In this way, the direction taken at each step is independent of the direction of the previous ones. The term was coined by K. Pearson in 1905 \cite{PEARSON1905}. It is a stochastic process and can be modeled by random variables, so if $x$ is the initial position and $S_{n}$ is its position after $n$ steps (or equivalent, after time $n$) then the random walk can be described as
\begin{equation}
    S_{n} = x+X_{1}+\dots+X_{n},
\end{equation}
where $X_{j}=\pm1$ is a random discrete variable. Besides a pure probabilistic treatment, a random walk can be modeled by a Markov chain.

The most notorious example of random walks in nature is the Brownian motion, first observed by Robert Brown as the movement of particles floating in a medium \cite{brown_2015}. In the continuous-time limit, it is proved to follow a diffusion equation \cite{sethna2006statistical_diffusion_solved,Klafter_2005_brownian_motion}
\begin{equation} \label{eq:diffusion}
    \frac{\partial p}{\partial t}=D\frac{\partial^2p}{\partial x^2}.
\end{equation}
Here $p=p(x,t)$ is the probability density of finding the particle in the $x$ position at time $t$, $D$ is the diffusion coefficient. If the value of $x$ is not bounded positively and negatively, the equation can be solved using the Fourier transform or Green functions, considering the initial conditions of $P(x,0)=\delta(x)$ the solution is a normal (Gaussian) distribution \cite{kampen2007stochastic}
\begin{equation}
    p(x,t)=\frac{1}{\sqrt{4\pi Dt}}e^{-\frac{x^2}{4Dt}},
\end{equation}
where the corresponding standard deviation is $\sigma=\sqrt{2Dt}$, a relation that implies that the mean-squared displacement is proportional to time \cite{Klafter_2005_brownian_motion}. This linear dependency is considered as a threshold to determine if the process is subdiffusive, normal, or superdiffusive, the first and third qualities are considered anomalous diffusion.

\subsection{Lévy flights}\label{subsec:levy}

A discrete random walk has a fixed step length, but we can have a dynamical process that includes variable step size, in particular its distribution $p(x)$ can be chosen as a power law with infinite variance, of the form \cite{FractionalBook2019} 
\begin{equation}
    p(x)\sim |x|^{-1-\alpha}
\end{equation}
for large $x$, and where $\alpha$ is a real parameter known as the Lévy index \cite{sandev2019fractional_levy} that takes values between $0<\alpha<2$.

In this stochastic process, given that the variance of the distribution is infinite, there is a high probability of making large jumps \cite{klages2008anomalous_selfsimilar}. So we have two types of dynamics, local for the smaller steps and non-local for larger steps where the walker gets far away from clusters formed when the step-length is shorter for consecutive times \cite{Metzler_2004_diffusion_levy}. Another consequence of infinite variance is that the Central Limit Theorem is not valid, instead the random variable $y=\sum x_{i}$ follows an $\alpha$-stable distribution \cite{PhysRevLett.91.018302_clt_levy}.

\begin{figure}[!t]
    \centering
    \includegraphics*[width=0.55\textwidth]{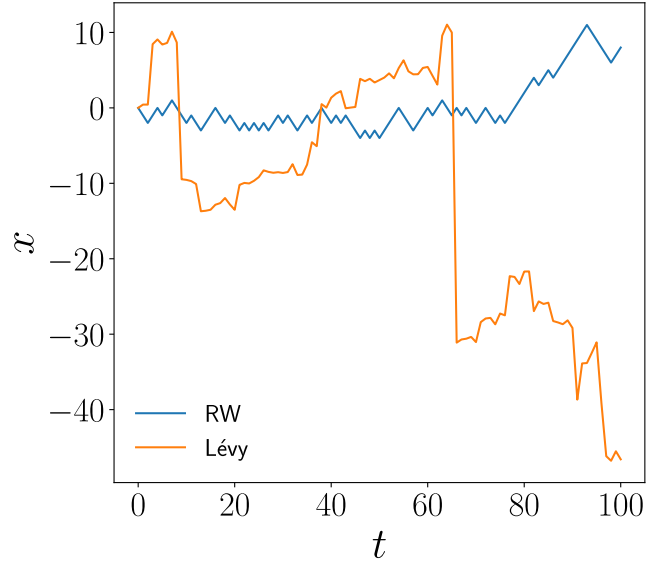}
    \caption{Illustrative comparison of a single realization of a Lévy flight with $\alpha=1.5$ and a random walk in one dimension. As we can see, the explored distance is greater for the Lévy flights.}\label{fig:Levy_RW_1D}
\end{figure}

%\begin{figure}[h!]
%    \centering
%    \includegraphics*[width=0.5\textwidth]{Images/Chap_Intro/Levy_2D.eps}
%    \caption{Illustrative comparison of a single realization of a Lévy flight $\alpha=1.5$ and a random walk in two dimensions for the same number of steps. As we can see, the explored distance is greater for the Lévy flight.}\label{fig:Levy_RW_2D}
%\end{figure}
%\FloatBarrier

When a Lévy flight is performed, a larger area is covered for a single realization compared to the random walk. As we can see in Figure \ref{fig:Levy_RW_1D}, the blue line represents a random walk which shows a more local behavior, while the orange line corresponds to a Lévy flight. This realization shows a characteristic long step that allows to reach further positions. We can guess that due to a Lévy flight, the process diffuses more rapidly producing anomalous transport. Its trajectories show a self-similar structure and its fractal dimension is $\alpha$ \cite{klages2008anomalous_selfsimilar}.

In order to model anomalous diffusion, a fractional calculus approach has been proposed \cite{Metzler_2004_diffusion_levy,sandev2019fractional_levy}. Particularly, for one dimensional Lévy flights with continuous waiting times, the diffusion equation needs to be modified as \cite{METZLER20001_fractional_levy}
\begin{equation}
    \frac{\partial W}{\partial t}=K^{\alpha}\nabla^{\alpha}W,
\end{equation}
where $W$ is the propagator, a distribution which depends on time $t$ and position $x$, constructed with the distribution of jump length $p(x)$ and a Poissonian distribution that models the waiting time between jumps, $\nabla^{\alpha}$ is the Riesz operator as defined in \cite{Compte1996_Riesz_operator} and $K^{\alpha}$ is a generalized diffusion constant. Its analytical solution is given in terms of the Fox functions \cite{METZLER20001_fractional_levy} and the normal random walk behavior is recovered in the limit $\alpha\to2$. Asymptotically
\begin{equation}
    W(x,t)\sim\frac{K^{\alpha}t}{|x|^{1+\alpha}}.
\end{equation}
Calculating the mean-square displacement with this distribution, it is easy to see that it is infinite $\langle x^{2}(t)\rangle\to\infty$. Meanwhile, the fractional moment, calculated as \cite{METZLER20001_fractional_levy}
\begin{equation}
    \langle |x|^{\delta}\rangle=2\int_{0}^{\infty}x^{\delta}W(x,t)dx\propto t^{\delta/\alpha}
\end{equation}

where $0<\delta<\alpha<2$, is not linear with time, characteristic of anomalous diffusion.
\section{Networks and graph theory}
Often, some systems present in nature can be represented by a network which is a connected graph. In this section, basic definitions from graph theory are reviewed, as well as some examples of graphs and their properties.

A graph $\mathcal{G}=(\mathcal{V},\mathcal{E})$ is composed by a set of vertices or nodes $\mathcal{V}$ and a set of edges $\mathcal{E}$ that link pairs of nodes \cite{FractionalBook2019}. We denote the number of nodes as $N=|\mathcal{V}|$. If the incoming or outgoing direction of the vertices is considered, we have a $\textit{directed}$ graph, if such constriction doesn't exist we have an $\textit{undirected}$ graph, as we will see, this property has implications in its mathematical description. A graph is said to be \textit{weighted} if the edges or nodes have a value.

\begin{figure}[!t]
    \centering
    \includegraphics*[width=0.6\textwidth]{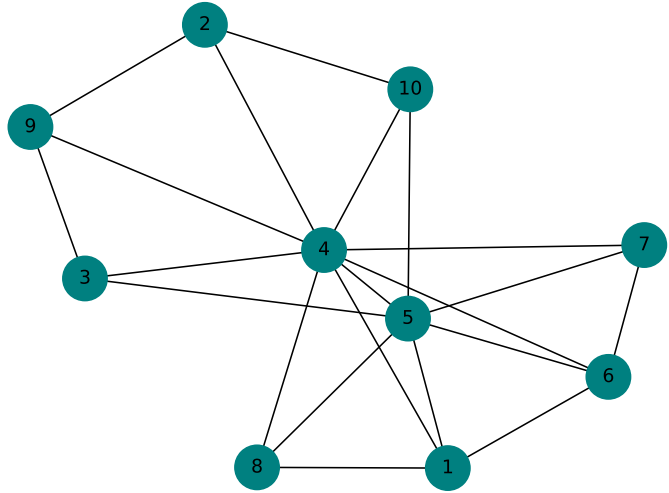}
    \caption{Example of an undirected unweighted network with $N=10$ nodes}\label{fig:random}
\end{figure}

Complex systems can be modeled by a graph whose nodes can represent the constitutive parts of the system and the edges  interactions between the elements. In Figure \ref{fig:random} a graphical representation of an undirected network is shown, the nodes are the colored circles with their respective numbers and the lines represent the edges.

The connectivity of the graph can be encoded in matrix form, called the \textit{adjacency matrix} $\mathbf{A}$. For a network with $N$ nodes, $\mathbf{A}$ has $N\times N$ entries, each $A_{ij}$ can be one or zero, depending if the nodes $i$ and $j$ are connected or not on the network. Therefore, the structure of the network is stored in an abstract mathematical object that can be manipulated easily. The diagonal of $\mathbf{A}$ is zero since there are no self-loops. For undirected networks, the adjacency matrix is symmetric, i.e., $A_{ij}=A_{ji}$ and therefore this matrix is diagonalizable. The \textit{degree} of node $i$, $k_{i}$, defined as the number of connections of the node, can be calculated using the elements in $\mathbf{A}$ with the expression $ k_{i}=\sum_{l=1}^{N}A_{il}$, this relation is valid for undirected networks \cite{FractionalBook2019}. If all the nodes have the same degree, i.e. $k_{i}=k$ for all $i$, we have a \textit{regular} or \textit{k-regular} network. The network diameter $\mathcal{D}$ is the maximum degree of separation between all pairs of vertices \cite{claudius2008complex_small_world}.

We can construct a \textit{path} in a network with a sequence of nodes and edges \cite{FractionalBook2019} in which consecutive nodes are connected by an edge. If for any pair of nodes there is a connecting path, the graph is said to be \textit{connected}. A particular type of path where the initial and final nodes are linked is called a \textit{cycle}. In addition, we can define a \textit{distance} between nodes as the number of edges in the shortest path, in this manner we are determining a metric in terms of the network structure.

The \textit{coordination number} $z$ is defined as the average number of edges per node \cite{claudius2008complex_small_world}, if it is related to the network diameter $\mathcal{D}$ and number of nodes as $z^{\mathcal{D}}\approx N$, then the diameter increases as the logarithm of $N$, and the network has the property of \textit{small-world}. The name comes from a social experiment by S. Milgram \cite{milgram}, where he tried to prove that two individuals in a social network could be connected by a short sequence of consecutive friends or acquaintances, this implies that any two nodes can reach each other in a number of steps much smaller than the total number of nodes.

In the next subsections, we review the main properties of the networks used in this work. Since for their description, a convenient tool is the spectral theory of stochastic matrices, the Dirac notation is adopted for the rest of this document.

\subsection{Circulant graphs} \label{subsec:circulant_graphs}

A special type of graphs with interesting properties are the circulant graphs, where the adjacency matrix is circulant \cite{balakrishnan2004graph_intro}. This type of square matrix with $n\times n$ entries has the following structure \cite{mieghem2011graph_spectra_circulant}
\begin{equation}
\mathbf{C}=
    \begin{pmatrix}
    c_{0} & c_{n-1} & c_{n-2} & \dots & c_{1}\\
    c_{1} & c_{0} & c_{n-1} & \dots &c_{2}\\
    c_{2} & c_{1} & c_{0} & \ddots & c_{3} \\
    \vdots & \vdots & \ddots &\ddots&\vdots \\
    c_{n-1} & c_{n-2} & c_{n-3} & \dots &c_{0}
\end{pmatrix}.
\end{equation}
As we can see, the elements are repeated in each column but shifted one index down so that the last entry becomes the first in the next column.
 
We can write a circulant matrix as \cite{mieghem2011graph_spectra_circulant}
\begin{equation} \label{eq:circulant_poli}
     \mathbf{C} =c_{0}\mathbf{I}+c_{1}\mathbf{E}+\dots + c_{n-1}\mathbf{E}^{n-1} =  \sum_{i=0}^{n-1}c_{i}\mathbf{E}^{i},
\end{equation}
where $\mathbf{E}$ is a circulant matrix called elementary matrix in which $c_{1}=1$ and all the entries are zero and $\mathbf{E}^{0}=\mathbf{I}$ is the identity matrix.
 
The eigenvalues of $\mathbf{C}$ are the Fourier transform of the entries of the matrix and given only in terms of the coefficients $c_{i}$ as \cite{aldrovandi2001special_circulant,mieghem2011graph_spectra_circulant}
\begin{equation}\label{spect_circulant}
    \lambda_{m}=\sum_{q=0}^{n-1}c_{q}e^{(2\pi \mathrm{i}/n)(m-1)q},  
\end{equation}
where $\mathrm{i}=\sqrt{-1}$ is the unit imaginary number. For any circulant matrix $\mathbf{C}$, the eigenvectors are the same as those of the matrix $\mathbf{E}$. Let us denote the $i$-th canonical right eigenvector as $|i\rangle$, a column vector where the $i$-th entry is one and the rest are zero, and the $j$-th canonical left eigenvector as $\langle j|$ which is a row vector with the $j$-th entry one and rest zero. With this notation the components of the eigenvectors of $\mathbf{E}$ are $\langle i |\phi_l\rangle=\frac{1}{\sqrt{n}}e^{-\mathrm{i}\varphi_{l}(i-1)}$ and $\langle \bar{\phi}_{l} |j\rangle=\frac{1}{\sqrt{n}}e^{\mathrm{i}\varphi_{l}(j-1)}$, with $\varphi_{l}=2\pi(l-1)/n$ (see \cite{mieghem2011graph_spectra_circulant} for details).

\begin{figure}[!t]
    \centering
    \includegraphics*[width=0.55\textwidth]{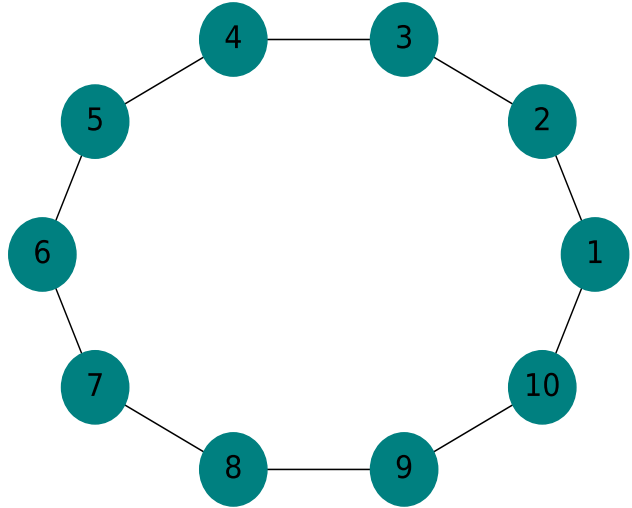}
    \caption{Ring with $N=10$ nodes.}\label{fig:ring}
\end{figure}

One of the simplest circulant graphs is the ring that consists of nodes connected just to its two nearest neighbors in a closed-form, this particular structure is shown in Figure \ref{fig:ring}. The application of the result in Eq. (\ref{spect_circulant}) with $c_1=c_{N-1}=1/2$ allows obtaining the eigenvalues of the transition matrix $\mathbf{W}$ for a random walker in a ring with $N$ nodes \cite{FractionalBook2019}
\begin{equation} \label{eq:ring_eigenvalues}
    \lambda_{l} = \cos\left[\frac{2\pi(l-1)}{N}\right].
\end{equation}

\section{Dynamical processes on networks}
On a network, if we have the rules to move from one node to another we obtain a good representation of transport since the vertices can be locations while edges can model roads. We define a \textit{walk} in the graph as a sequence of vertices in a graph, starting in node $i$  and finishing in node $j$ such that consecutive vertices in the sequence are adjacent \cite{chartrand2012a_walk_def}. If $i=j$ then we have a closed walk, if it ends on a different node the walk is open. The length of the walk is the number of edges encountered. A walk is a \textit{trail} if all of its edges are different and an open trail is a \textit{path} if all of its vertices are different \cite{thulasiraman1992graphs_path_alt}.

Dynamical processes that occur in a network-like structure can be studied by the formalism of the Laplacian matrix $\mathbf{L}$. Its entries are defined as $L_{ij}= k_{i}\delta_{ij}-A_{ij}$ (see Ref. \cite{FractionalBook2019} for a detailed discussion). With this expression, we observe that the non-diagonal elements are negative. An equivalent, more compact expression in terms of matrices is
\begin{equation} \label{eq:def_laplacian}
    \mathbf{L}=\mathbf{K}-\mathbf{A},
\end{equation}
where $\mathbf{K}$ is a diagonal matrix with the respective degree of the nodes. Since it preserves the symmetry property of $\mathbf{A}$ and its entries are real, $\mathbf{L}$ is Hermitian, $\mathbf{L}=(\mathbf{L}^{T})^{*}$. Also, for every row the magnitude of the diagonal entry is equal to the sum of the magnitudes of the off-diagonal entries, this means that $\mathbf{L}$ is a diagonally dominant matrix. This two properties imply that it is a semi-definite positive matrix. The eigenvalues of $\mathbf{L}$ are real, non-negative and their eigenvectors mutually orthogonal \cite{symm_eigen}. Following Dirac notation, let  $\{|\varphi_{j}\rangle\}_{j=1}^{N}$ be the set of eigenvectors and $\{\mu_{j}\}_{j=1}^{N}$ the set of eigenvalues of $\mathbf{L}$, then the spectral form of the Laplacian matrix is
\begin{equation}
    \mathbf{L}=\sum_{j=1}^{N}\mu_{j}|\varphi_{j}\rangle\langle \varphi_{j}|.
\end{equation}
Since the eigenvalues are non-negative, zero is a lower bound for the eigenvalues of $\mathbf{L}$.
%
%% Rev: Connecting paragraph
\subsection{Random walks}
A random walk on a finite network follows a master equation. If the walker is in node $i$, and the probability to jump to any of its neighbors is the same, then the transition probability is $w_{i\to j}=A_{ij}/k_{i}$ \cite{NohRieger2004}. A Markovian random walk is described by a master equation \cite{NohRieger2004,Hughes}
\begin{equation} \label{eq:RW_master}
    P_{ij}(t+1)=\sum_{l=1}^{N}P_{il}(t)\frac{A_{lj}}{k_{l}}=\sum_{l=1}^{N}P_{il}(t)w_{l\to j},
\end{equation}
where $P_{ij}(t)$ is the probability to find the walker at node $j$ at time $t$ starting in node $i$ at time $t=0$ \cite{NohRieger2004}. In the last equality, the transition probability is expressed in a more generalized way as an element of the time-independent transition matrix $\mathbf{W}$. Its entries $w_{i\to j}$ are the conditional probability of visiting node $j$ given the condition of having visited $i$ in the step earlier. Observe that for regular networks, the transition probability matrix is symmetric because all the nodes have the same degree $k_{i}$. In more general networks, $\mathbf{W}$ is not symmetric. The time evolution of $P_{ij}(t)$ for a Markov process can be given in terms of powers of the transition matrix
\begin{equation} \label{eq:trans_prob_eigenrep}
    P_{ij}(t)=\langle i |\mathbf{W}^{t}|j \rangle,
\end{equation}
where $\{|i\rangle\}_{i=1}^{N}$ is the canonical base of $\mathbb{R}^{N}$. This matrix is stochastic and satisfies
\begin{equation}
    \sum_{j=1}^{N}P_{ij}(t)=1,
\end{equation}
this means that the walker stays on the network at all times \cite{FractionalBook2019}.

Additionally, we want the random walker to be able to reach any node of the network, this condition is called \textit{ergodicity}. Strictly, a random walk will be ergodic if there exist a finite number of time-steps $t_{ij}$ for any pair of nodes $i$ and $j$ such that
\begin{equation}
    P_{ij}(t=t_{ij})=\langle i|\mathbf{W}^{t_{ij}}|j\rangle>0,
\end{equation}
in other terms, any node has a non-zero probability of being visited. Moreover, the stationary distribution is obtained given the limit \cite{NohRieger2004}
\begin{equation}
    P_{j}^{\infty}=\lim_{T\to\infty}\frac{1}{T}\sum_{t^\prime=0}^{T}P_{ij}(t^\prime).
\end{equation}
For a random walk in undirected networks, $P_{j}^{\infty}$ is independent of the starting position and completely determined by the degree of the node \cite{NohRieger2004}
\begin{equation} \label{eq:Pinf_no_reset}
    P_{j}^{\infty}=\frac{k_{j}}{\sum_{l=1}^{N}k_{l}}.
\end{equation}
This result coincides with the intuitive idea that a node with various connections will be visited more often.

%**At some point I will have to talk about the intuition behind the mean first passage time**
In addition to the stationary distribution, the mean first passage time (MFPT) is an important way of characterizing a random walk \cite{NohRieger2004}. It is denoted as $\langle T_{ij}\rangle$, and intuitively can be interpreted as the average time it takes a random walker to reach $j$ for the first time, starting from $i$. We can obtain its value from the first-passage probability $F_{ij}(t)$ which is the probability that starting from $i$, the first transition to $j$ occurs at time $t$. This quantity is closely related to $P_{ij}(t)$ through the relation \cite{NohRieger2004}
\begin{equation}
    P_{ij}(t)=\delta_{t0}\delta_{ij}+\sum_{t^{'}=0}^{t}P_{jj}(t-t^{'})F_{ij}(t').
\end{equation}
Applying the discrete Laplace transform $\tilde{f}(s)=\sum_{t=0}^\infty f(t)e^{-st}$, we can obtain $\tilde{F}_{ij}(s)$ with simple algebraic manipulations \cite{NohRieger2004}
\begin{equation} \label{eq:Laplace_trans_F}
    \tilde{F}_{ij}(s)=\frac{\tilde{P}_{ij}(s)-\delta_{ij}}{\tilde{P}_{jj}(s)}.
\end{equation}
On the other hand, by definition, the MFPT can be written in terms of the $F_{ij}(t)$ as
\begin{equation}
    \langle T_{ij}\rangle = \sum_{t=0}^{\infty}tF_{ij}(t),
\end{equation}
if we derive \eqref{eq:Laplace_trans_F} with respect to the parameter $s$ in the limit $s\to 0$ we find that
\begin{equation}
    \langle T_{ij}\rangle=-\frac{d\tilde{F}_{ij}}{ds}\bigg\rvert_{s=0}.
\end{equation}
Expanding $\tilde{P}_{ij}(s)$ in powers of $s$ and substituting in \eqref{eq:Laplace_trans_F}, the final result is \cite{NohRieger2004}
\begin{equation} \label{eq:MFTP_moments}
    \langle T_{ij}\rangle=\frac{\mathcal{R}_{jj}^{(0)}-\mathcal{R}_{ij}^{(0)}+\delta _{ij}}{P_j^{\infty}},
\end{equation}
where $\mathcal{R}_{ij}^{(n)}$ are the respective moments given by
\begin{equation} \label{eq:R_moments}
    \mathcal{R}_{ij}^{(n)}=\sum_{t=0}^{\infty}t^{n}(P_{ij}(t)-P_{j}^{\infty}).
\end{equation}
Alternatively, the eigenvalues and eigenvectors of $\mathbf{W}$ can be used to find the MFPT as shown in \cite{asesor_tesis}. Assume that we know the eigenvalues $\lambda_{i}$ as well as the right $|\phi_{i}\rangle$ and left $\langle \bar{\phi}_{i}|$ eigenvectors which satisfy
\begin{equation}
    \mathbf{W}|\phi_{i}\rangle=\lambda_{i}|\phi_{i}\rangle, \qquad
    \langle \bar{\phi}_{i}|\mathbf{W}=\lambda_{i}\langle \bar{\phi}_{i}|.
\end{equation}
Since $\mathbf{W}$ is stochastic, its eigenvalues can be ordered such that $\lambda_{1}=1$ is the maximum value. In this way, we can obtain the spectral representation of the transition matrix and when substituted in \eqref{eq:trans_prob_eigenrep} %it was found that \cite{asesor_tesis}
\begin{equation} \label{eq:trans_prob_left_right}
    P_{ij}(t)= \sum_{l=1}^{N}\lambda_{l}^{t}\langle i|\phi_{l}\rangle \langle\bar{\phi}_{l}|j\rangle,
\end{equation}
therefore the stationary distribution is $P_{j}^{\infty}=\langle i|\phi_{1}\rangle \langle\bar{\phi}_{1}|j\rangle$. Substituting \eqref{eq:trans_prob_left_right} in \eqref{eq:R_moments}, summing over $t$ and finally substituting in \eqref{eq:MFTP_moments} we obtain \cite{RiascosMateos2012}
\begin{equation} \label{eq:MFPT_vecs}
    \langle T_{ij}\rangle=\frac{\delta_{ij}}{\langle i|\phi_{1}\rangle
    	\langle \bar{\phi}_{1}|j\rangle}+\sum_{l=2}^{N}\frac{1}{1-\lambda_{l}}\frac{
    \langle j|\phi_{l}\rangle
    \langle \bar{\phi}_{l}|j\rangle-
    \langle i|\phi_{l}\rangle
    \langle \bar{\phi}_{l}|j\rangle
    }{\langle i|\phi_{1}\rangle
    \langle \bar{\phi}_{1}|j\rangle}.
\end{equation}
%
%MENTION HOW WE CAN USE COMPUTER TO CALCULATE THIS AND ITS BENEFICES
Observe that this expression, unlike \eqref{eq:MFTP_moments}, does not involve an infinite sum, now it only depends on the eigenvalues, left and right eigenvectors of the transition matrix $\mathbf{W}$. This is extremely useful because its calculation is straightforward with a computer program. For networks with a few nodes, the diagonalization process might take little time, but for bigger systems, the time of execution grows algebraically as $N^3$. If we want only one calculation, the diagonalization can take just a few seconds, but if we need to calculate it as a function of a parameter, the total time could add up considerably.
\subsection{Lévy flights}
%
%% Connection with circulant matrix

In subsection \ref{subsec:levy}, we explored the basic idea behind a Lévy flight in one dimension. In this section, we will assess this process when performed in a network. The main difference from a normal random walk is that now it is possible to jump to nodes that are not just first neighbors. L\'evy flights on an arbitrary graph can be generated by taking powers of the Laplacian matrix (defined in Eq. \eqref{eq:def_laplacian}). $\mathbf{L}^{\alpha}$ is called the fractional Laplacian of a graph with $0<\alpha<1$, this allows non-null transitions between sites in the network \cite{RiascosMateosFD2015}, even if they are not connected with an edge.

In this formalism, the transition probabilities are given by \cite{RiascosMateosFD2014}
\begin{equation}\label{wijfrac}
w_{i\to j}(\alpha)=\delta_{ij}-\frac{(\mathbf{L}^\alpha)_{ij}}{(\mathbf{L}^\alpha)_{ii}}\qquad 0<\alpha< 1.
\end{equation}
In particular, for $\alpha\to 1$ one recovers the simple random walk with transitions to nearest-neighbor nodes. The transition probabilities in Eq. (\ref{wijfrac}) with $0<\alpha<1$ define a L\'evy flight. Particularly on rings, a circulant network we reviewed in section \ref{subsec:circulant_graphs}, we have that $w_{i\to j}(\alpha)\sim d_{ij}^{-(1+2\alpha)}$, where the distance $d_{ij}$ is the length of the shortest path  between $i$ and $j$, and where $d_{ij}\gg1$ (see Refs. \cite{RiascosMateos2012,RiascosMateosFD2015,FractionalBook2019} for a detailed discussion on L\'evy flights and fractional transport on networks).

If the Lévy flight is performed in a finite ring, the eigenvectors remain the same because the adjacency matrix is circulant and all circulant matrices have the same eigenvectors as a consequence of Eq. \eqref{eq:circulant_poli}, while the eigenvalues are modified as \cite{FractionalBook2019}
\begin{equation} \label{eq:eigen_levy_ring}
    \lambda_l(\alpha)=1-\frac{1}{k^{(\alpha)}}\, \left(2-2\cos\varphi_l\right)^\alpha
\end{equation}
where $\varphi_l=\frac{2\pi}{N}(l-1)$ and $k^{(\alpha)}$ is the fractional degree, defined as \cite{FractionalBook2019}
\begin{equation}\label{eq:degreeGLring}
k^{(\alpha)}=\frac{1}{N}\sum_{l=1}^N \left(2-2\cos\varphi_l\right)^\alpha.
\end{equation}
Since we have an exact expression for the eigenvalues and eigenvectors, we can use the ring structure to analyze the method proposed to calculate the MFPT in the following chapters and compare the results to test its validity.

Lévy flights have proved to be efficient in search and movement strategies, for example, to optimize in encounter rate for animals in a predator-prey dynamics  \cite{VISWANATHAN2002208_search,Bartumeus2002_encounter}, in animal and human foraging  \cite{Boyer2004_foraging,2018_human_foraging,VISWANATHAN20001_foraging}, movement patterns performed by boats while fishing \cite{Edwards2011_boats}, movement strategy to avoid extinction and maximize population \cite{Dannemann2018_extintion}, as a first approximation to the migration of chemokines within lymphoid tissues crucial to the optimal working mechanism of the immune system \cite{Harris2012_tcells}, and also it has been used as an improvement mechanism for the metaheuristic optimization bat algorithm \cite{sym11070925_bat}. Not all the consequences of Lévy flights are beneficial, since they also accelerate the propagation of diseases to a world-wide scale \cite{Hufnagel2004_disease,Janssen1999_disease}.

%% file: Chapters/Chap_Reset1.tex
\chapter{Random walks with resetting on networks}
\label{RefChapterReset1}
\section{Introduction}

As stated in Chapter \ref{Chapter_StoNet}, a random walker can explore a network and eventually return to the initial node. However, it is possible to relocate randomly the walker to a particular node with probability $\gamma$ at every step. This process is called \textit{resetting} and has a fundamental role in transport processes on networks. This type of dynamic was initially studied in \cite{evans2011diffusion}, were the authors considered the diffusion of a particle which stochastically  resets to its initial position at a constant rate $r$ in continuous time using the master equation formalism. They found that the position does not follow a Gaussian distribution, as it would happen with a normal random walk, and that the MFPT becomes finite for $0<r<\infty$\footnote{For pure continuous random walks, we can use the first passage time in \cite{redner2001a} pp. 23 and calculate the MFPT as the first moment, the result is divergent.}, these results are direct effects of the resetting process. A natural resetting dynamic is found on birth-death processes when there is an absorbing state or queueing systems where an event sets the queue to zero length \cite{EvansReview2019}.

\begin{figure}[!t]
    \centering
    \includegraphics*[width=0.6\textwidth]{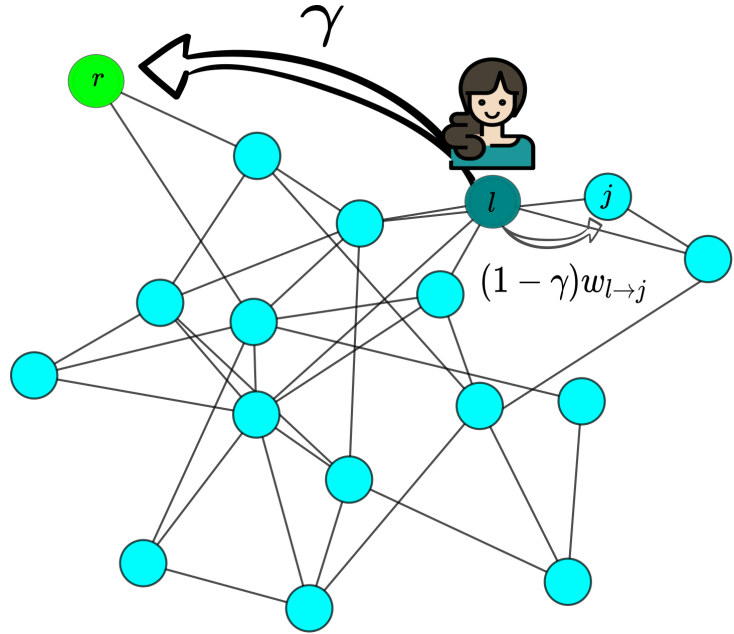}
    \caption{A tourist wants to explore the particular places in a city but returns to the hotel with probability $\gamma$. How can the exploration of the network be optimized?}\label{fig:network_one}
\end{figure}

Particularly, as shown in Figure \ref{fig:network_one}, let us imagine a tourist visiting a city and who wants to explore certain important historical places in the span of a week, but can return to the hotel with probability $\gamma$ at each time step. A question is: how does this resetting affect the efficiency of network exploration?

The importance of this problem is not restricted to tourism strategies, it is more general and linked to processes where it is necessary to explore the space and find particular targets such as in animal foraging \cite{ViswaBook2011}, public transportation \cite{RiascosMateosSciRep2020}, ranking and searching in databases \cite{LeskovecBook2014}, target search of proteins on DNA molecules \cite{PhysRevLett.103.138102}, label propagation in machine learning algorithms \cite{Bautista2019}, defining the relevance score between two nodes in graph mining \cite{Tong2007}, Brownian motion \cite{Evans2011JPhysA,MajumdarPRE2015,EvansReview2019}, models of anomalous diffusion \cite{kusmierz2014first,metzler_anomalous}, processes with a drift \cite{montero2013monotonic}, among many others. These problems, to some degree, can be modeled as a classical random walk. On the other hand, quantum walks have been in the spotlight the past few years because of the progress in the implementation of a hybrid quantum-classical Page-Rank \cite{6975444_quantum_page_rank,SnchezBurillo2012} and quantum search algorithms \cite{Magniez2011_quantum_searh}.

Local classical random walks on complex networks were studied by Noh and Rieger \cite{NohRieger2004}, establishing a formalism that considers the structure of the network, which are key to understanding for example, human mobility \cite{Szell2010HumanMobility}, spread of diseases in epidemics \cite{PU2015230_epidemics_noh_citation,Satorras2015_epidemic,Meloni2012ModelingEpidemics} and information diffusion in social networks \cite{Wang2014InformationDI,InformationEntropyDiffusion2014}. Only recently, stochastic resetting was introduced in the context of quantum walks on networks in \cite{PhysRevE.103.012122_quantum_resetting_networks}, where the authors propose a Hamiltonian with a parameter that interpolates between a classical and quantum behavior.

\FloatBarrier

Considering these reasons, we proceed to analyze the effect of resetting in a classical random walk on networks. In the next section, we will present the theory following the results introduced in \cite{ResetNetworks_PRE2020} to show how quantities like the stationary distribution and the MFPT are affected by resetting to the initial node, applying the formalism to rings. In this structure, we will consider local random walks and Lévy flights, emphasizing the differences. Later on, we will explore the limit when the number of nodes  $N\to\infty$ recovering the infinite one-dimensional lattice and the corresponding asymptotic behavior.
\section{Random walks with stochastic resetting to one node}
\label{RW_reset1}
As stated before, a random walk in a network follows a master equation \eqref{eq:RW_master}, where $w_{l\to j}$ are elements of the transition matrix, independent of time. If the process involves the return to node $r$ with a resetting probability $\gamma$ at each time step, the corresponding master equation is
\begin{equation}
 \label{mastereq1}
 P_{ij}(t+1;r,\gamma) = (1-\gamma)\sum_{l=1}^N P_{il}(t;r,\gamma)w_{l\to j}+\gamma\delta_{rj},
\end{equation}
here $P_{ij}(t;r,\gamma)$ denotes the probability to find the walker in $j$ at time $t$, given the initial position $i$, resetting node $r$ and resetting probability $\gamma$. It is directly verifiable that for $\gamma=0$, the master equation (\ref{eq:RW_master}) is recovered. In addition, due to the second term, it is not possible to use \eqref{eq:MFPT_vecs} directly to calculate the MFPT. Instead, in \cite{ResetNetworks_PRE2020}, the master equation is rewritten in terms of a new transition matrix $\mathbf{\Pi}(r;\gamma)$
\begin{equation}\label{mastermarkov}
P_{ij}(t+1;r,\gamma) = \sum_{l=1}^N  P_{il} (t;r,\gamma) \pi_{l\to j}(r;\gamma),
\end{equation}
where the matrix $\mathbf{\Pi}(r;\gamma)$  is constructed with two other matrices, the transition matrix without resetting, $\mathbf{W}$, and $\mathbf{\Theta}(r)$ with elements $\Theta_{lm}(r)=\delta_{mr}$ \cite{ResetNetworks_PRE2020} which represent the resetting part. The complete expression for  $\mathbf{\Pi}(r;\gamma)$ is
\begin{equation}\label{Def_MatPi_R1}
\mathbf{\Pi}(r;\gamma)=(1-\gamma)\mathbf{W}+\gamma \mathbf{\Theta}(r).
\end{equation}
This matrix is stochastic, as well as $\mathbf{W}$. In this form, Eq. \eqref{eq:MFPT_vecs} can be applied, but it is not always the best option since for large networks, the direct calculation for each value of $\gamma$ of the eigenvalues $\zeta_l(r;\gamma)$, right $\left|\psi_l(r;\gamma)\right\rangle$ and left $\left\langle\bar{\psi}_l(r;\gamma)\right|$ eigenvectors can be computationally expensive, and if the resetting probability or resetting node is changed, the diagonalization must be redone. Instead, in \cite{ResetNetworks_PRE2020}, the description of the random walker with resetting is made in terms of the eigenvalues $\lambda_{l}$, right eigenvectors $|\phi_{l}\rangle$ and left eigenvectors $\langle\bar{\phi}_{l}|$ of the random walk process without resetting $\mathbf{W}$.

Following the procedure in \cite{ResetNetworks_PRE2020}, the eigenvalues of the matrix $\mathbf{\Pi}(r;\gamma)$ are
\begin{equation}\label{eigvals_zeta}
\zeta_l(r;\gamma)=
\begin{cases}
1 \qquad &\mathrm{for}\qquad l=1,\\
(1-\gamma)\lambda_l \qquad &\mathrm{for}\qquad l=2,3,\ldots, N,
\end{cases}
\end{equation}
independent of the resetting node $r$. The left eigenvector corresponding to the first eigenvalue is a linear combination of the left eigenvectors of $\mathbf{W}$
\begin{equation}\label{psil1}
\left\langle\bar{\psi}_1(r;\gamma)\right|=\left\langle\bar{\phi}_1\right|
+\sum_{m=2}^N\frac{\gamma}{1-(1-\gamma)\lambda_m}\frac{\left\langle r|\phi_m\right\rangle}{\left\langle r|\phi_1\right\rangle}\left\langle\bar{\phi}_m\right|,
\end{equation}
whereas the others are equal $\left\langle\bar{\psi}_l(r;\gamma)\right|=\left\langle\bar{\phi}_l\right|$ for $l=2,\ldots,N$. In the case of the right eigenvectors, the correspondence is different, because now the linear combination is for $l=2,\dots,N$
\begin{equation}\label{psirl_reset}
\left|\psi_l(r;\gamma)\right\rangle=
\left|\phi_l\right\rangle-\frac{\gamma}{1-(1-\gamma)\lambda_l}\frac{\left\langle r|\phi_l\right\rangle }{\left\langle r|\phi_1\right\rangle}\left|\phi_1\right\rangle,
\end{equation}
whereas the equality holds for $|\psi_{1}(r;\gamma)\rangle=|\phi_{1}\rangle$. This set of eigenvectors are orthonormal and satisfy the completeness relation \cite{ResetNetworks_PRE2020}. Consequently, the spectral representation of the transition matrix is
\begin{equation}\label{Pi_prob_powert}
\mathbf{\Pi}(r;\gamma)=\sum_{l=1}^N\zeta_l(r;\gamma)\left|\psi_l(r;\gamma)\right\rangle\left\langle\bar{\psi}_l(r;\gamma)\right|,
\end{equation}
which is particularly useful for calculating the occupation probability $P_{ij}(t;r,\gamma)$ by substituting in Eq. \eqref{eq:trans_prob_eigenrep}, the expression obtained is \cite{ResetNetworks_PRE2020}
\begin{equation}
P_{ij}(t;r,\gamma)=P_j^\infty(r;\gamma)+\sum_{l=2}^N(1-\gamma)^t\lambda_l^t\left[\left\langle i|\phi_l\right\rangle \left\langle\bar{\phi}_l|j\right\rangle-\gamma\frac{\left\langle r|\phi_l\right\rangle \left\langle\bar{\phi}_l|j\right\rangle}{1-(1-\gamma)\lambda_l} \right], \label{Pijspect}
\end{equation}
where the first term is the stationary distribution with resetting
\begin{equation}\label{Pinfvectors_1}
P_j^\infty(r;\gamma)=\langle i|\psi_{1}(r;\gamma)\rangle \langle \bar{\psi}_{1}(r;\gamma)|j\rangle=P_j^\infty+\gamma\sum_{l=2}^N\frac{\left\langle r|\phi_l\right\rangle \left\langle\bar{\phi}_l|j\right\rangle}{1-(1-\gamma)\lambda_l},
\end{equation}
and $P_j^\infty=\langle i|\phi_{1}\rangle\langle\bar{\phi}_{1}|j\rangle$ is the stationary distribution without resetting. For $\gamma=0$ the second term cancels out and we recover the original stationary distribution without resetting. Now, the MFPT can be calculated with the same expression as in Eq. \eqref{eq:MFTP_moments}, using as starting point the expression %
\begin{equation}\label{Tij_R}
\langle T_{ij}(r;\gamma)\rangle=\frac{\mathcal{R}_{jj}^{(0)}(r;\gamma)-\mathcal{R}_{ij}^{(0)}(r;\gamma)+\delta _{ij}}{P_j^{\infty}(r;\gamma)},
\end{equation}
considering the resetting node $r$ and resetting probability $\gamma$ as parameters, where the moments are
\begin{equation}\label{Rmoments_def}
\mathcal{R}^{(n)}_{ij}(r;\gamma)\equiv \sum_{t=0}^{\infty} t^n ~ \{P_{ij}(t;r,\gamma)-P_j^\infty(r;\gamma)\}.
\end{equation}
Manipulating Eqs. \eqref{Pinfvectors_1} to \eqref{Rmoments_def} and simplifying we get
\begin{equation} \label{MFPT_resetSM}
\left\langle T_{ij}(r;\gamma)\right\rangle=\frac{\delta_{ij}}{P_j^\infty(r;\gamma)}\\+
\frac{1}{P_j^\infty(r;\gamma)}\sum_{\ell=2}^N\frac{
\left\langle j|\phi_\ell\right\rangle \left\langle\bar{\phi}_\ell|j\right\rangle-\left\langle i|\phi_\ell\right\rangle \left\langle\bar{\phi}_\ell|j\right\rangle
}{1-(1-\gamma)\lambda_\ell}.
\end{equation}
There is something important to note in this equality, compared to Eq. \eqref{eq:MFPT_vecs}, we see that the expression and dependencies change, since now the eigenvalues appear explicitly and the value of $\left\langle T_{ij}(r;\gamma)\right\rangle$ depends also of the resetting node $r$ and the resetting probability $\gamma$ as expected.

In the next section, we will apply this procedure to a simple ring network, and compare the results for a walker performing a normal random walk and a Lévy flight.

\section{Dynamics with resetting on rings}

In the first chapter, subsection \ref{subsec:circulant_graphs}, we reviewed the general properties of ring networks, the corresponding adjacency matrix is circulant \cite{mieghem2011graph_spectra_circulant}, their exact eigenvalues (Eq. \eqref{eq:ring_eigenvalues}) and their left and right eigenvectors have analytical known expressions. This is relevant to our study because we can calculate exact expressions of the stationary distribution and MFPT, and compare them with the numerical results, therefore verifying the proposed method.
\\

\subsection{Random walks}
Substituting the eigenvalues and considering the projections $\langle i |\phi_l\rangle=\frac{1}{\sqrt{N}}e^{-\mathrm{i}\varphi_{l}(i-1)}$ and $\langle \bar{\phi}_{l} |j\rangle=\frac{1}{\sqrt{N}}e^{\mathrm{i}\varphi_{l}(j-1)}$, $\varphi_{l}=2\pi(l-1)/N$ in Eq. \eqref{Pinfvectors_1}, the stationary distribution for the ring with resetting to the initial node $r=i$ is \cite{ResetNetworks_PRE2020}
\begin{equation} \label{eq:stat_one_ring}
    P_{j}^{\infty}(i;\gamma)= \frac{1}{N}+\frac{\gamma}{N}\sum_{l=2}^{N}\frac{\cos(\varphi_{l}d_{ij})}{1-(1-\gamma)\cos(\varphi_{l})},
\end{equation}
where $d_{ij}$ is the distance between nodes $i$, $j$ and only the real parts of the projections are considered, since imaginary parts are canceled in the sum. We can also calculate the exact expression for the MFPT using Eq. \eqref{MFPT_resetSM}, the final result is \cite{ResetNetworks_PRE2020}
\begin{equation} \label{eq:MFPT_one_ring}
    \left\langle T_{ij}(i;\gamma)\right\rangle=\frac{1}{P_j^\infty(i;\gamma)}\Bigg[\delta_{ij}+\sum_{l=2}^N\frac{1-\cos(d_{ij}\varphi_{l})}{1-(1-\gamma)\cos(\varphi_{l})}\Bigg].
\end{equation}
as we can see both Eqs. \eqref{eq:stat_one_ring} and \eqref{eq:MFPT_one_ring} have an explicit dependence on the distance between nodes. Observe that for $i=j$, which corresponds to the mean first return time, the MFPT is the inverse of the stationary distribution just like in the case without resetting, this is due to the Kac's lemma on the mean recurrence time of discrete processes \cite{kac1947notion}.

Interesting results can be obtained when the limit $N\to\infty$ is taken, where the equivalent structure is the infinite one-dimensional lattice. In this scenario, the sums become integrals over a continuous variable with differential $d\varphi=\frac{2\pi}{N}$, therefore the stationary distribution is \cite{ResetNetworks_PRE2020}
\begin{equation} \label{eq:stat_one_ring_limit}
    P_j^{\infty}(i;\gamma)=\frac{\gamma}{2\pi}\int_{0}^{2\pi}\frac{1-\cos(d_{ij}\varphi)}{1-(1-\gamma)\cos(\varphi)}d\varphi=\sqrt{\frac{\gamma}{2-\gamma}}\Bigg[\frac{\sqrt{(2-\gamma)\gamma}+1}{1-\gamma}\Bigg]^{-d_{ij}}.
\end{equation}
For small $\gamma$, Eq. \eqref{eq:stat_one_ring_limit} follows an exponential distribution \cite{ResetNetworks_PRE2020}
\begin{equation} \label{eq:ring_one_limit_RW}
    P_j^{\infty}(i;\gamma)\approx\frac{\sqrt{2\gamma}}{2}e^{-\sqrt{2\gamma}d_{ij}},
\end{equation}
using the aproximations $\sqrt{\frac{\gamma}{2-\gamma}}\approx \frac{\sqrt{2\gamma}}{2}$ and $\log((\sqrt{(2-\gamma)\gamma}+1)/(1-\gamma))\approx \sqrt{2\gamma}$. The result in Eq. \eqref{eq:ring_one_limit_RW} is     similar to the non-equilibrium steady state of Brownian motion in one dimension \cite{evans2011diffusion}. Following a completely  analogous procedure, the MFPT can be obtained in the same limit
\begin{equation}
    \left\langle T_{ij}(i;\gamma)\right\rangle=
    \begin{cases}
        \sqrt{\frac{2-\gamma}{\gamma}}&\text{if $i=j$}\\
        \frac{1}{\gamma}\Big(\frac{\sqrt{(2-\gamma)\gamma}}{1-\gamma}\Big)^{d_{ij}}-\frac{1}{\gamma}&\text{if $i\neq j$},
    \end{cases}
\end{equation}
now, the MFPT has an exponential dependency of $d_{ij}$. Considering the limiting case $\gamma<<1$ and $d_{ij}>0$, the MFPT takes the form \cite{ResetNetworks_PRE2020} $\langle T_{ij}\rangle\approx \frac{1}{\gamma}[e^{\sqrt{2\gamma}d_{ij}}-1]$, where we can find the critical point taking the partial derivative of the asimptotic approximation of $\langle T_{ij}\rangle$ with respect to $\gamma$ and making it equal to zero, finding that $\gamma^{*}\simeq 1.26982/d_{ij}^{2}$ in the limit $d_{ij}>>1$ \cite{ResetNetworks_PRE2020}.

\begin{figure}[!t]
    \centering
    \includegraphics*[width=\textwidth]{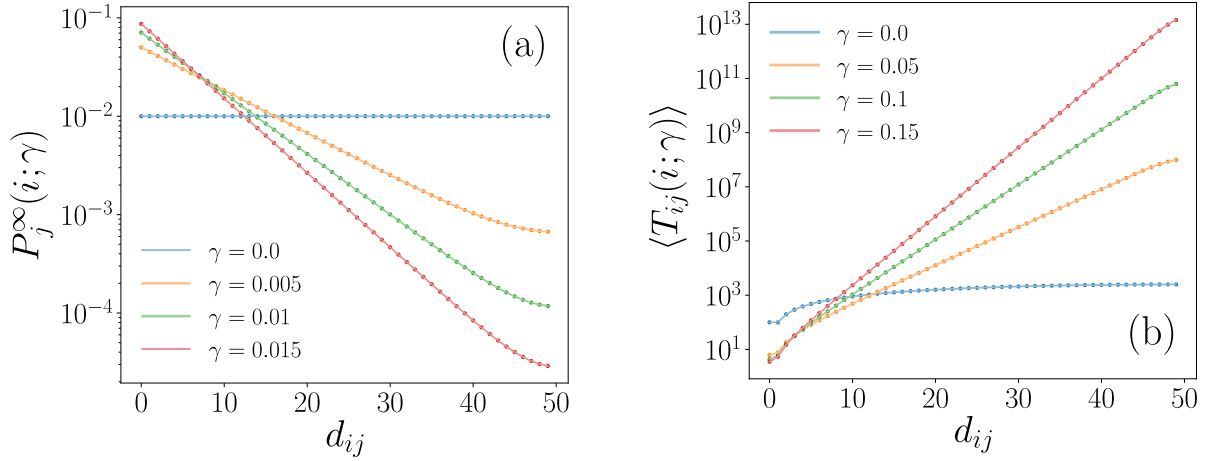}
    \caption{(a) Stationary distribution  and (b) MFPT for a normal random walk performed in a ring with $N=100$ nodes and resetting to the initial node $r=i$ for $\gamma=0.0,0.05,0.1,0.15$ as a function of the distance $d_{ij}$}\label{fig:ring_RW_one}
\end{figure}

In Figure \ref{fig:ring_RW_one} we can see the results as a function of the distance $d_{ij}$. The network structure is a simple ring with $N=100$ nodes and different resetting probabilities $\gamma$ to the initial node $r=i$. We observe two different calculations in this figure, the continuous line represents the result from the calculation using the eigenvalues, left and right of the $\mathbf{\Pi}(r;\gamma)$ substituting them in Eq. \eqref{eq:MFPT_vecs}, while the dots are calculated with the left and right eigenvectors of $\mathbf{W}$ and using Eq. \eqref{MFPT_resetSM}.

Observing Eq. \eqref{eq:Pinf_no_reset} it is clear that for a regular network, the stationary distribution without resetting is a constant value. This corresponds to the blue line in Figure \ref{fig:ring_RW_one}(a), recovering effectively the normal walk for $\gamma=0$. Since we are using a semi-logarithmic scale, the straight lines in the rest of the curves indicate exponential behavior \cite{ResetNetworks_PRE2020}. For Figure \ref{fig:ring_RW_one}(b) we observe that the MFPT increases as the distance increases, contrary to when $\gamma=0$ where there's no resetting, where the MFPT stays in the same magnitude order.

\subsection{Lévy flights}
In the past subsection, we showed the results of a normal random walk where the walker can pass from node $i$ to $j$, one of its neighboring nodes, as long as there is a connection, that is, if $A_{ij}=1$. For Lévy flights in the same structure, the walker can hop with long-range displacements following an inverse power-law distribution for the length of each step.

\begin{figure}[!t]
    \centering
    \includegraphics*[width=\textwidth]{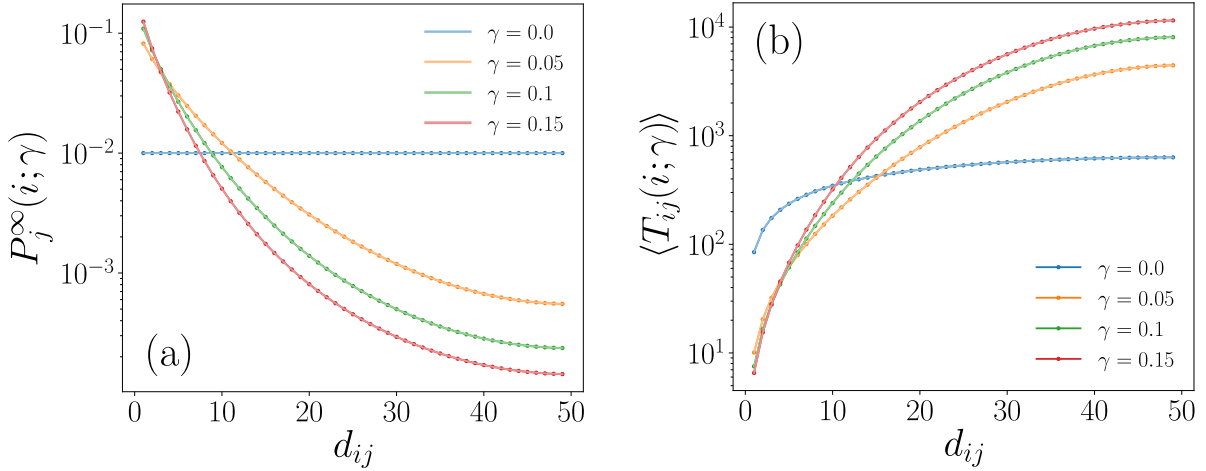}
    \caption{(a) Stationary distribution  and (b) MFPT for a Lévy flight $\alpha=0.75$ performed in a ring with $N=100$ nodes and resetting to $i=0$ for $\gamma=0.0,0.05,0.1,0.15$ as a function of the distance $d_{ij}$}\label{fig:ring_Levy_one}
\end{figure}

Since the structure of the ring is the same and the adjacency matrix is defined by a circulant matrix, the eigenvectors remain the same but the eigenvalues are now given in terms of Eqs. \eqref{eq:eigen_levy_ring} and \eqref{eq:degreeGLring}. We can directly replace these values in Eqs. \eqref{Pinfvectors_1} and \eqref{MFPT_resetSM} obtaining respectively
\begin{equation} \label{eq:stat_one_ring_levy}
    P_{j}^{\infty}(i;\gamma)= \frac{1}{N}+\frac{\gamma}{N}\sum_{l=2}^{N}\frac{\cos(\varphi_{l})d_{ij}}{1-(1-\gamma)\lambda_{l}(\alpha)}
\end{equation}
and
\begin{equation} \label{eq:MFPT_one_ring_levy}
    \left\langle T_{ij}(i;\gamma)\right\rangle=\frac{1}{P_j^\infty(i;\gamma)}\Bigg[\delta_{ij}+\sum_{l=2}^N\frac{1-\cos(d_{ij}\varphi_{l})}{1-(1-\gamma)\lambda_{l}(\alpha)}\Bigg],
\end{equation}
which is valid for  $0<\alpha\leq 1$. Figure \ref{fig:ring_Levy_one} displays the results for a Lévy flight with $\alpha=0.75$ in a simple ring with $N=100$ nodes and for different values of the resetting probability $\gamma$. This figure is analogous to Figure \ref{fig:ring_RW_one} in the sense that the curves are calculated varying the same parameter $\gamma$. Comparing, the stationary distribution (a) encompasses fewer orders than for the normal random walk and does not follow a straight line. In the case of the MFPT (b) something similar happens, the curves span over fewer orders and flatten out for more distant nodes.

The analysis in the limit $N\to\infty$ is more complicated than in the previous case, but simplifies considering that the fractional degree takes the form \cite{RiascosMateosFD2015}
\begin{equation}\label{eq:fractional_degree_levy_infinite}
    k^{(\alpha)}=-\frac{\Gamma(-\alpha)\Gamma(1+2\alpha)}{\pi \Gamma(1+\alpha)}\sin(\pi\alpha),
\end{equation}
in the limit, where $\Gamma(x)$ is the gamma function. This result is obtained analytically, beginning with the $ij$ element of the fractional Laplacian matrix for a finite ring \cite{RiascosMateosFD2015} 
\begin{equation}
    (\mathbf{L}^{\alpha})_{ij}=\frac{1}{N}\sum_{l=1}^{N}\bigg(2-2\cos\bigg[\frac{2\pi}{N}(l-1)\bigg]\bigg)^{\alpha}e^{\frac{2\pi\mathrm{i}}{N}(l-1)d_{ij}}.
\end{equation}
In the limit, the sum is transformed into a an integral in terms of the variable $\theta=\frac{2\pi}{N}(l-1)$, evaluating using a result in \cite{Zoia2007} we get that \cite{RiascosMateosFD2015}
\begin{equation}
    (\mathbf{L}^{\alpha})_{ij}= -\frac{\Gamma(d_{ij}-\alpha)\gamma(1+2\alpha)}{\pi\Gamma(1+\alpha+d_{ij})}\sin(\pi\alpha).
\end{equation}

The fractional degree corresponds to $k^{(\alpha)}=(\mathbf{L}^{\alpha})_{ii}$, obtaining Eq. \eqref{eq:fractional_degree_levy_infinite}.

We are particularly interested in the dependency of the stationary distribution $P_{j}^{\infty}(i;\gamma)$ with the distance to the resetting node $d_{ij}$.

First, for the stationary distribution, we take the limit and convert the sum in Eq. \eqref{eq:stat_one_ring_levy} into an integral with the differential $d\varphi=2\pi/N$, obtaining
\begin{equation}
    P_j^{\infty}(i;\gamma)=\frac{\gamma}{2\pi}\int_{0}^{2\pi}\frac{\cos(d_{ij}\varphi)}{1-(1-\gamma)[1-\frac{(2-2\cos(\varphi))^{\alpha}}{k^{(\alpha)}}]}d\varphi.
\end{equation}
Now, using the trigonometric identity $1-\cos(\varphi)=2\sin^2(\varphi/2)$, this expression transforms into
\begin{equation}
    P_j^{\infty}(i;\gamma)=\frac{\gamma}{2\pi}\int_{0}^{2\pi}\frac{\cos(d_{ij}\varphi)}{1-(1-\gamma)[1-\frac{2^{2\alpha}\sin^{2\alpha}(\varphi/2)}{k^{(\alpha)}}]}d\varphi.
\end{equation}
Changing variables $\theta=\varphi/2$ and the integration limits derived from this, expanding and canceling out terms in the denominator, we get
\begin{equation}
    P_j^{\infty}(i;\gamma)=\frac{1}{2\pi}\int_{0}^{\pi}\frac{\cos(2d_{ij}\theta)}{1+D_{\alpha}\sin^{2\alpha}(\theta)}d\theta,
\end{equation}
where $D_{\alpha}=\frac{2^{2\alpha}}{k^{(\alpha)}}\frac{(1-\gamma)}{\gamma}$ is a constant independent of $\theta$. The denominator can be expanded into a infinite sum, therefore
\begin{equation}
    P_j^{\infty}(i;\gamma)=\frac{1}{2\pi}\sum_{n=0}^{\infty}(-D_{\alpha})^{n}\int_{0}^{\pi}\cos(2d_{ij}\theta)\sin^{2n\alpha}(\theta)d\theta.
\end{equation}
It seems we have a more complicated equation than before because of the infinite sum and its convergence, but we have gained an expression without a denominator and just in terms of trigonometric functions. Considering the limit $x=d_{ij}\gg 1$, this integral form is analytic
\begin{align} \label{eq:integral_sin_cos}
    \int_{0}^{\pi}\cos(2x\theta)\sin^{2n\alpha}(\theta)d\theta&=\frac{2^{-2\alpha n}\pi\cos(\pi x)\Gamma(1+2\alpha n)}{\Gamma(1+\alpha n-x)\Gamma(1+\alpha n+x)}\\&=-2^{-2\alpha n}\sin(\pi \alpha n)\Gamma(1+2\alpha n)\frac{\Gamma(x-\alpha n)}{\Gamma(1+\alpha n+x)},
%\end{equation}
%\end{multline}
\end{align}
where in the last equality, the property $\Gamma(1-z)\Gamma(z)=\pi/\sin(\pi z)$ was used, taking $z=x-\alpha n$. Since we are in the limiting case, for $d_{ij}=x\gg1$ the gamma function can be approximated as $\Gamma(x+b)\approx\Gamma(x)x^{b}$ and the fraction in the last equality of Eq. \eqref{eq:integral_sin_cos} is
\begin{equation}
    \frac{\Gamma(x-\alpha n)}{\Gamma(1+\alpha n+x)}\approx\frac{1}{x^{1+2\alpha n}}
\end{equation}
so finally, the stationary distribution approximates to the leading term
\begin{equation}\label{levy_statP_infring}
    P_{j}^{\infty}(i;\gamma)\approx-\bigg(\frac{1-\gamma}{\gamma}\bigg)\frac{\Gamma(1+\alpha)}{\Gamma(-\alpha)}\frac{1}{d_{ij}^{1+2a}}
\end{equation}
in the asymptotic limit $d_{ij}\gg 1$. Here we have replaced $D_{\alpha}$ to obtain this last equality. So now, it is clear that the asymptotic behavior of the stationary distribution for Lévy flights with resetting in a ring is ruled by a power-law with the distance to the node. This result is consistent with  previous results for continuous Lévy flights on the infinite line \cite{kusmierz2015optimal}.

In addition, applying the same approach to the analysis of the MFPT in Eq. (\ref{eq:MFPT_one_ring_levy}) for L\'evy flights in the limit $N\to \infty$, we have
\begin{equation}\left\langle T_{ij}(i;\gamma)\right\rangle=\frac{1}{\gamma}+\frac{1}{P_j^\infty(i;\gamma)}\left[\delta_{ij}+\frac{1}{2\pi}\int_0^{2\pi}\frac{d\varphi}{1-(1-\gamma)\left[1-\frac{2^{2\alpha}}{k^{(\alpha)}}\sin(\varphi/2)^{2\alpha}\right]}\right].
\end{equation}
However
\begin{align*}&\frac{1}{2\pi}\int_0^{2\pi}\frac{d\varphi}{1-(1-\gamma)\left[1-\frac{2^{2\alpha}}{k^{(\alpha)}}\sin(\varphi/2)^{2\alpha}\right]}=\frac{1}{\pi}\int_0^{\pi}\frac{d\theta}{1-(1-\gamma)\left[1-\frac{2^{2\alpha}}{k^{(\alpha)}}\sin(\theta)^{2\alpha}\right]}\\
&=\frac{1}{\gamma\pi}\int_0^{\pi}\frac{d\theta}{1+\frac{(1-\gamma)}{\gamma}\frac{2^{2\alpha}}{k^{(\alpha)}}\sin(\theta)^{2\alpha}}=\frac{1}{\gamma\pi}\int_0^{\pi}\frac{d\theta}{1+D_\alpha\sin(\theta)^{2\alpha}}=\frac{1}{\gamma}\mathcal{G}(\gamma,\alpha),
\end{align*}
where we have defined $\mathcal{G}(\gamma,\alpha)$ that depends on $\gamma$ and $\alpha$ but is independent of the distance between $i$ and $j$. Therefore
\begin{equation}\left\langle T_{ij}(i;\gamma)\right\rangle=\frac{1}{\gamma}+\frac{1}{P_j^\infty(i;\gamma)}\left[\delta_{ij}+\frac{\mathcal{G}(\gamma,\alpha)}{\gamma}\right].
\end{equation}
In this way, for  $0<\gamma<1$, $1/2\leq\alpha<1$
\begin{equation}\label{levy_Tij_infring}
\left\langle T_{ij}(i;\gamma)\right\rangle\sim d_{ij}^{1+2\alpha},\qquad d_{ij}\gg 1.
\end{equation}
A relation that agrees with the result reported in Ref. \cite{kusmierz2015optimal}.
\begin{figure}[!t]
    \centering
    \includegraphics*[width=0.7\textwidth]{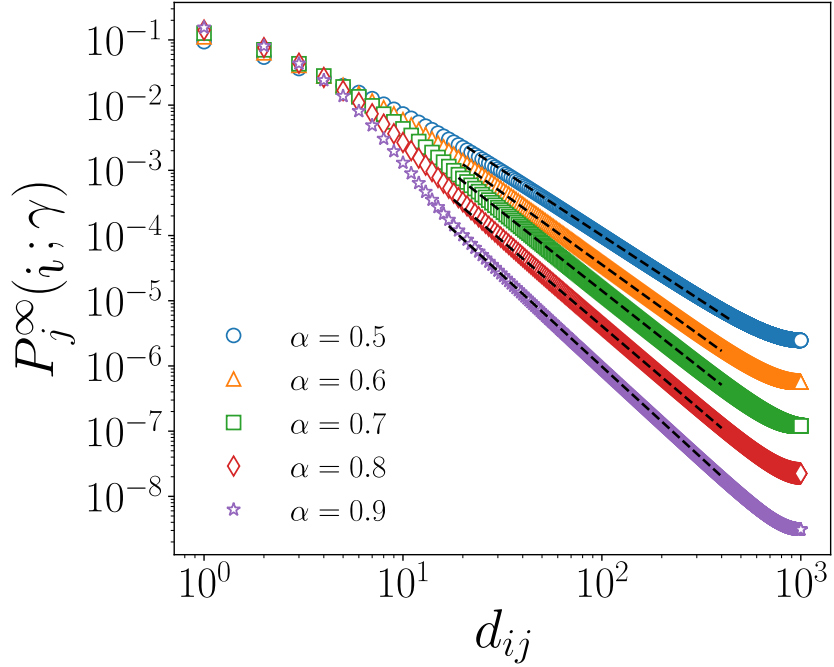}
    \caption{Stationary distribution for Lévy flight performed in a ring with $N=2000$ nodes and resetting to the initial node $i$ for different $\alpha$ as a function of the distance $d_{ij}$, $\gamma=0.2$. The dashed lines are the corresponding power-laws showing that $P_{j}^{\infty}\propto d^{-(1+2\alpha)}$.} \label{fig:ring_Levy_one_pstat}
\end{figure}
\begin{figure}[!t]
    \centering
    \includegraphics*[width=0.7\textwidth]{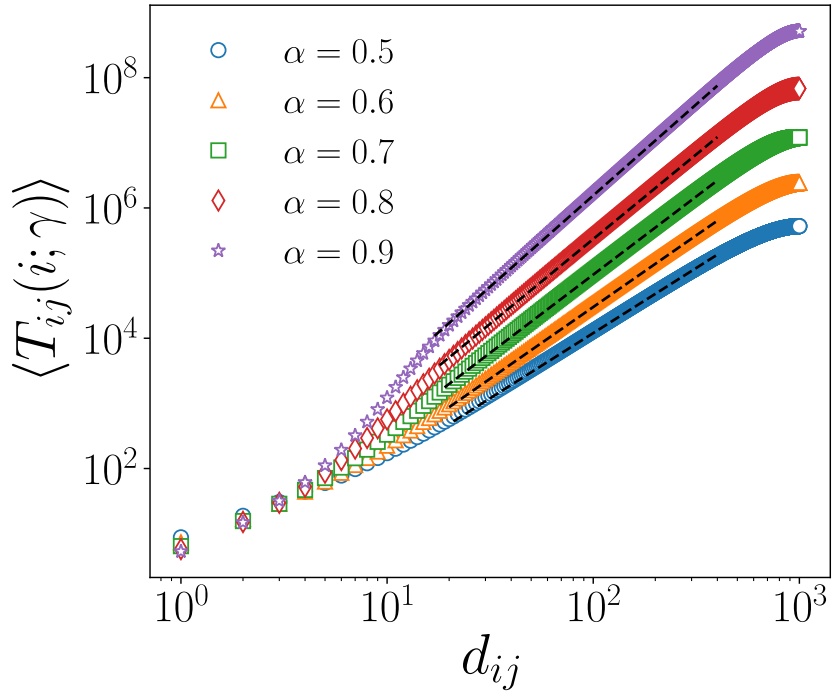}
    \caption{MFPT for Lévy flight performed in a ring with $N=2000$ nodes and resetting to the initial node $i$ for different $\alpha$ as a function of the distance $d_{ij}$, $\gamma=0.2$. The dashed lines are the corresponding power-laws showing that $\langle T_{ij}(i;\gamma)\rangle \propto d^{1+2\alpha}$.} \label{fig:ring_Levy_one_mfpt}
\end{figure}
\\[2mm]
Finally, applying the method to a ring with  $N=2000$ and resetting to the initial node $r=i$ we obtain results when the ring is large.
In Figure \ref{fig:ring_Levy_one_pstat}, the stationary distribution $P_{j}^{\infty}(r;\gamma)$ (Eq. \eqref{eq:stat_one_ring_levy}) is shown with different markers for different values of the Lévy index $\alpha$ while the resetting probability $\gamma=0.2$ is the same for all. The black dashed line shows the power-law $\propto d^{-(1+2\alpha)}$ corresponding to the limit $N\to\infty$. The curves show some variations from the exact line of the power-law, particularly for nodes closest to the resetting node and the farthest due to the periodicity of the ring structure. Now the behavior is different compared to Figure \ref{fig:ring_RW_one} where the dependency is exponential (Eq. \eqref{eq:ring_one_limit_RW}). Under the same conditions, in Figure \ref{fig:ring_Levy_one_mfpt}, we display the corresponding MFPT for different values of $\alpha$. Again, we notice the variations in the extreme values of the curves. It is important to observe that the interval of validity of the approximation is the same for both Figures \ref{fig:ring_Levy_one_pstat} and \ref{fig:ring_Levy_one_mfpt}. To increase this interval, we need a network with more nodes, since the analytical results in Eqs. (\ref{levy_statP_infring}) and (\ref{levy_Tij_infring}) are valid in the limit $N\to \infty$.
\FloatBarrier
Since $\langle T_{ij}(i;\gamma)\rangle$ is the average time it takes for a walker to arrive for the first time from node $i$ to node $j$, its value helps to understand the network exploration. For instance, for the hypothetical case where we could find a minimum value in Figures \ref{fig:ring_RW_one}(b) and \ref{fig:ring_Levy_one} (b), this would imply that the walker takes less time on average to arrive at that node. Now, to the actual figures, in comparison to the curve with $\gamma=0$, we observe in both cases that for nodes near the resetting point the MFPT decreases, but for distances $d_{ij}\gg 10$ the value is greater than the case without stochastic resetting. In the next chapter, we present the formalism for resetting to two different nodes and analyze the possible generalization of the method.  

%% file: Chapters/Chap_Reset2.tex
\chapter{Random walks with resetting to two nodes} \label{ch:two_nodes}

\section{Introduction}
In the last chapter, we explored how the stationary distribution and MFPT are affected by resetting of the random walker to a particular node $r$. Now, we want to use a similar approach but with resetting to two different nodes and generalize the notation and methods so that this procedure can be extended to include any number of resetting nodes.

\begin{figure}[!t]
    \centering
    \includegraphics*[width=0.7\textwidth]{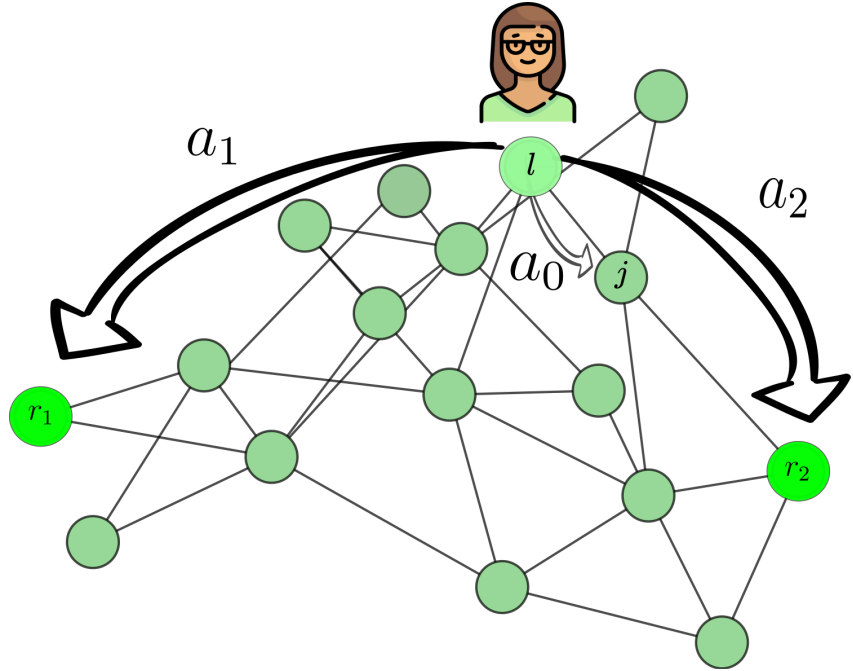}
    \caption{Random walker in a network with two different resetting nodes $r_{1}$ and $r_{2}$ and their corresponding probabilities $a_{1}$ and $a_{2}$. Here $a_0=1-a_1-a_2$ is the probability to perform a random walk step to a nearest neighbor.} \label{fig:network_two}
\end{figure}

As an example, this problem adapts perfectly to the idea of a person that explores a city (the network) but has to return constantly to its house and its workplace, as we can see in  Figure \ref{fig:network_two}. In this case, the walker has two different resetting nodes $r_{1}$ and $r_{2}$ and their corresponding resetting probabilities $a_{1}$ and $a_{2}$ at each time step. This situation appears naturally in human mobility, since this is a common pattern in human displacement. Of course, we can think ahead and consider an arbitrary number of resetting nodes, but that will be discussed in Chapter \ref{ch:M_reset}.

As we will see in this chapter, adding another resetting node and using a similar approach to calculate the eigenvalues, left and right eigenvectors, modifies considerably the analytical results for the stationary distribution and MFPTs. The formalism and notation will allow the possibility to use recursion methods in the calculation.

\newpage

\section{General approach}
\label{TwoNodes_General}
In this section, we explore the stationary distribution and MFPT for two resetting nodes $r_{1}$ and $r_{2}$, events that occur randomly with probabilities $a_{1}$ and $a_{2}$ at each time step. Following a similar procedure as the one presented in Chapter \ref{RefChapterReset1}, it is possible to write the transition matrix as
\begin{equation}\label{matPi_2reset}
\mathbf{\Pi}(r_1,r_2;a_1,a_2)=a_0\mathbf{W}+a_1\mathbf{\Theta}(r_1)+a_2\mathbf{\Theta}(r_2),
\end{equation}
where we have added two resetting matrices $\mathbf{\Theta}(r_1)$ and $\mathbf{\Theta}(r_2)$, the resetting probabilities and constant $a_{0}$ satisfy $a_{0}+a_{1}+a_{2}=1$. Reorganizing  $a_{0}+a_{1}$ in the first two terms, we have
\begin{equation} \label{eq:two_reset_calcs}
\mathbf{\Pi}(r_1,r_2;a_1,a_2)=\left(a_0+a_1\right)\Big[\frac{a_0}{a_0+a_1}\mathbf{W}
+\frac{a_1}{a_0+a_1}\mathbf{\Theta}(r_1)\Big]+a_2\mathbf{\Theta}(r_2),
\end{equation}
defining $\gamma_1=\frac{a_1}{a_0+a_1}$, and simplifying we obtain
\begin{align}\nonumber
\mathbf{\Pi}(r_1,r_2;a_1,a_2)&=\left(a_0+a_1\right)\left[(1-\gamma_1)\mathbf{W}+\gamma_1\mathbf{\Theta}(r_1)\right]+a_2\mathbf{\Theta}(r_2)\\ \nonumber
&=\left(a_0+a_1\right)\mathbf{\Pi}(r_1;\gamma_1)+a_2\mathbf{\Theta}(r_2)\\ \nonumber
&=\left(a_0+a_1+a_2\right)\Big[\frac{a_0+a_1}{a_0+a_1+a_2}\mathbf{\Pi}(r_1;\gamma_1)+\frac{a_2}{a_0+a_1+a_2}\mathbf{\Theta}(r_2)\Big].
\end{align}
In the last equality, we factor out $a_{0}+a_{1}+a_{2}$. This might seem redundant due to the condition $a_{0}+a_{1}+a_{2}=1$, but will be useful for the general case. Defining $\gamma_{2}=a_{2}$ the final expression for the transition matrix is with resetting to two nodes $\mathbf{\Pi}(r_1,r_2;a_1,a_2)$ is
\begin{equation}\label{matPi_iterative2}
\mathbf{\Pi}(r_1,r_2;a_1,a_2)=(1-\gamma_2)\mathbf{\Pi}(r_1;\gamma_1)+\gamma_2\mathbf{\Theta}(r_2).
\end{equation}
We can immediately recognize the form, since it is equal to Eq. \eqref{Def_MatPi_R1}, where $\mathbf{W}$ is replaced by $\mathbf{\Pi}(r_1;\gamma_1)$. This is convenient because we know the exact form of their eigenvalues (Eq. \eqref{eigvals_zeta}) and right and left eigenvectors (Eqs. \eqref{psil1} and \eqref{psirl_reset}), therefore we can get directly the eigenvalues, for $\zeta_1(r_1,r_2;\gamma_1,\gamma_2)=1$ and for $l=2,\dots,N$, we have
\begin{equation}
\zeta_l(r_1,r_2;\gamma_1,\gamma_2)=
(1-\gamma_2)(1-\gamma_1)\lambda_l.\,\,
\end{equation}
The eigenvectors are obtained using the analogous expressions, for the left eigenvector corresponding to the first eigenvalue $\zeta_1(r_1,r_2;\gamma_1,\gamma_2)$ we use the form of Eq. \eqref{psil1}
\begin{multline}
\label{eq:left_1_complete}
\left\langle\bar{\psi}_1(r_1,r_2;\gamma_1,\gamma_2)\right|=\left\langle\bar{\psi}_1(r_1;\gamma_1)\right|
\\
+\sum_{m=2}^N\frac{\gamma_2}{1-(1-\gamma_2)\zeta_m(r_1;\gamma_1)}\frac{\left\langle r_2|\psi_m(r_1;\gamma_1)\right\rangle}{\left\langle r_2|\psi_1(r_1;\gamma_1)\right\rangle}\left\langle\bar{\psi}_m(r_1;\gamma_1)\right|,
\end{multline}
where $\left\langle\bar{\psi}_m(r_1;\gamma_1)\right|$ denotes the left eigenvector of $\mathbf{\Pi}(r_1;\gamma_1)$ and $|\psi_m(r_1;\gamma_1)\rangle$ the right eigenvector. Substituting the eigenvectors and eigenvalues known for resetting to one node, we obtain
\begin{multline} \label{eq:left_1_gammas}
\left\langle\bar{\psi}_1(r_1,r_2;\gamma_1,\gamma_2)\right|=\left\langle\bar{\phi}_1\right|
+\sum_{m=2}^N\frac{\gamma_1}{1-(1-\gamma_1)\lambda_m}\frac{\left\langle r_1|\phi_m\right\rangle}{\left\langle r_1|\phi_1\right\rangle}\left\langle\bar{\phi}_m\right|\\
+\sum_{m=2}^N\frac{\gamma_2}{1-(1-\gamma_2)(1-\gamma_1)\lambda_m}\Big[\frac{\left\langle r_2|\phi_m\right\rangle }{\left\langle r_2|\phi_1\right\rangle}-\frac{\gamma_1}{1-(1-\gamma_1)\lambda_m}\frac{\left\langle r_1|\phi_m\right\rangle }{\left\langle r_1|\phi_1\right\rangle}\Big]\left\langle\bar{\phi}_m\right|.
\end{multline}
Here, the first two terms correspond to $\langle \bar{\psi}_{1}(r_{1};\gamma_{1})|$ and the last one is obtained considering that $\langle \bar{\psi}_{m}(r_{1};\gamma_{1})|=\langle \bar{\phi}_{m}|$, $\langle r_2|\psi_1(r_1;\gamma_1)\rangle=\langle r_2|\phi_{1}\rangle$ for $m=2,3,\ldots,N$ and the corresponding
\begin{equation} \label{eq:right_one_m}
    |\psi_{m}(r_{1};\gamma_{1})\rangle=|\phi_{m}\rangle-\frac{\gamma_{1}}{1-(1-\gamma_{1})\lambda_{m}}\frac{\langle r_{1}|\phi_{m}\rangle}{\langle r_{1}|\phi_{1}\rangle}|\phi_{1}\rangle
    \qquad m=2,3,\ldots,N.
\end{equation}
Rearranging the terms, we have the equivalent expression
\begin{multline}\label{EigenL1_M2}
\left\langle\bar{\psi}_1(r_1,r_2;\gamma_1,\gamma_2)\right|=\left\langle\bar{\phi}_1\right|
+\sum_{m=2}^N\frac{\gamma_1}{1-(1-\gamma_1)\lambda_m}\frac{\left\langle r_1|\phi_m\right\rangle}{\left\langle r_1|\phi_1\right\rangle}
\bigg[1-\frac{\gamma_2}{1-(1-\gamma_2)(1-\gamma_1)\lambda_m}\bigg]\left\langle\bar{\phi}_m\right|\\
+\sum_{m=2}^N\frac{\gamma_2}{1-(1-\gamma_2)(1-\gamma_1)\lambda_m}\frac{\left\langle r_2|\phi_m\right\rangle }{\left\langle r_2|\phi_1\right\rangle}\left\langle\bar{\phi}_m\right| .
\end{multline}
It is convenient to have a more compact notation, therefore we define the parameters
\begin{equation}\label{eq:nu}
\nu_{m}=\frac{\gamma_2}{1-(1-\gamma_2)(1-\gamma_1)\lambda_m}
\end{equation}
and
\begin{equation}
\label{eq:kappa}
\kappa_{m}=\frac{\gamma_1}{1-(1-\gamma_1)\lambda_m}(1-\nu_{m}).
\end{equation}
 It is important to keep in mind that both $\nu_{m}$ and $\kappa_{m}$ depend on $\gamma_{1}$, $\gamma_{2}$ and the spectrum of eigenvalues $\lambda_{m}$. Therefore, the left eigenvector for $l=1$ in terms of these new parameters is
\begin{equation} \label{eq:left_1_kappa_nu}
\left\langle\bar{\psi}_1(r_1,r_2;\gamma_1,\gamma_2)\right|=\left\langle\bar{\phi}_1\right|
+\sum_{m=2}^N\left(\kappa_{m}\frac{\left\langle r_1|\phi_m\right\rangle}{\left\langle r_1|\phi_1\right\rangle}
+\nu_{m}\frac{\left\langle r_2|\phi_m\right\rangle }{\left\langle r_2|\phi_1\right\rangle}\right)\left\langle\bar{\phi}_m\right|,
\end{equation}
while for the other values of $l=2,\dots,N$ we get
\begin{equation} \label{eq:left_l}
\left\langle\bar{\psi}_l(r_1,r_2;\gamma_1,\gamma_2)\right|=\left\langle\bar{\phi}_l(r_1;\gamma_1)\right|=\left\langle\bar{\phi}_l\right|,
\end{equation}
which means that once we calculate the left eigenvectors of $\mathbf{W}$ we also have the eigenvectors of $\mathbf{\Pi}(r_{1},r_{2};\gamma_{1},\gamma_{2})$. In particular, for the first right eigenvector, we have
\begin{equation} \label{eq:right_1}
    |\psi_{1}(r_{1},r_{2};\gamma_{1},\gamma_{2})\rangle = |\psi_{1}(r_{1};\gamma_{1})\rangle = |\phi_{1}\rangle
\end{equation}
while, following a similar procedure, we find that the right eigenvectors for $l=2,\dots,N$ are
\begin{multline*}
    |\psi_{l}(r_{1},r_{2};\gamma_{1},\gamma_{2})\rangle = |\psi_{l}(r_{1};\gamma_{1})\rangle
    -\frac{\gamma_{2}}{1-(1-\gamma_{2})(1-\gamma_{1})\lambda_{l}}\frac{\langle r_{2}|\psi_{l}(r_{1};\gamma_{1})\rangle}{\langle r_{2}|\psi_{1}(r_{1};\gamma_{1}) \rangle}|\psi_{1}(r_{1};\gamma_{1})\rangle.
\end{multline*}
Now, expressing $ |\psi_{l}(r_{1},r_{2};\gamma_{1},\gamma_{2})\rangle$ directly in terms of the eigenvectors of $\mathbf{W}$, using Eq. \eqref{eq:right_one_m}, we obtain
\begin{multline}
    |\psi_{l}(r_{1},r_{2};\gamma_{1},\gamma_{2})\rangle =
    |\phi_{l}\rangle-\frac{\gamma_{1}}{1-(1-\gamma_{1})\lambda_{l}}\frac{\langle r_{1}|\phi_{l}\rangle}{\langle r_{1}|\phi_{1}\rangle}|\phi_{1}\rangle\\-\frac{\gamma_{2}}{1-(1-\gamma_{2})(1-\gamma_{1})\lambda_{l}}
    \bigg[\frac{\langle r_{2}|\phi_{l}\rangle}{\langle r_{2}|\phi_{1}\rangle}-\frac{\gamma_{1}}{1-(1-\gamma_{1})\lambda_{l}}\frac{\langle r_{1}|\phi_{l}\rangle}{\langle r_{1}|\phi_{1}\rangle}\bigg]|\phi_{1}\rangle
\end{multline}
and, regrouping terms we get the expression
\begin{multline}
    |\psi_{l}(r_{1},r_{2};\gamma_{1},\gamma_{2})\rangle = |\phi_{l}\rangle+ \bigg[\frac{\gamma_{1}}{1-(1-\gamma_{1})\lambda_{l}}\frac{\langle r_{1}|\phi_{l}\rangle}{\langle r_{1}|\phi_{1}\rangle}
    \bigg(\frac{\gamma_{2}}{1-(1-\gamma_{2})(1-\gamma_{1})\lambda_{l}}-1\bigg)\\
    -\frac{\gamma_{2}}{1-(1-\gamma_{2})(1-\gamma_{1})\lambda_{l}}\frac{\langle r_{2}|\phi_{l}\rangle}{\langle r_{2}|\phi_{1}\rangle}\bigg]|\phi_{1}\rangle.
\end{multline}
And finally, substituting the $\kappa_{l}$ and $\nu_{l}$ parameters we get
\begin{equation}
 \label{eq:right_l_kappa_nu}
    |\psi_{l}(r_{1},r_{2};\gamma_{1},\gamma_{2})\rangle = |\phi_{l}\rangle
- \bigg(\kappa_{l}\frac{\langle r_{1}|\phi_{l}\rangle}{\langle r_{1}|\phi_{1}\rangle}
    +\nu_{l}\frac{\langle r_{2}|\phi_{l}\rangle}{\langle r_{2}|\phi_{1}\rangle}\bigg)|\phi_{1}\rangle.
\end{equation}
This equation shows that the right eigenvectors are a linear combination of two right eigenvectors of $\mathbf{W}$. This is a direct result of substituting the already known eigenvectors for one resetting node (Eqs. \eqref{psil1} and \eqref{psirl_reset}).

Now that we have the left and right eigenvectors, we can calculate the stationary distribution directly in terms of the first left and right eigenvectors as follows
\begin{align*}
P_j^\infty(r_1,r_2;\gamma_1,\gamma_2)
&=\left\langle i\left|\psi_1(r_1,r_2;\gamma_1,\gamma_2)\right\rangle \left\langle\bar{\psi}_1(r_1,r_2;\gamma_1,\gamma_2)\right|j\right\rangle\\
&=P_i^\infty+ \sum_{m=2}^N\frac{\gamma_1}{1-(1-\gamma_1)\lambda_m}\frac{\left\langle r_1|\phi_m\right\rangle}{\left\langle r_1|\phi_1\right\rangle}\times\\
&\hspace{7mm}\bigg[1-\frac{\gamma_2}{1-(1-\gamma_2)(1-\gamma_1)\lambda_m}\bigg]\langle i|\phi_{1}\rangle\left\langle\bar{\phi}_m\right|j\rangle\\
&\hspace{3mm}+\sum_{m=2}^N\frac{\gamma_2}{1-(1-\gamma_2)(1-\gamma_1)\lambda_m}\frac{\left\langle r_2|\phi_m\right\rangle }{\left\langle r_2|\phi_1\right\rangle}\langle i|\phi_{1}\rangle\left\langle\bar{\phi}_m\right|j\rangle.
\end{align*}
Using the parameters $\kappa_{m}$ and $\nu_{m}$, we obtain
\begin{equation}
\label{eq:Pj_infty_final_two}
P_j^\infty(r_1,r_2;\gamma_1,\gamma_2)=P_i^\infty\\+ \sum_{m=2}^N\left(\kappa_{m}\left\langle r_1|\phi_m\right\rangle+\nu_{m}\left\langle r_2|\phi_m\right\rangle\right)\left\langle\bar{\phi}_m\right|j\rangle.
\end{equation}
As we can see, the first term $P_i^\infty$ is the stationary distribution without resetting and the sum in Eq. (\ref{eq:Pj_infty_final_two})  considers the effect of  resetting to two nodes. Another advantage of having the eigenvectors and eigenvalues in an exact form is that we can calculate the transition probabilities with the spectral representation of $\mathbf{\Pi}(r_{1},r_{2};\gamma_{1},\gamma_{2})$, using Eq. \eqref{Pi_prob_powert} and also have its time evolution, as in Eq. \eqref{eq:trans_prob_eigenrep}. For finite time $t$, the occupation probability is
%
%
%\begin{align}\nonumber
%P_{ij}(t,r_1,r_2;\gamma_1,\gamma_2)=P_j^\infty(r_1,r_2;\gamma_1,\gamma_2)\\\nonumber
%&\hspace{3mm}+\sum_{l=2}^N\zeta_l(r_1,r_2;\gamma_1,\gamma_2)^t\times\\
%&\hspace{3mm}\left\langle i\left|\psi_l(r_1,r_2;\gamma_1,\gamma_2)\right\rangle \left\langle\bar{\psi}_l(r_1,r_2;\gamma_1,\gamma_2)\right|j\right\rangle.
%\end{align}
%%
\begin{multline}
P_{ij}(t,r_1,r_2;\gamma_1,\gamma_2)=P_j^\infty(r_1,r_2;\gamma_1,\gamma_2)\\+\sum_{l=2}^N\zeta_l(r_1,r_2;\gamma_1,\gamma_2)^t\left\langle i\left|\psi_l(r_1,r_2;\gamma_1,\gamma_2)\right\rangle \left\langle\bar{\psi}_l(r_1,r_2;\gamma_1,\gamma_2)\right|j\right\rangle,
\end{multline}

where the first term is the stationary distribution as given by Eq. \eqref{eq:Pj_infty_final_two}. This expression for $P_{ij}(t,r_1,r_2;\gamma_1,\gamma_2)$ is useful to calculate the moments, substituting directly in
\begin{align}\nonumber
\mathcal{R}_{ij}^{(0)}(r_{1},r_{2};\gamma_{1},\gamma_{2})
&=\sum_{t=0}^\infty(P_{ij}(t,r_1,r_2;\gamma_1,\gamma_2)-P_j^\infty(r_1,r_2;\gamma_1,\gamma_2))\\
&=\sum_{l=2}^{N}\frac{\langle i|\psi_{l}(r_{1},r_{2};\gamma_{1},\gamma_{2})\rangle \langle \bar{\psi}_{l}(r_{1},r_{2};\gamma_{1},\gamma_{2})|j\rangle }{1-(1-\gamma_{2})(1-\gamma_{1})\lambda_{l}},
 \label{eq:R_ij}
\end{align}
%
%REV: AGREGAR MAS PASOS AQUI
we obtain a compact expression. The moments $\mathcal{R}_{ij}^{(0)}(r_{1},r_{2};\gamma_{1},\gamma_{2})$ can be used further to calculate the MFPT given by
\begin{equation}
 \label{eq:MFPT_general}
    \langle T_{ij}(r_{1},r_{2};\gamma_{1},\gamma_{2})\rangle=\frac{\delta_{ij}}{P_{j}^{\infty}(r_{1},r_{2};\gamma_{1},\gamma_{2})}
+\frac{\mathcal{R}_{jj}^{(0)}(r_{1},r_{2};\gamma_{1},\gamma_{2})-\mathcal{R}_{ij}^{(0)}(r_{1},r_{2};\gamma_{1},\gamma_{2})}{P_{j}^{\infty}(r_{1},r_{2};\gamma_{1},\gamma_{2})}.
\end{equation}

In Eq. \eqref{eq:right_l_kappa_nu} we show that the right eigenvector $|\psi_{l}(r_{1},r_{2};\gamma_{1},\gamma_{2})\rangle$ is a linear combination of $|\phi_{l}\rangle$ and $|\phi_{1}\rangle$  and that the left eigenvector with resetting for $l=2,\dots,N$ is equal to the left eigenvector without resetting $\langle\bar{\phi}_{l}|=\langle \bar{\psi}_{l}(r_{1};\gamma_{1})|$ so we can simplify Eq. \eqref{eq:MFPT_general} noticing that
\begin{align}\label{RijM2part_1}
    \langle i|\psi_{l}(r_{1},r_{2};\gamma_{1},\gamma_{2})\rangle&=\langle i | \phi_{l}\rangle -\kappa_{l}\langle r_{1}|\phi_{l}\rangle -\nu_{l}\langle r_{2}|\phi_{l}\rangle
\end{align}
and
\begin{equation}\label{RijM2part_2}
    \langle \bar{\psi}_{l}(r_{1},r_{2};\gamma_{1},\gamma_{2})|j\rangle = \langle \bar{\phi}_{l}|j\rangle.
\end{equation}
Therefore
{\small
\begin{align*}
	&{\mathcal{R}_{jj}^{(0)}(r_{1},r_{2};\gamma_{1},\gamma_{2})-\mathcal{R}_{ij}^{(0)}(r_{1},r_{2};\gamma_{1},\gamma_{2})}\\
	=&\sum_{l=2}^{N}\frac{\langle j|\psi_{l}(r_{1},r_{2};\gamma_{1},\gamma_{2})\rangle \langle \bar{\psi}_{l}(r_{1},r_{2};\gamma_{1},\gamma_{2})|j\rangle-\langle i|\psi_{l}(r_{1},r_{2};\gamma_{1},\gamma_{2})\rangle \langle \bar{\psi}_{l}(r_{1},r_{2};\gamma_{1},\gamma_{2})|j\rangle }{1-(1-\gamma_{2})(1-\gamma_{1})\lambda_{l}}\\
	=&\sum_{l=2}^{N}\frac{\langle j|\phi_{l}\rangle \langle \bar{\phi}_{l}|j\rangle - \langle i | \phi_{l}\rangle \langle \bar{\phi}_{l}|j\rangle}{1-(1-\gamma_{2})(1-\gamma_{1})\lambda_{l}}.
\end{align*}
}
Introducing this result in  Eq. \eqref{eq:MFPT_general}, we obtain a simpler expression for the MFPT 
\begin{equation}
\label{eq:MFPT_vectors_two}
    \langle T_{ij}(r_{1},r_{2};\gamma_{1},\gamma_{2})\rangle =\frac{1}{P_{j}^{\infty}(r_{1},r_{2};\gamma_{1},\gamma_{2})}
    \bigg[\delta_{ij}+\sum_{l=2}^{N}\frac{\langle j|\phi_{l}\rangle \langle \bar{\phi}_{l}|j\rangle - \langle i | \phi_{l}\rangle \langle \bar{\phi}_{l}|j\rangle}{1-(1-\gamma_{2})(1-\gamma_{1})\lambda_{l}}\bigg].
\end{equation}
Observe that this equation is very similar to Eqs. \eqref{eq:MFPT_vecs} and \eqref{MFPT_resetSM}, the only part affected is the denominator inside the sum, where the eigenvalues have been modified accordingly.

\section{Dynamics with resetting on rings}

Just as we did with one resetting node, now we calculate the stationary distribution and MFPT for the dynamics with resetting to two nodes on rings. We compare the results between the two types of walkers with local and non-local transitions between nodes.

\subsection{Random walks}

Since we know the analytic eigenvalues and eigenvectors for a ring (Eq. \eqref{eq:ring_eigenvalues}), we can use the expression  in Eqs. \eqref{eq:Pj_infty_final_two} and \eqref{eq:MFPT_vectors_two} to obtain the exact expressions for the stationary distribution and MFPT. Substituting $\lambda_{l}=\cos(\varphi_l)$ and using the respective eigenvectors, we obtain
\begin{multline} \label{eq:stat_two_ring_RW}
    P_{j}^{\infty}(r_{1},r_{2};\gamma_{1},\gamma_{2})=\frac{1}{N}\\+\frac{1}{N}\sum_{m=2}^{N}\frac{\gamma_{1}\cos(\varphi_{m}d_{jr_{1}})}{1-(1-\gamma_{1})(1-\gamma_{2})\cos(\varphi_{m})}\bigg(1-\frac{\gamma_{2}}{1-(1-\gamma_{2})(1-\gamma_{1})\cos(\varphi_{m})}\bigg)\\
    +\frac{1}{N}\sum_{m=2}^{N}\frac{\gamma_{2}\cos(\varphi_{m}d_{jr_{2}})}{1-(1-\gamma_{2})(1-\gamma_{1})\cos(\varphi_m)}.
\end{multline}
The corresponding expression for the MFPT is
\begin{equation} \label{eq:MFPT_two_ring_RW}
    \langle T_{ij}(r_{1},r_{2};\gamma_{1},\gamma_{2})\rangle = \frac{1}{P_{j}^{\infty}(r_{1},r_{2};\gamma_{1},\gamma_{2})}
    \bigg[\delta_{ij}+\sum_{l=2}^{N}\frac{1-\cos(\varphi_{l}d_{ij})}{1-(1-\gamma_{2})(1-\gamma_{1})\cos(\varphi_{l})}\bigg].
\end{equation}

\begin{figure}[!t]
    \centering
    \includegraphics*[width=\textwidth]{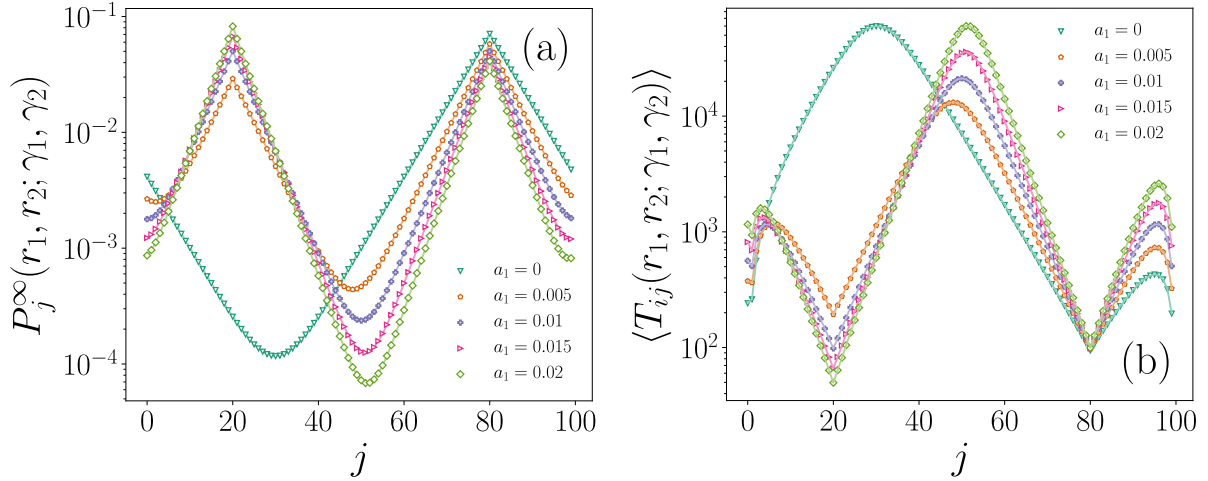}
    \caption{(a) Stationary distribution  and (b) MFPT for a random walk performed in a ring with $N=100$ nodes and resetting to $r_{1}=20$, $r_{2}=80$, $i=0$. The resetting probability $a_{2}=0.01$ is constant and $a_{1}=0,0.005,\ldots,0.02$. The results are presented as a function of the node $j$. }\label{fig:ring_RW_two}
\end{figure}

In Figure \ref{fig:ring_RW_two}, we show the results obtained for a random walk with two resetting nodes. We calculate the stationary distribution using Eq. \eqref{eq:Pj_infty_final_two} and the MFPT with Eq. \eqref{eq:MFPT_vectors_two} keeping the resetting probability $a_{2}=0.01$ fixed and changing the value of $a_{1}$ for each curve. An important and necessary step is to verify that the results are consistent, so we made the calculations using both methods, the first one obtaining the eigenvalues, left and right eigenvectors of matrix $\mathbf{\Pi}(r_{1},r_{2};\gamma_{1},\gamma_{2})$ and directly use Eq. \eqref{eq:MFPT_vecs} for the MFPT, the result is shown as a continuous line, while the other is calculated diagonalizing
$\mathbf{W}$ and evaluate numerically Eqs. (\ref{eq:stat_two_ring_RW}) and (\ref{eq:MFPT_two_ring_RW}), the results are shown with different markers. As we can see, both results overlap exactly, validating the general result in Eqs. \eqref{eq:Pj_infty_final_two} and \eqref{eq:MFPT_vectors_two}.

To investigate further the accuracy of the method, we calculated as well the relative error between the two values which is a measure of how far off the numerical approximation is relative to the actual value, expressed as \cite{noda2005introduccion}
\begin{equation}
    err =\frac{\delta x}{x(n)}= \frac{|x_{n}-x(n)|}{|x(n)|}
\end{equation}
where $x_{n}$ is the numerical value of our method (using the eigenvalues and eigenvectors of $\mathbf{W}$) and $x(n)$ is the actual value, in this case, we considered the value obtained with the numerical calculation of the eigenvalues and eigenvectors of the matrix $\mathbf{\Pi}(r_{1},r_{2};\gamma_{1},\gamma_{2})$.

\begin{figure}[!t]
    \centering
    \includegraphics*[width=\textwidth]{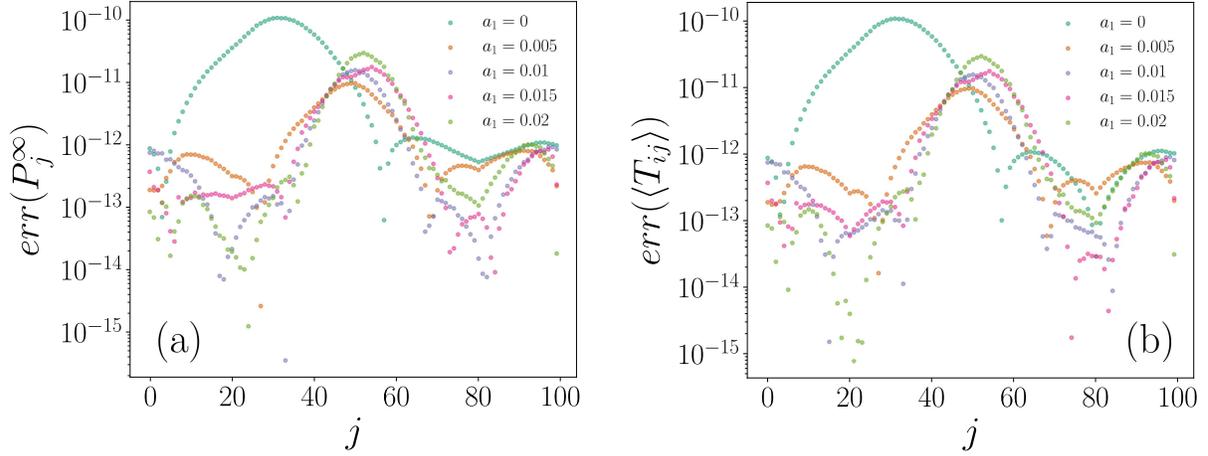}
    \caption{(a) Relative error for the stationary distribution  and (b) MFPT for a random walk performed in a ring with $N=100$ nodes and resetting to $r_{1}=20$, $r_{2}=80$ leaving the resetting probability $a_{2}=0.01$ fixed varying $a_{1}$, both as a function of the node $j$.}
    \label{fig:ring_RW_two_err}
\end{figure}

As we can observe, in Figure \ref{fig:ring_RW_two_err}, for the stationary distribution and the MFPT most of the curves stay bounded between $10^{-15}$ and $10^{-10}$ which is close to the machine's precision, showing that our analytical approach is correct. We plot these values to see the general behaviour of the relative error.

An analysis for the limit $N\to\infty$ is possible using Eq. \eqref{eq:stat_two_ring_RW} and \eqref{eq:MFPT_two_ring_RW}, following an analogous procedure to that of resetting to one node, turning the discrete variable $\varphi_{m}$ into a continuous with differential $d\varphi=2\pi/N$ in Eq. \eqref{eq:stat_two_ring_RW}, therefore we obtain
\begin{multline} \label{eq:stat_dist_two_infty}
    P_{j}^{\infty}(r_{1},r_{2};\gamma_{1},\gamma_{2})=\frac{\gamma_{1}}{2\pi}\int_{0}^{2\pi}\frac{\cos(\varphi d_{jr_{1}})}{1-b_{1}\cos(\varphi)}d\varphi+
    \frac{\gamma_{2}}{2\pi}\int_{0}^{2\pi}\frac{\cos(\varphi d_{jr_{2}})}{1-b_{12}\cos(\varphi)}d\varphi\\
    -\frac{\gamma_{1}\gamma_{2}}{2\pi}\int_{0}^{2\pi}\frac{\cos(\varphi d_{jr_{1}})}{(1-b_{1}\cos(\varphi))(1-b_{12}\cos(\varphi))}d\varphi.
\end{multline}
Here, we have introduced a more compact notation, defining the auxiliary variables $b_{1}=1-\gamma_{1}$ and $b_{12}=(1-\gamma_{1})(1-\gamma_{2})$. The first two integrals can be obtained exactly using Eq. \eqref{eq:stat_one_ring_limit} since the dependency with $\varphi$ is the same. For the last term, we have to operate further to get a similar expression. The integral we need to calculate has the form
\begin{equation} \label{eq:integral2}
    \mathcal{I}_{2}=\frac{1}{2\pi}\int_{0}^{2\pi}\frac{\cos(\varphi x)}{(1-y\cos(\varphi))(1-z\cos(\varphi))}d\varphi
\end{equation}
which can be expressed as the sum of two fractions. Applying partial fraction decomposition, we have
\begin{equation}
    \frac{1}{(1-y\cos(\varphi))(1-z\cos(\varphi))}=\frac{1}{y-z}\bigg[\frac{y}{1-y\cos(\varphi)}-\frac{z}{1-z\cos(\varphi)}\bigg].
\end{equation}
Since this is the denominator of the integral $\mathcal{I}_{2}$, substituting in Eq. \eqref{eq:integral2}, we obtain
\begin{equation}
    \mathcal{I}_{2}=\frac{1}{2\pi(y-z)}\bigg[\int_{0}^{2\pi}\frac{y\cos(\varphi x)}{1-y\cos(\varphi)}d\varphi-\int_{0}^{2\pi}\frac{z\cos(\varphi x)}{1-z\cos(\varphi)}d\varphi\bigg],
\end{equation}
which are integrals that we have previously calculated (Eq. \eqref{eq:stat_one_ring_limit}). Substituting the exact values considering $x=d_{jr_{1}}$, $y=b_{1}$ and $z=b_{12}$ we obtain four terms
\begin{multline}
    P_{j}^{\infty}(r_{1},r_{2};\gamma_{1},\gamma_{2})=\frac{\gamma_{1}}{\sqrt{1-b_{1}^2}}\bigg(\frac{1+\sqrt{1-b_{1}^2}}{b_{1}}\bigg)^{-d_{jr_{1}}}+\frac{\gamma_{2}}{\sqrt{1-b_{12}^2}}\bigg(\frac{1+\sqrt{1-b_{12}^2}}{b_{12}}\bigg)^{-d_{jr_{2}}}\\
    -\frac{\gamma_{1}\gamma_{2}}{b_{1}-b_{2}}\frac{b_{1}}{\sqrt{1-b_{1}^2}}\bigg(\frac{1+\sqrt{1-b_{1}^2}}{b_{1}}\bigg)^{-d_{jr_{1}}}+\frac{\gamma_{1}\gamma_{2}}{b_{1}-b_{2}}\frac{b_{12}}{\sqrt{1-b_{12}^2}}\bigg(\frac{1+\sqrt{1-b_{12}^2}}{b_{12}}\bigg)^{-d_{jr_{1}}}.
\end{multline}
We can simplify further because $b_{1}-b_{12}=\gamma_{2}(1-\gamma_{1})$, $b_{1}=1-\gamma_{1}$ and $b_{12}=(1-\gamma_{1})(1-\gamma_{2})$, so the first term cancels out with the third and the remaining can be expressed as
\begin{multline} \label{eq:stat_two_ring_limit}
    P_{j}^{\infty}(r_{1},r_{2};\gamma_{1},\gamma_{2})=\frac{\gamma_{1}(1-\gamma_{2})}{\sqrt{1-b_{12}^2}}\bigg(\frac{1+\sqrt{1-b_{12}^2}}{b_{12}}\bigg)^{-d_{jr_{1}}}\\
    +\frac{\gamma_{2}}{\sqrt{1-b_{12}^2}}\bigg(\frac{1+\sqrt{1-b_{12}^2}}{b_{12}}\bigg)^{-d_{jr_{2}}}.
\end{multline}
%%%
In this result, we can see that in both terms appears, $b_{12}$ which depends on both $\gamma_{1}$ and $\gamma_{2}$. If either is zero, we recover the exact expression for resetting to one node in the limit $N\to\infty$ (Eq. \eqref{eq:stat_one_ring_limit}). In addition, to analyze asymptotic behavior, we define the parameter
\begin{equation}
    \chi=\log\bigg(\frac{1+\sqrt{1-b_{12}^2}}{b_{12}}\bigg).
\end{equation}
Substituting in Eq. \eqref{eq:stat_two_ring_limit} the stationary distribution is
\begin{equation} \label{eq:stat_two_ring_limit_chi}
    P_{j}^{\infty}(r_{1},r_{2};\gamma_{1},\gamma_{2})=\frac{\gamma_{1}(1-\gamma_{2})e^{-\chi d_{jr_{1}}}+\gamma_{2}e^{-\chi d_{jr_{2}}}}{\sqrt{1-(1-\gamma_{1})^2(1-\gamma_{2})^2}}.
\end{equation}
As we observe, the distribution for nodes far from the resetting nodes falls as an exponential regulated by the parameter $\chi$. For the MFPT we follow an analogous procedure, changing the summation for an integral and turning the discrete variable $\varphi_{l}$ into the continuous $\varphi$, the resulting expression is
\begin{equation}
    \langle T_{ij}(r_{1},r_{2};\gamma_{1},\gamma_{2})\rangle = \frac{\delta_{ij}}{P_{j}^{\infty}(r_{1},r_{2};\gamma_{1},\gamma_{2})}+\frac{1}{P_{j}^{\infty}(r_{1},r_{2};\gamma_{1},\gamma_{2})}\int_{0}^{2\pi}\frac{1-\cos(\varphi d_{ij})}{1-b_{12}\cos(\varphi)}d\varphi
\end{equation}
\begin{figure}[!t]
	\centering
	\includegraphics*[width=\textwidth]{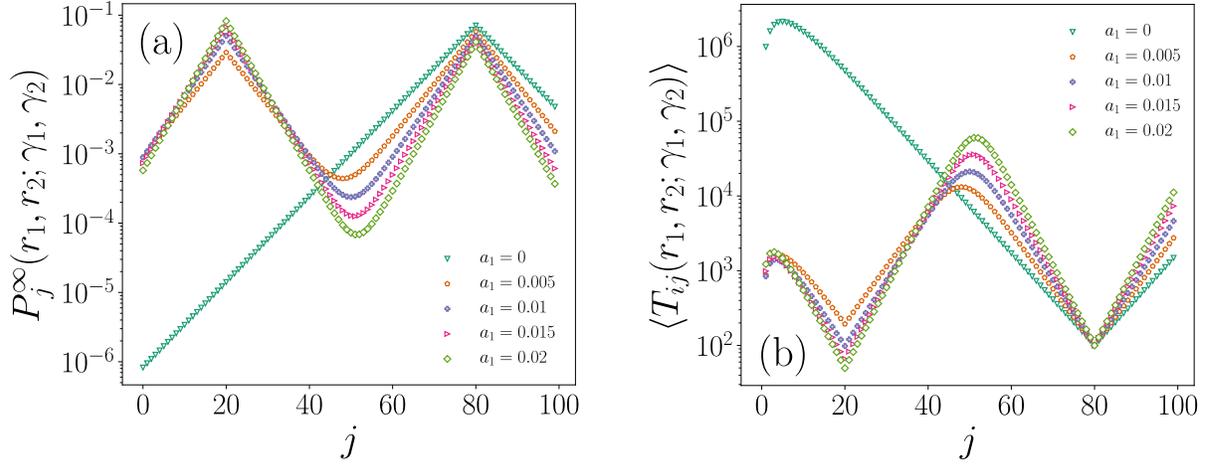}
	\caption{(a) Stationary distribution  and (b) MFPT  for a random walk performed in a ring in the limit $N\to \infty$ for the first 100 nodes, resetting to $r_{1}=20$, $r_{2}=80$ leaving the resetting probability $a_{2}=0.01$ fixed varying $a_{1}$, both as a function of the node $j$, $i=0$. We used Eq. \eqref{eq:stat_two_ring_limit_chi} and \eqref{eq:MFPT_ring_limit_chi} considering the same parameters as in Figure \ref{fig:ring_RW_two}. }\label{fig:ring_two_infty}
\end{figure}
which can be directly integrated to obtain
\begin{multline}
    \langle T_{ij}(r_{1},r_{2};\gamma_{1},\gamma_{2})\rangle = \frac{\delta_{ij}}{P_{j}^{\infty}(r_{1},r_{2};\gamma_{1},\gamma_{2})}\\
    +\frac{1}{P_{j}^{\infty}(r_{1},r_{2};\gamma_{1},\gamma_{2})}\frac{1}{\sqrt{1-b_{12}^2}}\bigg[1-\bigg(\frac{1+\sqrt{1-b_{12}^2}}{b_{12}}\bigg)^{-d_{ij}}\bigg],
\end{multline}
when $i=j$ we have $d_{ij}=0$ and the second term vanishes. In this way, the remaining expression is just the inverse of the stationary distribution in agreement with Kac's Lemma \cite{kac1947notion}. In the case $i\neq j$ we can write $\langle T_{ij}(r_{1},r_{2};\gamma_{1},\gamma_{2})\rangle$ in terms of $\chi$
\begin{equation} \label{eq:MFPT_ring_limit_chi}
    \langle T_{ij}(r_{1},r_{2};\gamma_{1},\gamma_{2})\rangle=\frac{1-e^{-\chi d_{ij}}}{\gamma_{1}(1-\gamma_{2})e^{-\chi d_{jr_{1}}}+\gamma_{2}e^{-\chi d_{jr_{2}}}}.
\end{equation}
As we can see, $\chi$ is a parameter that regulates the behavior with respect to $d_{ij}$, $d_{jr_{1}}$ and $d_{jr_{2}}$. In Figure \ref{fig:ring_RW_two} the values for $b_{12}$ are $0.99,0.985,0.98,0.975$ and $0.97$ with the corresponding $\chi$'s $0.142,0.174,0.201,0.225$ and $0.248$.

To analyze how the stationary distribution and MFPT would behave in the limit $N\to\infty$, we calculated the values with Eqs. \eqref{eq:stat_two_ring_limit_chi} and \eqref{eq:MFPT_ring_limit_chi} as shown in Figure \ref{fig:ring_two_infty}. We observe straight lines in the semi-logarithmic scale due to the exponential dependence. In comparison with the results in Figure \eqref{fig:ring_RW_two}, the border effects are avoided and are bound between the same orders, except for the case $a_{1}=0$ where the curves are very different.
%%%

\subsection{Lévy flights}
\begin{figure}[!b]
	\centering
	\includegraphics*[width=\textwidth]{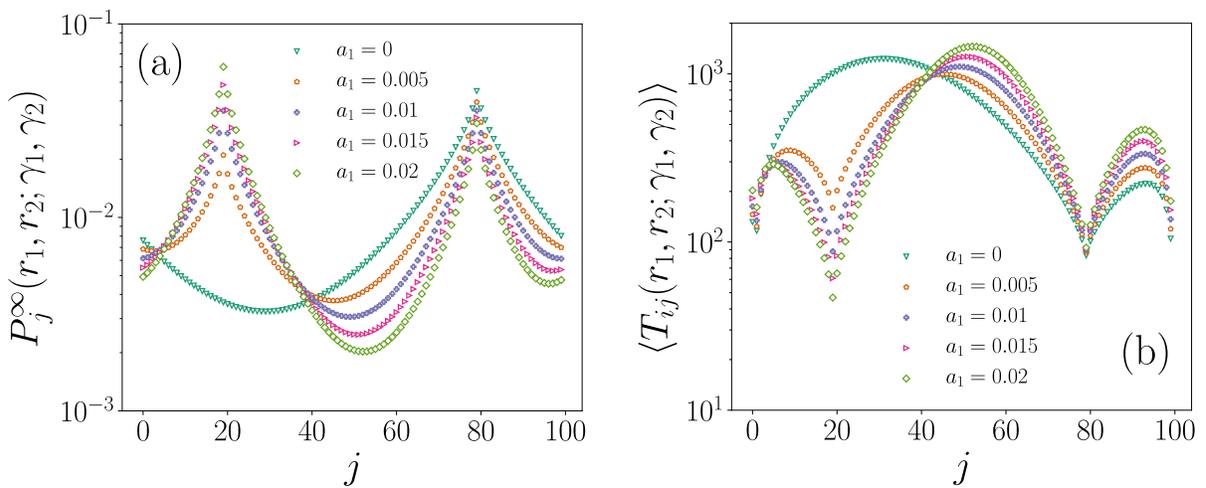}
	\caption{(a) Stationary distribution  and (b) MFPT for a Lévy flight $\alpha=0.75$ performed in a ring with $N=100$ nodes and resetting to $r_{1}=20$, $r_{2}=80$ leaving the resetting probability $a_{2}=0.01$ fixed varying $a_{1}$, both as a function of the node $j$, $i=0$.}\label{fig:ring_Levy_two}
\end{figure}
Now, we apply the formalism to L\'evy flight dynamics on a ring with $N$ nodes. In comparison with the results in Eqs. \eqref{eq:stat_two_ring_RW} and \eqref{eq:MFPT_two_ring_RW}, the study of L\'evy flights only requires modifications of the eigenvalues using the result in Eq. \eqref{eq:eigen_levy_ring} 
\begin{equation*} 
	\lambda_l(\alpha)=1-\frac{1}{k^{(\alpha)}}\, \left(2-2\cos\varphi_l\right)^\alpha, \qquad k^{(\alpha)}=\frac{1}{N}\sum_{l=1}^N \left(2-2\cos\varphi_l\right)^\alpha.
\end{equation*}
Then, we have for L\'evy flights with resetting to nodes $r_1$ and $r_2$
\begin{multline}
	P_{j}^{\infty}(r_{1},r_{2};\gamma_{1},\gamma_{2})=\frac{1}{N}\\+\frac{1}{N}\sum_{m=2}^{N}\frac{\gamma_{1}\cos(\varphi_{m}d_{jr_{1}})}{1-(1-\gamma_{1})(1-\gamma_{2})\lambda_m(\alpha)}\bigg(1-\frac{\gamma_{2}}{1-(1-\gamma_{2})(1-\gamma_{1})\lambda_m(\alpha)}\bigg)\\
	+\frac{1}{N}\sum_{m=2}^{N}\frac{\gamma_{2}\cos(\varphi_{m}d_{jr_{2}})}{1-(1-\gamma_{2})(1-\gamma_{1})\lambda_m(\alpha)}.
\end{multline}
The corresponding expression for the MFPT is
\begin{equation} 
	\langle T_{ij}(r_{1},r_{2};\gamma_{1},\gamma_{2})\rangle = \frac{1}{P_{j}^{\infty}(r_{1},r_{2};\gamma_{1},\gamma_{2})}
	\bigg[\delta_{ij}+\sum_{l=2}^{N}\frac{1-\cos(\varphi_{l}d_{ij})}{1-(1-\gamma_{2})(1-\gamma_{1})\lambda_l(\alpha)}\bigg].
\end{equation}
The results are shown in Figure \ref{fig:ring_Levy_two}, for the stationary distribution (a) and the MFPT (b). It is observed that the order of magnitude of the stationary distribution and MFPT changes in comparison to the results for the normal random walk in Figure \ref{fig:ring_RW_two}. As we can see the curves are bounded by one and two orders respectively.

With Chapters \ref{Chapter_StoNet}, \ref{RefChapterReset1} and \ref{ch:two_nodes} we conclude the presentation of the method, its validation, and application to simple structures. In the next chapter we will explore a generalization of the equations considering a resetting to an arbitrary number of nodes, this framework will be applied to networks with more complex structures.

%% file: Chapters/Chap_ResetM.tex
\chapter{Dynamics with resetting to $\mathcal{M}$ nodes} \label{ch:M_reset}
\section{Introduction}

In the previous chapters, we explored the effects of resetting on networks considering one and two nodes. In this chapter, we will describe a generalization of random walks with resetting to an arbitrary number of nodes $\mathcal{M}$. In order to make the equations clear and understandable, it is necessary to introduce a more compact notation. Once the theory is fully explained and implemented, we will apply this generalization to more complex structures than the simple ring, particularly to Cayley trees, random distribution of points in a continuous space, and interacting cycles. For this last type of networks, we apply the \textit{Google} search strategy where the dynamics resets to all nodes.

We introduce the total resetting probability $\beta$ and the global mean first passage time $\mathcal{T}$, which is the average of the MFPT over all the target and source nodes, consequently its value does not depend on the initial condition of the random walker. The main objective is to show how the parameter $\beta$ affects $\mathcal{T}$ and in some cases optimizes it.

\section{General approach}
\label{RW_resetM_general}
We consider a network with $N$ nodes and a random walker defined by a transition matrix $\mathbf{W}$. Now the resetting is performed to $\mathcal{M}$ nodes, respectively $r_1,r_2,\ldots,r_\mathcal{M}$, and the corresponding restart probabilities
$0 \leq a_s\leq 1$ (where $s=1,\ldots, \mathcal{M}$). The transition matrix is
\begin{equation}
	\label{matPi_Mreset}
	\mathbf{\Pi}(r_1,r_2,\ldots,r_\mathcal{M};a_1,a_2,\ldots,a_\mathcal{M})=a_0\mathbf{W}+\sum_{s=1}^\mathcal{M} a_s\mathbf{\Theta}(r_s).
\end{equation}
We have the constriction $ a_0\equiv 1-\sum_{s=1}^\mathcal{M}a_s$ due to conservation of probability. We also require $0\leq \sum_{s=1}^\mathcal{M}a_s\leq 1$. As in previous sections, the resetting matrix to node $r_{s}$ will be denoted as $\mathbf{\Theta}(r_s)$.

Since now the transition matrix depends on various different parameters, we introduce a more compact notation for Eq. (\ref{matPi_Mreset}), thus
\begin{equation}
	\mathbf{\Pi}(r_1,r_2,\ldots,r_\mathcal{M};a_1,a_2,\ldots,a_\mathcal{M})=\mathbf{\Pi}_\mathcal{M}.
\end{equation}
The total transition matrix  $\mathbf{\Pi}_\mathcal{M}$ can be calculated iteratively, considering that at step $s$ we can substitute the transition matrix for step $s-1$ and the resetting matrix corresponding to $s$, $\mathbf{\Theta}(r_s)$, following Eq. \eqref{eq:two_reset_calcs} we get
\begin{equation} \label{eq:iter_transition_matrix}
	\mathbf{\Pi}_s=\frac{\sum_{l=0}^{s-1}a_l}{\sum_{l=0}^{s}a_l}\mathbf{\Pi}_{s-1}+\frac{a_s}{\sum_{l=0}^{s}a_l}\mathbf{\Theta}(r_s),
\end{equation}
for $s=1,2,\ldots, \mathcal{M}$ with $\mathbf{\Pi}_0\equiv\mathbf{W}$. We introduce the generalized parameter $\gamma_{s}$ in terms of the individual resetting probabilities $a_{s}$
\begin{equation}\label{eq:gamma_gendef}
	\gamma_s\equiv \frac{a_s}{\sum_{l=0}^{s}a_l},\qquad s=1,2,\ldots,\mathcal{M},
\end{equation}
which follows that $\frac{\sum_{l=0}^{s-1}a_l}{\sum_{l=0}^{s}a_l}=1-\gamma_s$. We directly substitute this relation to obtain the final expression for the transition matrix in step $s$ of the iteration
\begin{equation}
	\mathbf{\Pi}_s=(1-\gamma_s)\mathbf{\Pi}_{s-1}+\gamma_s\mathbf{\Theta}(r_s).
\end{equation}
Let us now denote the right and left eigenvectors of $\mathbf{\Pi}_s$ as $|\psi^{(s)}_{l}\rangle$  and  $\langle \bar{\psi}^{(s)}_{l}|$, and their corresponding eigenvalues as $\zeta_l^{(s)}$. According to the result in Eq. (\ref{eigvals_zeta}),
the first eigenvalue satisfies $\zeta_1^{(s)}=1$ and the others are obtained iteratively
\begin{align}\nonumber
	\zeta_l^{(s)}&=(1-\gamma_s)\zeta_l^{(s-1)}=(1-\gamma_s)(1-\gamma_{s-1})\zeta_l^{(s-2)}\\
	&=\cdots=\lambda_l\prod_{m=1}^s(1-\gamma_m),\qquad l=2,3,\ldots, N,\label{eigval_reset_ws}
\end{align}
where we use $\zeta_l^{(0)}=\lambda_l$, and $\lambda_l$ is the eigenvalue of $\mathbf{W}$, the transition matrix without resetting. For the right eigenvector corresponding to the first eigenvalue, we have
\begin{equation}
	\label{rpsi_reset_1}
	|\psi^{(s)}_{1}\rangle = |\psi^{(s-1)}_{1}\rangle= |\psi^{(s-2)}_{1}\rangle=\cdots=|\psi^{(0)}_{1}\rangle=|\phi_{1}\rangle,
\end{equation}
and, for the left eigenvectors with index $m=2,\ldots,N$, we get
\begin{equation}
	\label{lpsi_reset_s}
	\langle\bar{\psi}^{(s)}_m|=\langle\bar{\psi}^{(s-1)}_m|=\cdots
	=\langle\bar{\psi}^{(0)}_m|=\langle\bar{\phi}_m|.
\end{equation}
Observe that, in these two cases, the eigenvectors remain unaltered with the introduction of resetting. On the other hand, from Eqs. (\ref{psirl_reset}) and (\ref{rpsi_reset_1}), for $l=2,\ldots,N$ the right eigenvectors are
\begin{equation}\label{psi_s_Mnodes}
	|\psi^{(s)}_{l}\rangle = |\psi^{(s-1)}_{l}\rangle-\frac{\gamma_s}{1-\zeta_l^{(s)}}\frac{\langle r_{s}|\psi^{(s-1)}_{l}\rangle}{\langle r_{s}|\phi_{1}\rangle}|\phi_{1}\rangle.
\end{equation}
Finally, for $\langle\bar{\psi}^{(s)}_1|$, combining the results in Eq. (\ref{psil1}) with (\ref{rpsi_reset_1}) and (\ref{lpsi_reset_s}) we obtain the first left eigenvector
\begin{equation}\label{bar_psi_s}
	\langle\bar{\psi}^{(s)}_1|=\langle\bar{\psi}^{(s-1)}_1|
	+\sum_{m=2}^N\frac{\gamma_s}{1-\zeta_l^{(s)}}\frac{\langle r_s|\psi^{(s-1)}_m\rangle}{\left\langle r_s|\phi_1\right\rangle}\left\langle\bar{\phi}_m\right|.
\end{equation}
With these eigenvectors, the stationary distribution is
\begin{align}\nonumber \label{eq:Pstat_multiple2}
	P_{j}^{\infty}(\vec{r};\vec{\gamma})&\equiv P_{j}^{\infty}(r_{1},r_{2},\ldots,r_{\mathcal{M}};\gamma_{1},\gamma_{2},\ldots,\gamma_{\mathcal{M}})\\
	&=\langle j|\psi^{(\mathcal{M})}_{1}\rangle \langle\bar{\psi}^{(\mathcal{M})}_1| j\rangle,
\end{align} 
where we used the property $\langle i|\psi_{1}^{(\mathcal{M})}\rangle=\langle j|\psi_{1}^{(\mathcal{M})}\rangle$ for every $i$ and $j$. The expression in Eq. \eqref{eq:Pstat_multiple2} would be written explicitly in terms of the eigenvectors in past iterations $s-1,s-2,\dots,1$. Alternatively, we can calculate it from the eigenvalues and eigenvectors of the transition matrix $\mathbf{W}$ through the successive application of the Eqs. (\ref{eigval_reset_ws})-(\ref{bar_psi_s}) to obtain $\langle\bar{\psi}^{(\mathcal{M})}_1| j\rangle$. We introduce the notation $P_{j}^{\infty}(\vec{r};\vec{\gamma})$ for the stationary distribution, where $\vec{r}$ is a vector containing all the resetting nodes and $\vec{\gamma}$ has the parameters as presented in Eq. \eqref{eq:gamma_gendef}. This approach also allows to deduce iteratively all the eigenvalues and eigenvectors of $\mathbf{\Pi}_{\mathcal{M}}$ in Eq. (\ref{matPi_Mreset}). Introducing these results to the moments defined in Eq. \eqref{Rmoments_def}, we get
\begin{align}\nonumber
	\mathcal{R}^{(0)}(i,j,\vec{r},\vec{\gamma})&=\sum_{t=0}^\infty \left[ P_{ij}(\vec{r},\vec{\gamma};t)-P_j^\infty(\vec{r},\vec{\gamma}) \right]\\ \nonumber
	&=\sum_{t=0}^\infty \sum_{l=2}^N (\zeta_{l}^{(\mathcal{M})})^t \langle i |\psi^{(\mathcal{M})}_l\rangle\langle\bar{\psi}^{(\mathcal{M})}_l|j\rangle\\
	&=\sum_{l=2}^N  \frac{1}{1-\zeta_{l}^{(\mathcal{M})}} \langle i |\psi^{(\mathcal{M})}_l\rangle\langle\bar{\psi}^{(\mathcal{M})}_l|j\rangle.\label{Rij_Mnodes_Psi}
\end{align}
Hence, the difference is calculated as
\begin{equation}
	\mathcal{R}^{(0)}(j,j,\vec{r},\vec{\gamma})-\mathcal{R}^{(0)}(i,j,\vec{r},\vec{\gamma})
	=\sum_{l=2}^N  \frac{\langle j |\psi^{(\mathcal{M})}_l\rangle\langle\bar{\psi}^{(\mathcal{M})}_l|j\rangle-\langle i |\psi^{(\mathcal{M})}_l\rangle\langle\bar{\psi}^{(\mathcal{M})}_l|j\rangle}{1-\zeta_{l}^{(\mathcal{M})}}.
\end{equation}
However, from Eq. (\ref{psi_s_Mnodes}), we see that the components of second term (proportional to $|\phi_{1}\rangle$) in $ |\psi^{(s)}_{l}\rangle$ are constant for all the nodes. As a consequence
%
%
%\begin{align}\nonumber
%&\langle j|\psi^{(\mathcal{M})}_{l}\rangle \langle\bar{\psi}^{(\mathcal{M})}_l| j\rangle-\langle i|\psi^{(\mathcal{M})}_{l}\rangle \langle\bar{\psi}^{(\mathcal{M})}_l| j\rangle\\\nonumber
%&=\langle j|\psi^{(\mathcal{M}-1)}_{l}\rangle \langle\bar{\psi}^{(\mathcal{M}-1)}_l| j\rangle-\langle i|\psi^{(\mathcal{M}-1)}_{l}\rangle \langle\bar{\psi}^{(\mathcal{M}-1)}_l| j\rangle\\ \nonumber
%&=\langle j|\psi^{(\mathcal{M}-2)}_{l}\rangle \langle\bar{\psi}^{(\mathcal{M}-2)}_l| j\rangle-\langle i|\psi^{(\mathcal{M}-2)}_{l}\rangle \langle\bar{\psi}^{(\mathcal{M}-2)}_l| j\rangle\\\nonumber
%&=\cdots\\\nonumber
%&=\langle j|\psi^{(0)}_{l}\rangle \langle\bar{\psi}^{(0)}_l| j\rangle-\langle i|\psi^{(0)}_{l}\rangle \langle\bar{\psi}^{(0)}_l| j\rangle\\\nonumber
%&=\langle j|\phi_{l}\rangle \langle\bar{\phi}_l| j\rangle-\langle i|\phi_{l}\rangle \langle\bar{\phi}_l| j\rangle.
%\end{align}
%
%
\begin{align}\nonumber
	\langle j|\psi^{(\mathcal{M})}_{l}\rangle \langle\bar{\psi}^{(\mathcal{M})}_l| j\rangle-\langle i|\psi^{(\mathcal{M})}_{l}\rangle \langle\bar{\psi}^{(\mathcal{M})}_l| j\rangle \nonumber
	&=\langle j|\psi^{(\mathcal{M}-1)}_{l}\rangle \langle\bar{\psi}^{(\mathcal{M}-1)}_l| j\rangle-\langle i|\psi^{(\mathcal{M}-1)}_{l}\rangle \langle\bar{\psi}^{(\mathcal{M}-1)}_l| j\rangle\\ \nonumber
	&=\langle j|\psi^{(\mathcal{M}-2)}_{l}\rangle \langle\bar{\psi}^{(\mathcal{M}-2)}_l| j\rangle-\langle i|\psi^{(\mathcal{M}-2)}_{l}\rangle \langle\bar{\psi}^{(\mathcal{M}-2)}_l| j\rangle\\\nonumber
	&=\cdots\\\nonumber
	&=\langle j|\psi^{(0)}_{l}\rangle \langle\bar{\psi}^{(0)}_l| j\rangle-\langle i|\psi^{(0)}_{l}\rangle \langle\bar{\psi}^{(0)}_l| j\rangle\\\nonumber
	&=\langle j|\phi_{l}\rangle \langle\bar{\phi}_l| j\rangle-\langle i|\phi_{l}\rangle \langle\bar{\phi}_l| j\rangle.
\end{align}
In Appendix \ref{app:A} we present a more detailed calculation, demonstrating this equality with mathematical induction. In this manner, the difference between moments for different source node is
\begin{equation}
	\label{RjjRij_reset}
	\mathcal{R}^{(0)}(j,j,\vec{r},\vec{\gamma})-\mathcal{R}^{(0)}(i,j,\vec{r},\vec{\gamma})=
	\sum_{l=2}^N  \frac{1}{1-\zeta_{l}^{(\mathcal{M})}} \left[\langle j|\phi_{l}\rangle \langle\bar{\phi}_l| j\rangle-\langle i|\phi_{l}\rangle \langle\bar{\phi}_l| j\rangle\right],
\end{equation}
which is useful to calculate the MFPT in terms of the eigenvectors of the transition matrix without resetting $\mathbf{W}$.

As we can see, the net effect of the resetting relies on the eigenvalues that appear in the denominator. The application of the Eq. (\ref{eq:MFTP_moments}) is valid for ergodic random walks but now considering the resetting to  $\mathcal{M}$ nodes, the resulting MFPT can be generally expressed \cite{2021_moi}
\begin{equation}
	\label{MFPT_Mreset}
	\langle T_{ij}(\vec{r};\vec{\gamma})\rangle =\frac{\delta_{ij}}{P_{j}^{\infty}(\vec{r};\vec{\gamma})}+\frac{1}{P_{j}^{\infty}(\vec{r};\vec{\gamma})} \sum_{l=2}^{N}\frac{\langle j|\phi_{l}\rangle \langle \bar{\phi}_{l}|j\rangle - \langle i | \phi_{l}\rangle \langle \bar{\phi}_{l}|j\rangle}{1-z(\vec{\gamma})\lambda_{l}}
\end{equation}
with $z(\vec{\gamma})\equiv z(\gamma_1,\gamma_2,\ldots,\gamma_{\mathcal{M}})=\prod_{s=1}^\mathcal{M}(1-\gamma_s)$, where $\gamma_{s}$ are parameters linked to the resetting and $\lambda_{l}$ are the eigenvalues of the transition matrix $\mathbf{W}$.

\section{Examples of the dynamics with  multiple reset}

\subsection{Cayley trees}

\begin{figure*}[!b]
	\begin{center}
		\includegraphics*[width=0.85\textwidth]{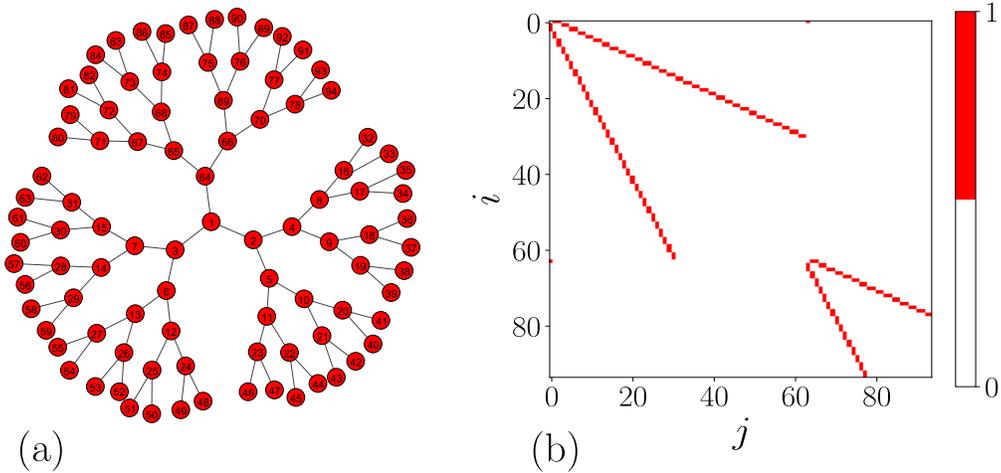}
	\end{center}
	\vspace{-5mm}
	\caption{(a) Cayley tree with $z=3$ and numbered nodes, with a total of $N=94$ and (b) the  adjacency matrix.}
	\label{fig:Cayley_number_nodes}
\end{figure*}

A tree is a simple connected undirected graph with no cycles  \cite{valiente2002algorithms_tree}. We can construct a Cayley tree starting from a root or central seed vertex \cite{OSTILLI20123417_cayley}, the next  generation of vertices is formed with $z$ sites connected with edges to the root, this composes the first shell. For the next generations of shells, each vertex is connected to other $z$ nodes, so that for finite trees, the last shell has degree one and all the others have a total degree $z$ considering the connection to the previous shell. In Figure \ref{fig:Cayley_number_nodes}(a) we observe a Cayley tree with $N=94$ nodes organized in five shells with coordination number $z=3$ and in (b) its adjacency matrix.

We define a global mean first passage time $\mathcal{T}$ to analyze the behavior of the random walker, where the MFPT is averaged over all starting nodes $i$ and target nodes $j$, so that
\begin{equation} \label{eq:global_MFPT}
	\mathcal{T}=\frac{1}{N^2}\sum_{i=1}^{N}\sum_{j=1}^{N}\langle T_{ij}\rangle.
\end{equation}
Its value is an alternative to quantify the capacity of the random walk to reach any node considering all possible initial conditions \cite{ResetNetworks_PRE2020}. A particular advantage of using the global MFPT is that it does not depend on the initial condition $i$. Besides, the average over all the source nodes has been proven useful studying the dynamics of the trapping problem where a trap that absorbs a random walker is set to a particular location in a graph, the global MFPT is used as an indicator of the trapping efficiency \cite{Lin2012_traps}.

\begin{figure*}[!t]
	\begin{center}
		\includegraphics*[width=0.95\textwidth]{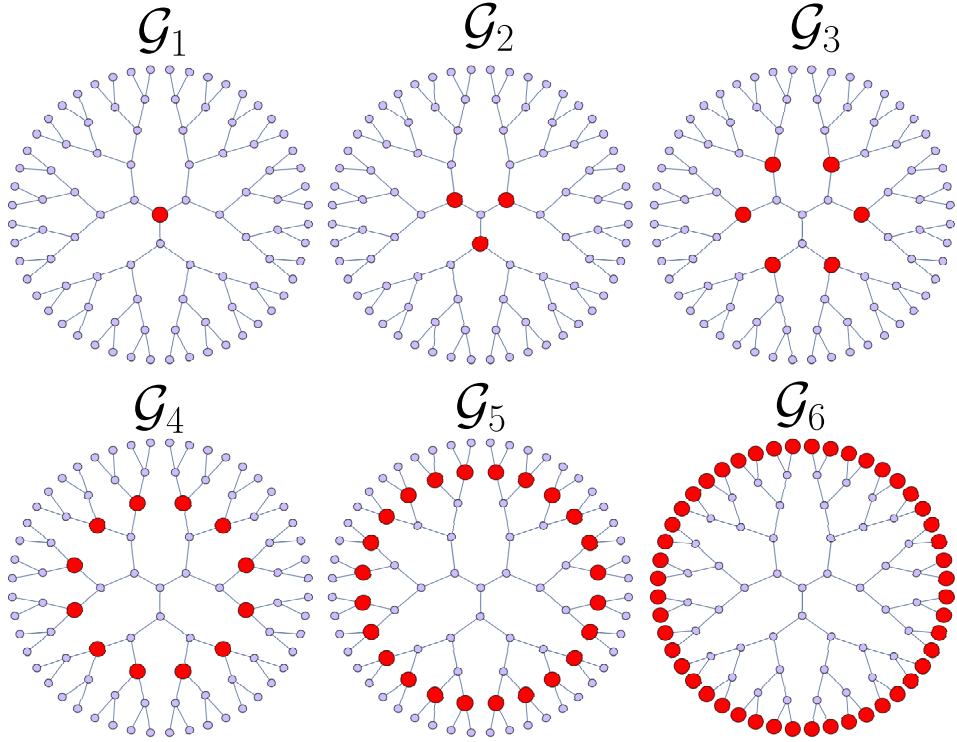}
	\end{center}
	\vspace{-5mm}
	\caption{Cayley trees with $N=94$. In the graphs $\mathcal{G}_s$ with $s=1,2,\ldots,6$, the red nodes represent the vertices where reset is produced.  
	}
	\label{fig:Cayley_multireset}
\end{figure*}

We calculate Eq. \eqref{eq:global_MFPT} as a function of the total resetting probability $\beta=\sum_{s=1}^{\mathcal{M}}a_{s}$ for different strategic distributions of the resetting nodes. For convenience, we divide $\beta$ equally between the resetting nodes, this means that if we have $\mathcal{M}$ resetting points, to each corresponds $a_{s}=\beta/\mathcal{M}$ as the resetting probability. As we can see in Figure \ref{fig:Cayley_multireset}, the resetting nodes are colored in red and were chosen symmetrically considering complete shells, in the first simulation $\mathcal{G}_{1}$ we have only one resetting node, for $\mathcal{G}_{2}$ there are three, in $\mathcal{G}_{3}$ we have six, then twelve for $\mathcal{G}_{4}$, twenty four for $\mathcal{G}_{5}$ and in the last one  $\mathcal{G}_{6}$ there are forty eight resetting nodes.

\begin{figure*}[!t]
	\begin{center}
		\includegraphics*[width=0.8\textwidth]{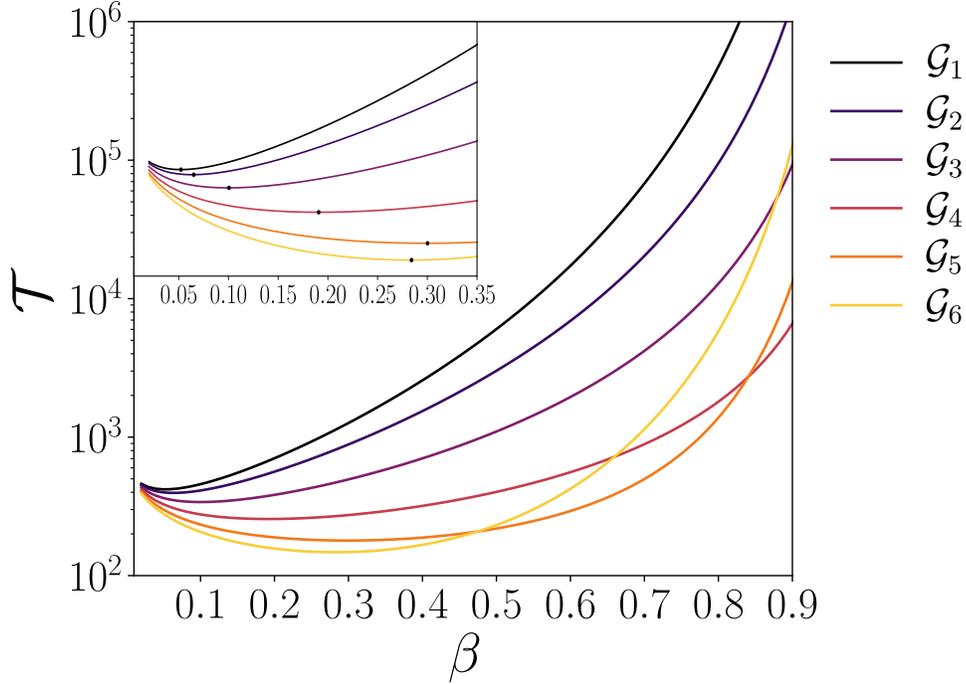}
	\end{center}
	\vspace{-5mm}
	\caption{Global mean first passage time $\mathcal{T}$ as a function of the total resetting probability $\beta$ for different number of resetting nodes and the inset shows a detail of the minimum of the respective curves. Precise values of $\beta_{min}$ can be found in Table \ref{tab:GMFPT_beta_min}}
	\label{fig:Cayley_MFPT}
\end{figure*}

\begin{table}[t!]
	\centering
	\begin{tabular}{ |c|c|c|}
		\hline
		& $\beta_{min}$ & $\mathcal{T}(\beta_{min})$ \\ \hline \hline
		$\mathcal{G}_{1}$ & 0.0551515151515151 & 419.674549615583 \\ \hline
		$\mathcal{G}_{2}$ & 0.0610101010101010 & 395.635100983603 \\ \hline
		$\mathcal{G}_{3}$ & 0.1020202020202020 & 339.787614563130 \\ \hline
		$\mathcal{G}_{4}$ & 0.1898989898989898 & 256.295634485661 \\ \hline
		$\mathcal{G}_{5}$ & 0.3012121212121212 & 179.134420720821 \\ \hline
		$\mathcal{G}_{6}$ & 0.2836363636363636 & 147.483657849399 \\ \hline
	\end{tabular}
	\\[3pt]
	\caption{Minimum values for the global mean first passage time and the corresponding $\beta_{min}$} \label{tab:GMFPT_beta_min}
\end{table}

\begin{figure*}[!t]
	\begin{center}
		\includegraphics*[width=\textwidth]{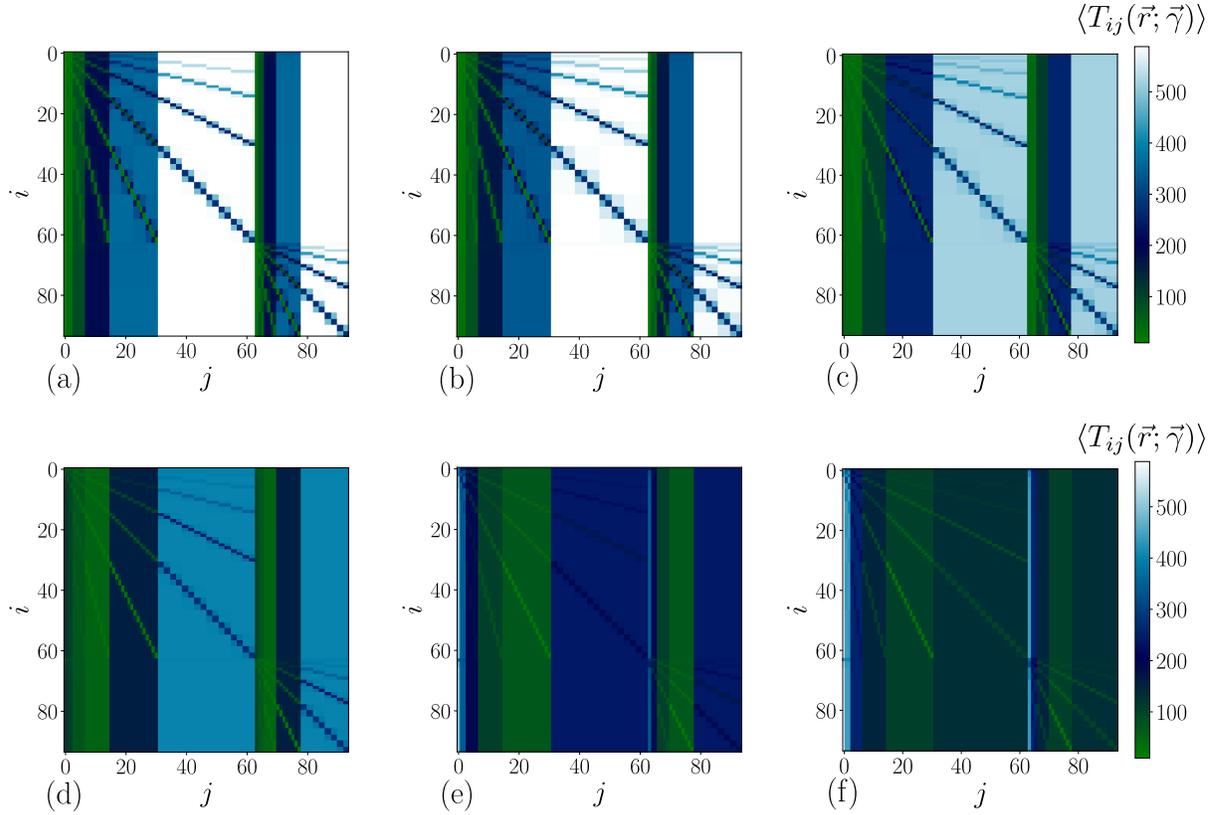}
	\end{center}
	\vspace{-5mm}
	\caption{Matrix visualization of $\langle T_{ij}(\vec{r};\vec{\gamma})\rangle$ for (a) $\mathcal{G}_{1}$, (b) $\mathcal{G}_{2}$, (c) $\mathcal{G}_{3}$, (d) $\mathcal{G}_{4}$, (e) $\mathcal{G}_{5}$ and (f) $\mathcal{G}_{6}$, calculated using the minimum value of $\beta$, denoted as $\beta_{min}$ that produced the minimal global time reported in Table \ref{tab:GMFPT_beta_min}.}
	\label{fig:Cayley_MFPT_mat}
\end{figure*}

The curves corresponding to the global mean first passage time are shown in Figure \ref{fig:Cayley_MFPT}. In the inset, we can see the minimum value of $\mathcal{T}$ highlighted as a black dot. For the first five, there is a tendency to increase its position in the horizontal axis while decreasing in the vertical but for $\mathcal{G}_6$, where the resetting is made to all the nodes in the outer shell, we can observe a significant change where the lowest point is settled before the curve obtained for $\mathcal{G}_5$. Even in the case $\mathcal{G}_6$, there is an optimal resetting $\beta_{min}$ that minimizes the global MFPT showing that the transport is more effective than the dynamics without resetting. For higher values of $\beta$ the curves corresponding to $\mathcal{G}_{5}$ and $\mathcal{G}_{6}$ intersect with others. In Table \ref{tab:GMFPT_beta_min} we can find the precise value of $\beta_{min}$ and its corresponding global MFPT for all the configurations analyzed in Figure \ref{fig:Cayley_MFPT}.

%For the pink one
%To further investigate the behavior of the MFPT, in Figure \ref{fig:Cayley_MFPT_mat} we plot  $\langle T_{ij}(\vec{r};\vec{\gamma})\rangle$ as a matrix, with node $i$, the source, varying in the vertical axis and $j$, the target node, in the horizontal, considering the node numeration just as shown in Figure \ref{fig:Cayley_multireset}. We use the value $\beta_{min}$ to see the specific behavior for the optimal reset. The most notorious property is that the majority of values are independent of  the source node, which is shown as vertical stripes of the same color. In addition, we observe that the overall general structure is the same for the six figures independent of the number of resetting nodes but the difference is clear in the values, since for (a) which has one resetting point we have predominantly dark colors in the order of 500, while for six resetting nodes in (c) we have lighter colors in areas that for (a). As the number of resetting nodes increases, we observe the  decrease in values which correspond to lighter tones  and finally for forty eight resetting nodes (f) the dark areas in the other figures are much lighter which indicates values of the order of 200 and less.

%for the blue one
To further investigate the behavior of the MFPT, in Figure \ref{fig:Cayley_MFPT_mat} we plot  $\langle T_{ij}(\vec{r};\vec{\gamma})\rangle$ as a matrix, with node $i$, the source, varying in the vertical axis and $j$, the target node, in the horizontal, considering the node numeration just as shown in Figure \ref{fig:Cayley_number_nodes}. We use the value $\beta_{min}$ to see the specific behavior for the optimal reset. The most notorious property is that the majority of values are independent of  the source node, which is shown as vertical stripes of the same color. In addition, we observe that the overall general structure is the same for the six figures independent of the number of resetting nodes but the difference is clear in the values, since for (a) which has one resetting point we have predominantly light blue and white colors in the order of 500 for some nodes, while for six resetting nodes in (c) we have darker colors in areas that for (a). As the number of resetting nodes increases, we observe the  decrease in values which correspond to darker tones and finally for forty eight resetting nodes (f) the light areas in the other figures are much darker which indicates values of the order of 200 and less.

As we can observe, the Cayley trees are very structured. Also, choosing a symmetrical distribution of resetting nodes produced similar patterns in the MFPT. In the next section, we will apply the method to a non-local random walker that visits points in a continuous space to see the effects of multiple reset in more complex dynamics.

\subsection{Dynamics on a distribution of points}

We now consider a set of $N$ points randomly distributed in a two-dimensional space and agglomerated in clusters around a specific center. The networks used in the past sections, such as rings and Cayley trees, the position of nodes and edges do not have actual relation to the distribution in space, but in this type of dynamics we will have a transition matrix that depends on the distance between nodes and a given radius $R$ used as a threshold for the transition probability. Following this premise, we designate that the probability of going from one point to another is proportional to a power $\alpha$ of the Euclidean distance $d_{ij}$ between them for $d_{ij}>R$ and independent of the distance for $d_{ij}\leq R$, such that \cite{RiascosMateosPlosOne2017}
\begin{equation} \label{eq:transition_points}
	w_{i\to j}^{(\alpha)}(R) = \frac{\Omega_{ij}^{(\alpha)}(R)}{\sum_{m=1}^{N}\Omega_{im}^{(\alpha)}(R)}
\end{equation}
where
\begin{equation}
	\Omega_{ij}^{(\alpha)}(R)=
	\begin{cases}
		1 &\quad\text{if } 0\leq d_{ij}\leq R\\
		\big(\frac{d_{ij}}{R}\big)^{-\alpha} &\quad\text{otherwise.} \\
	\end{cases}
\end{equation}
The sum in the denominator of Eq. \eqref{eq:transition_points} guarantees that  $\sum_{j}w_{i\to j}^{(\alpha)}(R)=1$. These transition probabilities generate a Lévy-like dynamics.

The procedure to generate the distribution of points (nodes in a spatial network) consists in choosing the centers and the number of nodes per cluster, selecting random positions around the center with a Gaussian distribution characterized by predefined standard deviations. A random node is selected from each cluster as a resetting node. Once the clusters are formed, we fill a matrix with distances $d_{ij}$ and operate to create the transition matrix according to Eq. \eqref{eq:transition_points}, finally we apply the multi-resetting formalism.

\begin{figure*}[!t]
	\begin{center}
		\includegraphics*[width=\textwidth]{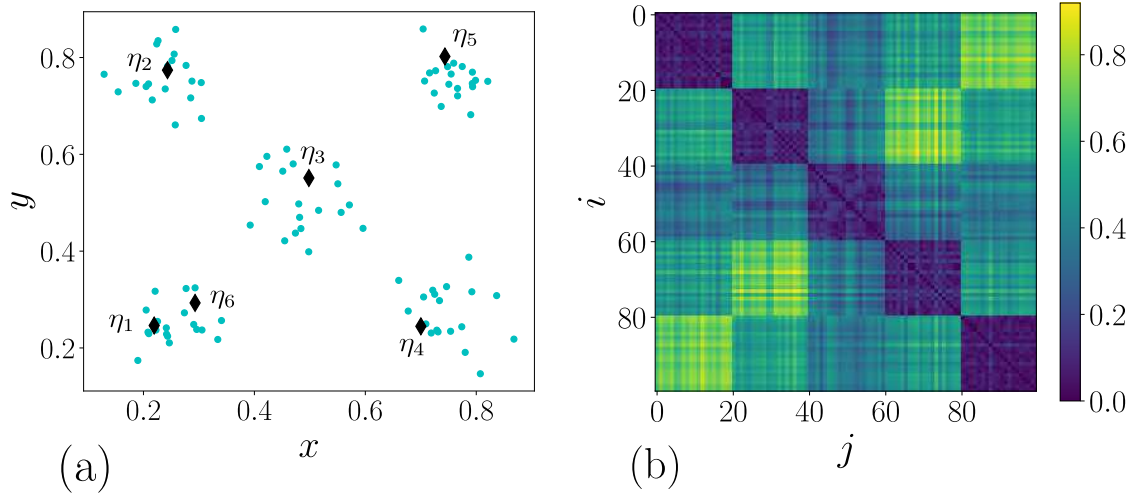}
	\end{center}
	\vspace{-5mm}
	\caption{(a) Random points distributed in space and resetting nodes. (b) Distance matrix used to calculate the transition matrix}
	\label{fig:points_D_mat}
\end{figure*}

In Figure \ref{fig:points_D_mat}(a) we observe five clusters formed by 20 nodes each, the Gaussian distribution that determines their positions has different values, the one in the center where $\eta_{3}$ is located has standard deviation $\sigma_{3}=0.055$, the bottom left and upper right have $\sigma_{1}=\sigma_{5}=0.04$ and bottom right and upper left have $\sigma_{2}=\sigma_{4}=0.05$. The resetting nodes $\eta_{1},\eta_{2},\eta_{3},\eta_{4},\eta_{5}$ and $\eta_{6}$ where randomly chosen and denoted as black diamonds. In Figure \ref{fig:points_D_mat}(b), we plot the distance matrix $D$ which contains the Euclidean distance $d_{ij}$ between nodes. We observe that the highest value is around the unit, since the points are inside the two-dimensional space $[0,1]\times [0,1]$. There is a global symmetry due to the numeration of nodes but the randomness is evident at a smaller scale.

\begin{figure*}[!t]
	\begin{center}
		\includegraphics*[width=0.7\textwidth]{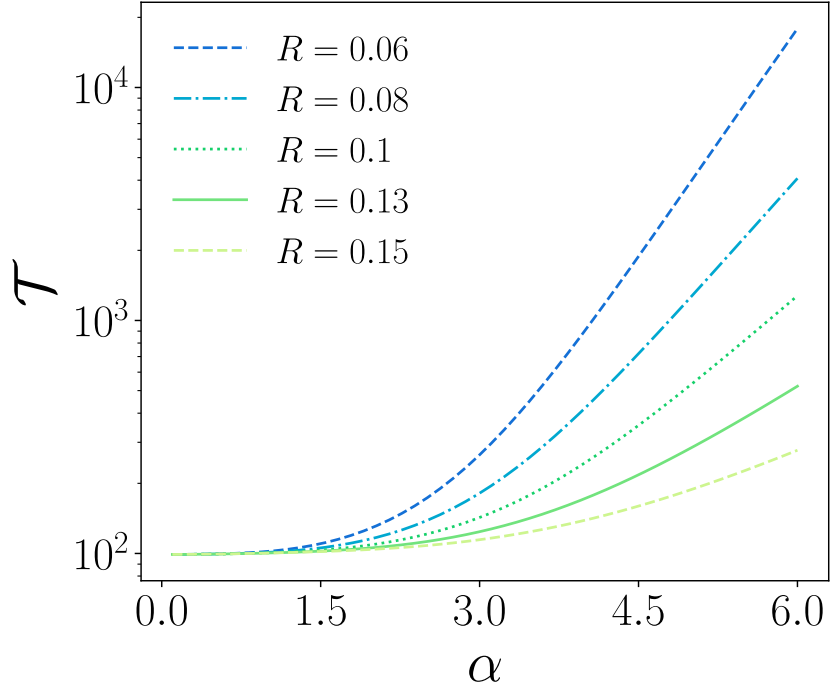}
	\end{center}
	\vspace{-5mm}
	\caption{Global MFPT as a function of  $\alpha$ for different values of $R$ for the L\'evy-like dynamics in Eq. (\ref{eq:transition_points}).}
	\label{fig:points_alpha}
\end{figure*}

Using this particular setting, we calculated the global MFPT for different $R$ for $\beta=0$, ie. without resetting, as a function of the parameter $\alpha$, the results are shown in Figure \ref{fig:points_alpha}. We choose the values of $R$ such that they were less than the percolation limit of a random geometric graph $\sqrt{\frac{\log(N)}{\pi N}}$ \cite{Dall2002}, since it is used as a reference length. We observe that for greater radius, the global MFPT decreases, this behavior is consistent because for greater $R$ the probability of transition is non-null for more nodes.

\begin{figure*}[!t]
	\begin{center}
		\includegraphics*[width=1.0\textwidth]{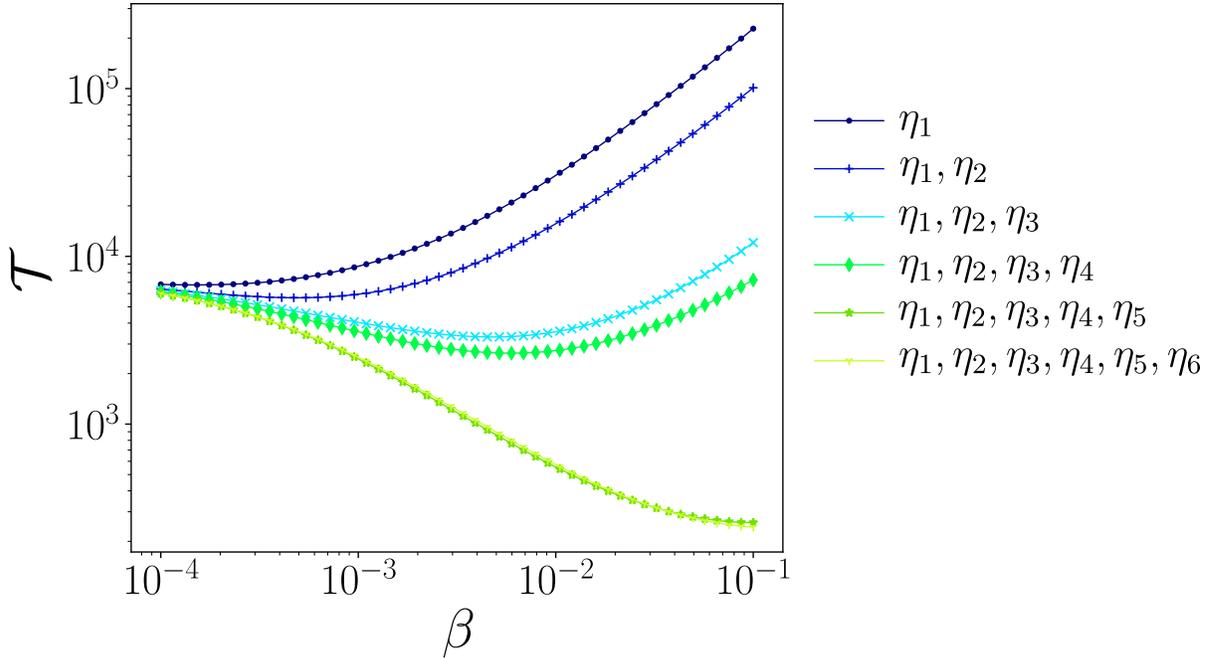}
	\end{center}
	\vspace{-5mm}
	\caption{Global MFPT as a function of total probability $\beta$ for a random distribution of points in space considering different number of resetting nodes $\eta_{i}$, $i=1,2,3,4,5,6$, with $\alpha=5$ and $R=0.05$}
	\label{fig:points_beta}
\end{figure*}

In Figure \ref{fig:points_beta} we present the global MFPT as a function of the total probability $\beta$ for $\alpha =5$ and $R=0.05$. For one resetting node, the first curve in dark blue is monotonically increasing, but for two and more resetting points the results show different behavior. For three and four resetting nodes, there is a minimum value and the change in curvature is clear, but five and six resetting points reduce significantly the global MFPT in comparison with the dynamics without resetting at $\beta=0$. Also, the result with $\eta_{i}$, $i=1,2,3,4,5,6$  shows that more than one resetting per cluster is redundant and does not further improve the value $\mathcal{T}$. With these results, we can see that for a small radius, the resetting process optimizes the general exploration of this spatial distribution of points, as expected.

\subsection{\textit{Google} strategy on interacting cycles}

A particular search method is the \textit{Google} random walk strategy, were a local search to nearest-neighbor nodes is combined with a stochastic relocation to any of the nodes, using a constant resetting probability \cite{Brin1998,ShepelyanskyRevModPhys2015}. In our formalism, this type of search requires using $\mathcal{M}=N$ reset nodes and defining $a_{1}=a_{2}=\dots=a_{N}$, considering a total probability $\beta$, the resetting probabilities are given by $a_{s}=\beta/N$. We directly obtain the eigenvalues with Eq. \eqref{eq:gamma_gendef} and \eqref{eigval_reset_ws}, and considering that $1=a_{0}+\beta$

\begin{equation}\label{eigvals_zeta_google}
	\zeta^{\mathrm{Google}}_l=
	\begin{cases}
		1 \qquad &\mathrm{for}\qquad l=1,\\
		(1-\beta)\lambda_l  &\mathrm{for}\qquad l=2,3,\ldots, N.
	\end{cases}
\end{equation}
Remembering that $\lambda_{l}$ is the corresponding eigenvector of $\mathbf{W}$ which is the transition matrix without resetting and carries the information about the structure of the network.

The uniform resetting in a regular network produces a constant stationary distribution, such that
\begin{equation}
	P_{j}^{\infty}(\vec{r},\vec{\gamma})=\frac{1}{N},
\end{equation}
so that the global MFPT in Eq. \eqref{eq:global_MFPT} for regular networks is
\begin{equation} \label{eq:gMFPT_google_vecs}
	\mathcal{T}=\frac{1}{N^2}\sum_{i=1}^N\sum_{j=1}^N \langle T_{ij}\rangle=\frac{1}{N}\sum_{i=1}^N\sum_{j=1}^N\left[\delta_{ij}+\sum_{l=2}^{N}\frac{\langle j|\phi_{l}\rangle \langle \bar{\phi}_{l}|j\rangle - \langle i | \phi_{l}\rangle \langle \bar{\phi}_{l}|j\rangle}{1-(1-\beta)\lambda_{l}}\right].
\end{equation}
We can  further simplify this expression. Considering just the first term, we have
\begin{equation} \label{eq:first_term}
	\frac{1}{N}\sum_{i=1}^N\sum_{j=1}^N\delta_{ij} = \frac{1}{N}\sum_{i=1}^N 1 = 1
\end{equation}

For the second term, we swap the sums and separate the subtraction such that
{\small
	\begin{multline} \label{eq:second_term}
		\frac{1}{N}\Bigg[\sum_{l=2}^{N}\sum_{i=1}^N\sum_{j=1}^N\frac{\langle j|\phi_{l}\rangle \langle \bar{\phi}_{l}|j\rangle}{1-(1-\beta)\lambda_{l}}-
		\sum_{l=2}^{N}\sum_{i=1}^N\sum_{j=1}^N\frac{\langle i | \phi_{l}\rangle \langle \bar{\phi}_{l}|j\rangle}{1-(1-\beta)\lambda_{l}}\Bigg] =\\
		\frac{1}{N}\Bigg[\sum_{l=2}^{N}\frac{1}{1-(1-\beta)\lambda_{l}}\sum_{j=1}^N\langle j|\phi_{l}\rangle \langle \bar{\phi}_{l}|j\rangle\sum_{i=1}^N 1 -
		\sum_{l=2}^{N}\frac{1}{1-(1-\beta)\lambda_{l}}\sum_{i=1}^{N}\langle i | \phi_{l}\rangle \sum_{j=1}^{N}\langle \bar{\phi}_{l}|j\rangle\Bigg],
	\end{multline}
}
where in the last equality, the terms independent of $i$ and $j$ were separated when possible. Since $\sum_{j=1}^N \langle \bar{\phi}_{l}|j\rangle=0$ for $l=2,3,\ldots,N$, the second term vanishes. To express the sum over $j$ in the remaining term, we have that
\begin{equation*}
	\sum_{j=1}^{N}\langle j|\phi_{l}\rangle \langle \bar{\phi}_{l}|j\rangle=\sum_{j=1}^{N}\langle \bar{\phi}_{l}|j\rangle\langle j|\phi_{l}\rangle = \langle \bar{\phi}_{l}|\bigg(\sum_{j=1}^{N}|j\rangle\langle j|\bigg)|\phi_{l}\rangle=\langle \bar{\phi}_{l}|\phi_{l}\rangle=1 .
\end{equation*}
In this relation, we explicitly used the completeness of the $N-$dimensional space and the orthonormality condition between $\langle \bar{\phi}_{l}|$ and $|\phi_{l}\rangle$. Substituting Eqs. \eqref{eq:first_term} and \eqref{eq:second_term} in Eq. \eqref{eq:gMFPT_google_vecs}, we finally obtain
\begin{equation}\label{tauglobal_google}
	\mathcal{T}(\beta)=1+\sum_{l=2}^N\frac{1}{1-(1-\beta)\lambda_l}.
\end{equation}
\begin{figure*}[!t]
	\begin{center}
		\includegraphics*[width=0.9\textwidth]{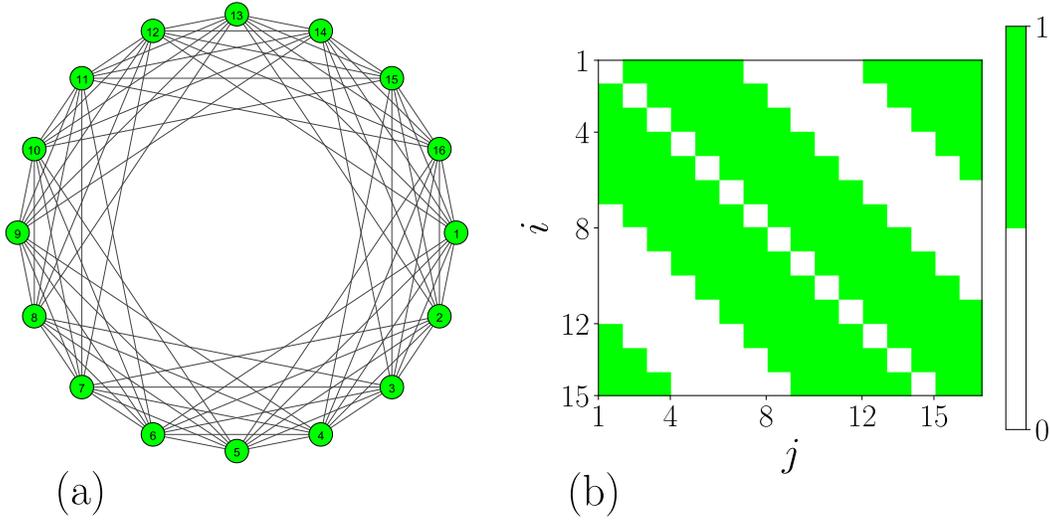}
	\end{center}
	\vspace{-5mm}
	\caption{(a) Interacting cycle with $N=16$ and $J=5$ (b) the corresponding adjacency matrix}
	\label{fig:interacting_cycles_adj}
\end{figure*}
Observe that this simple equation is valid for any regular network, since for its derivation the principles of the \textit{Google} search strategy and basic properties of the orthonormal bases were used, also due to the fact that the stationary distribution is constant for all nodes. The dependence of the structure is expressed in the eigenvalues $\lambda_{l}$.
\\[2mm]
In Section \ref{subsec:circulant_graphs} we studied circulant graphs whose adjacency and transition matrices are circulant \cite{mieghem2011graph_spectra_circulant}. A special case of this type of networks are interacting cycles and in this section we will apply the multi-resetting formalism to this structure using the \textit{Google} strategy.

\begin{figure}[!t]
	\begin{center}
		\includegraphics[width=0.9\textwidth]{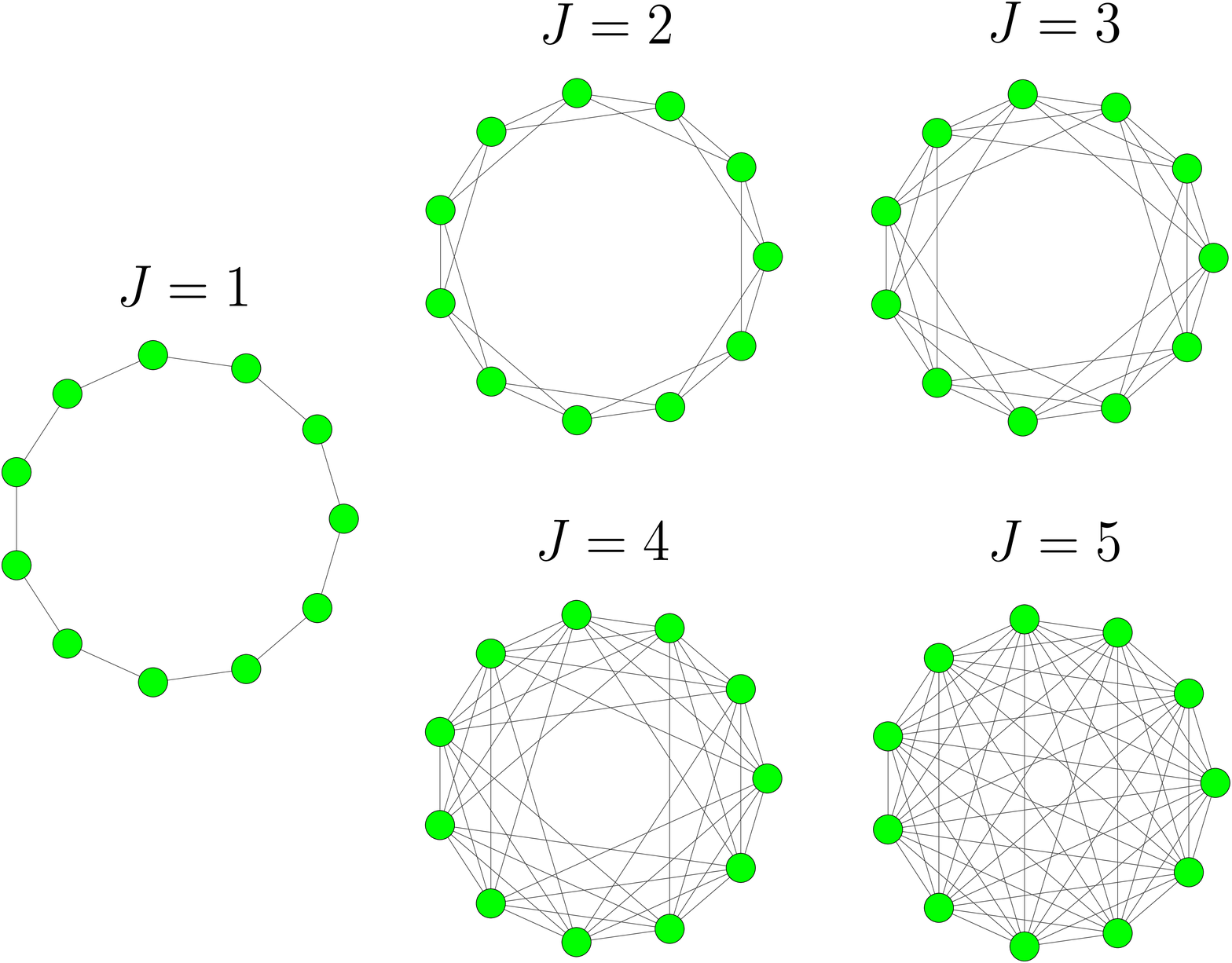}
	\end{center}
	\vspace{-4mm}
	\caption{
		\label{FigCycles} Interacting cycles with $N=11$. For $J=1$ we obtain the initial ring; in this network we add links in order to connect each node to its $J$ left and $J$ right nearest nodes. The value $J=5$ defines a fully connected graph.}
\end{figure}

An interacting cycle has a layout based in a simple ring with periodic boundary conditions, additionally, each node is connected to $J$ nearest neighbors to the left and $J$ to the right, therefore having degree $2J$ \cite{FractionalBook2019}. In Figure \ref{fig:interacting_cycles_adj}(a) we observe an interacting cycle with $N=16$ nodes and $J=5$ and to the right in (b) the adjacency matrix, where the colored region represents a link between nodes. The value of $J$ is called the interaction parameter, in order to have just one edge between two nodes $J$ is restricted to $1\le J \le (N-1)/2$. With $J=1$ we recover the simple ring and for the other extreme we have a fully connected graph as we can see in Figure \ref{FigCycles}.

The eigenvalues of $\mathbf{W}$, $\lambda_{m}$, can be obtained exactly, as shown in Appendix \ref{app:B}, considering properties of circulant matrices, the complete expression is
\begin{equation}\label{spectJNN}
	\lambda_m=\frac{1}{2J}\left[\frac{\sin\left[\frac{\pi}{N}(m-1)(2J+1)\right]}{\sin\left[\frac{\pi}{N}(m-1)\right]}-1\right]
	\qquad \mathrm{for}\qquad m=2,\ldots, N.
\end{equation}
From the particular value $J=1$, we obtain the spectra of a ring with $N$ nodes, using the trigonometric identities $\sin(3x)=3\cos^2(x)\sin(x)-\sin^{3}(x)$, $\sin^2(x)+\cos^{2}(x)=1$ and $2\cos^{2}(x/2)=1+\cos(x)$.  Consequently, the exact eigenvalues can be substituted in Eq. \eqref{tauglobal_google} and we can calculate directly the global MFPT.

\begin{figure}[!t]
	\begin{center}
		\includegraphics[width=0.85\textwidth]{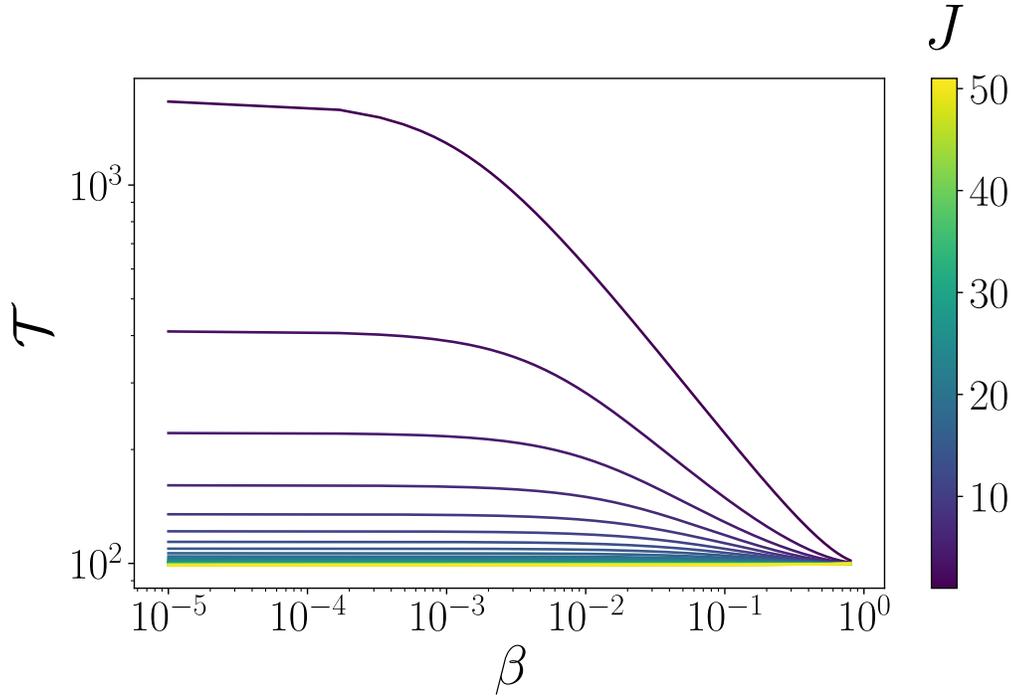}
	\end{center}
	\vspace{-4mm}
	\caption{Global time $\mathcal{T}$ as a function of total probability $\beta$ for the \textit{Google} search strategy on interacting cycles with $N=100$ and different values of the interaction parameter $J$.} \label{FigTglobalGoogle}
\end{figure}

In Figure \ref{FigTglobalGoogle} we observe several curves that correspond to the global MFPT as a function of $\beta$ for different values of $J$. The behavior of the lines are very different compared to the other networks and resetting strategies. The value of $\mathcal{T}(\beta)$ tends to decrease as $\beta$ approaches to 1. The curves do not show a global minimum, instead they tend to agglomerate around $10^{2}$. For low values of $J$ (purple and darker colors) we observe variation in the order, while for high $J$ (yellow and lighter colors) all curves remain bounded around $10^{2}$.

With these three different simulations, we have explored and applied the method to different strategies and structures, observing that generally, the resetting process might optimize the network exploration.

%% file: Chapters/Chap_Conc.tex
\chapter*{Conclusions}
\addcontentsline{toc}{chapter}{Conclusions}

In this work, we deduced analytical results for the stationary distribution and mean first passage time (MFPT) for Markovian random walks with local transitions and long-range dynamics with stochastic resetting to multiple nodes in different networks. For an arbitrary number of resetting nodes, the derivation of the eigenvalues and eigenvectors of the transition matrix for the dynamics with resetting can be calculated iteratively using the information of the dynamics without resetting $\mathbf{W}$. The introduction of resetting in the dynamics affects the stationary distribution and MFPTs, as is easily generalized in Eq. \eqref{MFPT_Mreset}.

For Lévy flights with parameter $\alpha$ on an infinite ring with one resetting, the MFPT behaves proportionally to the power $1+2\alpha$ of the distance between nodes, similar behaviors are observed for resetting to two nodes. For Cayley trees, we found that particular values of the total resetting probability $\beta$ optimizes the global MFPT. We obtained a similar result for the random distribution of points where for a small radius the exploration is benefited by the resetting process. For the interacting cycles with the \textit{Google} strategy, it is shown that for greater $J$ the global MFPT reduces having a lower bound at $\mathcal{T}\approx N^2$.

The methods developed in this research provide a general framework to study different dynamics with resetting to multiple nodes with applications in the modeling of routines in animal foraging, human mobility, among many others.

%% file: Chapters/appendix.tex
%\appendix
\begin{appendices}
%\section{Appendices}

\chapter{Calculating the MFPT for $\mathcal{M}$ resetting nodes by induction} \label{app:A}
We can prove by induction that the MFPT for $\mathcal{M}$ resetting nodes
\begin{multline}
 \label{eq:MFPT_general_M}
    \langle T_{ij}(\vec{r};\vec{\gamma})\rangle=\frac{\delta_{ij}}{P_{j}^{\infty}(\vec{r};\vec{\gamma})}
+\frac{\mathcal{R}_{jj}^{(0)}(\vec{r};\vec{\gamma})-\mathcal{R}_{ij}^{(0)}(\vec{r};\vec{\gamma})}{P_{j}^{\infty}(\vec{r};\vec{\gamma})}\\
=\frac{\delta_{ij}}{P_{j}^{\infty}(\vec{r};\vec{\gamma})}+\frac{1}{P_{j}^{\infty}(\vec{r};\vec{\gamma})}\sum_{l=2}^N  \frac{\langle j |\psi^{(\mathcal{M})}_l\rangle\langle\bar{\psi}^{(\mathcal{M})}_l|j\rangle-\langle i |\psi^{(\mathcal{M})}_l\rangle\langle\bar{\psi}^{(\mathcal{M})}_l|j\rangle}{1-\zeta_{l}^{(\mathcal{M})}}
\end{multline}
reduces to
\begin{equation} \label{eq:final_to_prove}
    \langle T_{ij}(\vec{r};\vec{\gamma})\rangle=
    \frac{\delta_{ij}}{P_{j}^{\infty}(\vec{r};\vec{\gamma})}+\frac{1}{P_{j}^{\infty}(\vec{r};\vec{\gamma})} \sum_{l=2}^{N}\frac{\langle j|\phi_{l}\rangle \langle \bar{\phi}_{l}|j\rangle - \langle i | \phi_{l}\rangle \langle \bar{\phi}_{l}|j\rangle}{1-z(\vec{\gamma})\lambda_{l}},
\end{equation}

where $\zeta_{l}^{(\mathcal{M})}$ are the eigenvalues, $|\psi^{(\mathcal{M})}_l\rangle$ are the right eigenvectors and $\langle\bar{\psi}^{(\mathcal{M})}_l|$ the left eigenvectors of $\mathbf{\Pi}_{\mathcal{M}}$. In addition, $\lambda_{l}$ are the eigenvalues, $|\phi_{l}\rangle$ are the right eigenvectors and $\langle \bar{\phi}_{l}|$ are the left eigenvectors of $\mathbf{W}$, the transition matrix without resetting.

The key of the problem is the numerator in the fraction inside the sum. In the following we prove that the equality
\begin{equation} \label{eq:nominator1}
    \langle j |\psi^{(\mathcal{M})}_l\rangle\langle\bar{\psi}^{(\mathcal{M})}_l|j\rangle-\langle i |\psi^{(\mathcal{M})}_l\rangle\langle\bar{\psi}^{(\mathcal{M})}_l|j\rangle=\langle j|\phi_{l}\rangle \langle \bar{\phi}_{l}|j\rangle - \langle i | \phi_{l}\rangle \langle \bar{\phi}_{l}|j\rangle
\end{equation}
holds for any $s$. In this manner, we deduce directly the result in Eq. \eqref{eq:final_to_prove}.  We start by rewriting the left eigenvectors for $m=2,3,\ldots,N$ (Eq. \eqref{lpsi_reset_s})
\begin{equation} \label{eq:left_s}
    \langle\bar{\psi}^{(s)}_m|=\langle\bar{\phi}_m|
\end{equation}
and right eigenvectors (Eq. \eqref{psi_s_Mnodes}) , for a given $s>1$ and $l=2,\ldots, N$
\begin{equation} \label{eq:right_s}
       |\psi^{(s)}_{l}\rangle = |\psi^{(s-1)}_{l}\rangle-\frac{\gamma_s}{1-\zeta_l^{(s)}}\frac{\langle r_{s}|\psi^{(s-1)}_{l}\rangle}{\langle r_{s}|\phi_{1}\rangle}|\phi_{1}\rangle.
\end{equation}
We have already shown that Eq. \eqref{eq:final_to_prove} holds for $s=1,2$ so we take the last as our base case.
We assume valid for arbitrary $s$, such that
\begin{equation} \label{eq:nominator2}
    \langle j |\psi^{(s)}_l\rangle\langle\bar{\psi}^{(s)}_l|j\rangle-\langle i |\psi^{(s)}_l\rangle\langle\bar{\psi}^{(s)}_l|j\rangle=\langle j|\phi_{l}\rangle \langle \bar{\phi}_{l}|j\rangle - \langle i | \phi_{l}\rangle \langle \bar{\phi}_{l}|j\rangle.
\end{equation}
Now, we demonstrate it holds for $s+1$. Directly from Eq. \eqref{eq:left_s} we get
\begin{equation} \label{eq:left_j_s1}
    \langle\bar{\psi}^{(s+1)}_l|j\rangle=\langle\bar{\phi}_l|j\rangle
\end{equation}
for the other factor, we use Eq. \eqref{eq:right_s} with $s+1$
\begin{multline}
    \langle j |\psi^{(s+1)}_l\rangle-\langle i |\psi^{(s+1)}_l\rangle= \langle j |\psi^{(s)}_l\rangle-\frac{\gamma_{s+1}}{1-\zeta_{l}^{(s+1)}}\frac{\langle r_{s+1}|\psi_{l}^{(s)}\rangle}{\langle r_{s+1}|\phi_{1} \rangle}\langle j|\phi_{1}\rangle\\
    -\langle i |\psi^{(s)}_l\rangle+\frac{\gamma_{s+1}}{1-\zeta_{l}^{(s+1)}}\frac{\langle r_{s+1}|\psi_{l}^{(s)}\rangle}{\langle r_{s+1}|\phi_{1} \rangle}\langle i|\phi_{1}\rangle.
\end{multline}
Observe that $\langle j|\phi_{1}\rangle=\langle i|\phi_{1}\rangle$ since $|\phi_{1}\rangle$ defines the stationary distribution independent of the initial condition. Therefore the second and fourth term cancel out to get
\begin{equation}
    \langle j |\psi^{(s+1)}_l\rangle-\langle i |\psi^{(s+1)}_l\rangle= \langle j |\psi^{(s)}_l\rangle
    -\langle i |\psi^{(s)}_l\rangle.
\end{equation}
Now, considering Eq. \eqref{eq:left_j_s1} and the induction step, it is direct to see that
\begin{equation}
    \langle j |\psi^{(s+1)}_l\rangle\langle\bar{\psi}^{(s+1)}_l|j\rangle-\langle i |\psi^{(s+1)}_l\rangle\langle\bar{\psi}^{(s+1)}_l|j\rangle=\langle j|\phi_{l}\rangle \langle \bar{\phi}_{l}|j\rangle - \langle i | \phi_{l}\rangle \langle \bar{\phi}_{l}|j\rangle,
\end{equation}
proving it valid for any $s$.

\chapter{Eigenvalues $\lambda_{m}$ for interacting cycles} \label{app:B}

Consider the adjacency matrix $\mathbf{A}$ for an interacting cycle with $N$ nodes and interaction parameter $J$. In Section \ref{subsec:circulant_graphs} we briefly introduced the structure of a circulant matrix, where it was presented as the sum of elementary matrices, as if it were a polynomial (Eq. \eqref{eq:circulant_poli}). Denoting it as $p(z)=\sum_{j=0}^{N}c_{j}z^{j}$,  the spectrum of the circulant matrix can be found evaluating  \cite{mieghem2011graph_spectra_circulant}
\begin{equation} \label{eq:eigenvals_poli}
    \lambda_{m}=p(\xi ^{1-m}),
\end{equation}
where $\xi=e^{-2\pi\mathrm{i}/N}$.

Since every node has degree $2J$, the transition $\mathbf{W}$ and adjacency $\mathbf{A}$ matrices are circulant and follow
\begin{equation} \label{eq:cycles_W_A}
    \mathbf{W}=\frac{1}{2J}\mathbf{A}.
\end{equation}

From now on, we are going to work with $\mathbf{A}$. Because there are no self loops (as shown in Figure \ref{fig:interacting_cycles_adj}(b)), particularly for the first node, $c_{0}=0$, splitting the terms in the polynomial according to $J$ and rearranging the indexes we have
\begin{equation}
    p(z)=\sum_{j=1}^{J}c_{j}z^{j}+\sum_{j=J+1}^{N-1}c_{j}z^{j}=\sum_{j=1}^{J}c_{j}z^{j}+\sum_{j=1}^{N-J-1}c_{N-j}z^{N-j}.
\end{equation}

Because the matrix is circulant, the coefficients satisfy $c_{j}=c_{N-j}$ and $c_{j}=1$ for $j\le J$ and zero for the rest, also due to the bound over $J$ we obtain
\begin{equation}
    p(z)=\sum_{j=1}^{J}z^{j}+z^{N}\sum_{j=1}^{J}z^{-j}
\end{equation}

We can calculate the terms in a closed form considering as a finite power sum \cite{fitzpatrick2009advanced_sum}

\begin{equation} \label{eq:geo_sum}
    \sum_{j=0}^{n}r^{j}=\frac{1-r^{n+1}}{1-r}
\end{equation}

To have the exact expression, we can factor out a $z$ in the first sum, $z^{-1}$ from the second and reordering the indexes we get

\begin{equation}
    p(z)=z\sum_{j=0}^{J-1}z^{j}+z^{N-1}\sum_{j=0}^{J-1}z^{-j},
\end{equation}
substituting Eq. \eqref{eq:geo_sum}, the result is
\begin{equation}
    p(z)=z\frac{1-z^{J}}{1-z}+z^{N-1}\frac{1-z^{-J}}{1-z^{-1}}.
\end{equation}

Evaluating $\xi^{(1-m)}=e^{-2\pi\mathrm{i}(1-m)/N}$ according to Eq. \eqref{eq:eigenvals_poli} we obtain
\begin{align}
    p(\xi^{1-m})= &\xi^{1-m}\frac{1-\xi^{J-Jm}}{1-\xi^{(1-m)}}+\xi^{(1-m)(N-1)}\frac{1-\xi^{-(J-Jm)}}{1-\xi^{-(1-m)}}\\
    =&e^{-2\pi\mathrm{i}(1-m)/N}\frac{1-e^{-2\pi\mathrm{i}(J-Jm)/N}}{1-e^{-2\pi\mathrm{i}(1-m)/N}}+e^{-2\pi\mathrm{i}(1-m)(N-1)/N}\frac{1-e^{2\pi\mathrm{i}(J-Jm)/N}}{1-e^{2\pi\mathrm{i}(1-m)/N}}\\
    =&e^{-2\pi\mathrm{i}(1-m)/N}\frac{1-e^{-2\pi\mathrm{i}(J-Jm)/N}}{1-e^{-2\pi\mathrm{i}(1-m)/N}}+e^{2\pi\mathrm{i}(1-m)/N}\frac{1-e^{2\pi\mathrm{i}(J-Jm)/N}}{1-e^{2\pi\mathrm{i}(1-m)/N}},
\end{align}
we observe that the second term is the complex conjugate of the first, then we get the double of the real part. Also, rewriting the complex fraction in terms of the sine function
\begin{equation}
    \frac{1-e^{2\pi\mathrm{i}(J-Jm)/N}}{1-e^{2\pi\mathrm{i}(1-m)/N}}=e^{\pi\mathrm{i}(1-m)(J-1)/N}\frac{\sin\big(\frac{\pi(1-m)J}{N}\big)}{\sin\big(\frac{\pi(1-m)}{N}\big)}.
\end{equation}

So, we get
\begin{align}
    p(\xi^{1-m})=2\frac{\sin\big(\frac{\pi(m-1)J}{N}\big)}{\sin\big(\frac{\pi(m-1)}{N}\big)}\cos\bigg(\frac{\pi (1-m)(J+1)}{N}\bigg)\\
    =2\frac{\sin\big(\frac{\pi(m-1)J}{N}\big)}{\sin\big(\frac{\pi(m-1)}{N}\big)}\cos\bigg(\frac{\pi (m-1)(J+1)}{N}\bigg)
\end{align}
using the formula
\begin{equation}
    \sin(a)\cos(b)=\frac{1}{2}[\sin(a+b)+\sin(a-b)]
\end{equation}
with $a=\frac{\pi(m-1)J}{N}$ and $b=\frac{\pi(m-1)(J+1)}{N}$ we get
\begin{equation}
    p(\xi^{1-m})= \frac{\sin\big(\frac{\pi(m-1)(2J+1)}{N}\big)-\sin\big(\frac{\pi(m-1)}{N}\big)}{\sin\big(\frac{\pi(m-1)}{N}\big)}
    .
\end{equation}

Remembering that this are the eigenvalues of $\mathbf{A}$, the eigenvalues of $\mathbf{W}$ are
\begin{equation}
    \lambda_{m}=\frac{1}{2J}\bigg[\frac{\sin(\frac{\pi}{N}(m-1)(2J+1))}{\sin(\frac{\pi}{N}(m-1))}-1\bigg]
\end{equation}
considering Eq. \eqref{eq:cycles_W_A}.
\end{appendices}